\documentclass[11pt,english]{article}

\usepackage[utf8]{inputenc} 

\usepackage{dsfont}

\usepackage[left=1in,right=1in,top=1in,bottom=1in]{geometry}
\usepackage{authblk}
\usepackage{enumitem}
\usepackage{ulem} 
\usepackage{empheq}

\usepackage{graphicx}
\usepackage{subcaption}
\usepackage{float}
\usepackage{wrapfig} 

\usepackage{xcolor}
\usepackage{color}
\definecolor{darkgreen}{rgb}{0,0.5,0}
\definecolor{darkblue}{rgb}{0,0,0.6}
\definecolor{purple}{rgb}{0.4,.2,0.7}

\usepackage{amsmath}
\usepackage{amssymb}

\usepackage{tensor} 
\usepackage{slashed}
\usepackage{mathtools}
\usepackage{leftidx}
\usepackage{esint}
\usepackage{amsfonts,amsthm,bm}

\usepackage{physics}
\usepackage{feynmf}
\usepackage{braket}

\usepackage{prettyref}
\usepackage[colorlinks=true,citecolor=darkgreen,linkcolor=purple,urlcolor=purple]{hyperref}

\usepackage[numbers,sort&compress]{natbib}

\usepackage{tikz}  
\newcommand \mathtikz[1] {\quad \vcenter{\hbox{\tikz{#1}}} \quad}

\newcommand\etaC[2] { 
\begin{scope}[xshift=#1,yshift=#2]
\filldraw[right color=white,left color=lightgray] (-0.25,0) to [out=90,in=180] (0,0.33) to [in=90,out=0] (0.25,0) to [in=-90,out=-90] (-0.25,0);
\draw[dotted] (0.25,0) arc (0:180:0.25 and 0.1);
\end{scope}
}

\newcommand{\nn}{\nonumber}

\def\p{\partial}
\def\half{\frac{1}{2}}
\def\be{\begin{equation}}
\def\ee{\end{equation}}
\def\ba{\begin{aligned}}
\def\ea{\end{aligned}}
\def\mc{\mathcal}
\def\eps{\varepsilon}
\def\sig{\sigma}
\def\lam{\lambda}
\def\weq{{\;\hat =\;}}
\def\Gam{\Gamma}

\newcommand{\inner}[2]{\langle \, #1 \mid #2 \, \rangle}

\numberwithin{equation}{section}
\numberwithin{figure}{section}
\numberwithin{table}{section}

\begin{document}

\title{\LARGE\textsc{Dynamical Edge Modes and Entanglement \\ in Maxwell Theory}}  


\author[a]{\vskip1cm \normalsize Adam Ball}
\affil[a]{\it \normalsize Perimeter Institute for Theoretical Physics, Waterloo, ON, Canada}
\author[b,c,d]{\normalsize Y.T.\ Albert Law}
\affil[b]{\it \normalsize Center for the Fundamental Laws of Nature, Harvard University, Cambridge, MA, USA}
\affil[c]{\it \normalsize Black Hole Initiative, Harvard University, Cambridge, MA, USA}
\affil[d]{\it \normalsize Stanford Institute for Theoretical Physics, Stanford, CA, USA}
\author[e,f]{Gabriel Wong}
\affil[e]{\it \normalsize Harvard CMSA, Cambridge, MA, USA}
\affil[f]{\it \normalsize Oxford University Math institute, Oxford, UK}
\date{}
\maketitle

\begin{center}
	\vskip-10mm
	{\footnotesize \href{mailto:aball1@pitp.ca}{aball1@pitp.ca}\,,\;  \href{mailto:ytalaw@stanford.edu}{ytalaw@stanford.edu}\,,\;  \href{mailto:wong@maths.ox.ac.uk}{wong@maths.ox.ac.uk} }
\end{center}

\vskip10mm

\thispagestyle{empty}

\begin{abstract}

Previous work on black hole partition functions and entanglement entropy suggests the existence of ``edge" degrees of freedom living on the (stretched) horizon. We identify a local and ``shrinkable'' boundary condition on the stretched horizon that gives rise to such degrees of freedom. They can be interpreted as the Goldstone bosons of gauge transformations supported on the boundary, with the electric field component normal to the boundary as their symplectic conjugate. Applying the covariant phase space formalism for manifolds with boundary, we show that both the symplectic form and Hamiltonian exhibit a bulk-edge split. We then show that the thermal edge partition function is that of a codimension-two ghost compact scalar living on the horizon.  In the context of a de Sitter static patch, this agrees with the edge partition functions found by Anninos et al. in arbitrary dimensions. It also yields a 4D entanglement entropy consistent with the conformal anomaly. Generalizing to Proca theory, we find that the prescription of Donnelly and Wall reproduces existing results for its edge partition function, while its classical phase space does not exhibit a bulk-edge split.

\end{abstract}

\newpage

\tableofcontents


\section{Introduction}

Edge modes are degrees of freedom associated with gauge transformations supported on the boundary. They have been studied extensively
\cite{Donnelly:2011hn, Donnelly:2012st, Radicevic:2014kqa, Donnelly:2014gva, Donnelly:2014fua, Huang:2014pfa, Ghosh:2015iwa, Hung:2015fla, Aoki:2015bsa, Donnelly:2015hxa, Radicevic:2015sza, Pretko:2015zva, Soni:2015yga, Donnelly:2016auv, Zuo:2016knh, Soni:2016ogt, Donnelly:2016jet, Donnelly:2016mlc, Donnelly:2016qqt, Agarwal:2016cir, Geiller:2017xad, Wong:2017pdm, Geiller:2017whh, Blommaert:2018rsf, Blommaert:2018oue, Barnich:2018zdg, Freidel:2018fsk, Lin:2018xkj, Gomes:2018dxs, Lin:2018bud, Donnelly:2018ppr, Freidel:2019ees, Gomes:2019xto, Geiller:2019bti, Hung:2019bnq, Barnich:2019qex, Freidel:2020svx, Freidel:2020ayo, Donnelly:2020teo, Donnelly:2020xgu, Ciambelli:2021vnn, Freidel:2021cjp, Carrozza:2021gju, GBarbero:2022gnx, Ciambelli:2022cfr, Mertens:2022ujr, Donnelly:2022kfs, Cheng:2023bcv, Cheng:2023cms, Chen:2023tvj, Mukherjee:2023ihb, Wong:2023bhs, Balasubramanian:2023dpj}
and are an indispensable ingredient in important phenomena including entanglement entropy and black hole entropy, but the understanding of edge modes in continuum quantum field theory (QFT) is rather incomplete. In this paper we push it forward on several fronts. We focus on pure Maxwell theory, i.e. QED with no matter, in $D\geq 2$ dimensions as a tractable arena. A key assumption in the usual definition of entanglement entropy in terms of von Neumann entropy is the factorization of the global Hilbert space into a product over subregions, i.e. if we partition our Cauchy surface as $Y = \Sigma \cup \overline \Sigma$ then na\"ively
\be \mc{H}_Y \stackrel{?}{=} \mc{H}_\Sigma \otimes \mc{H}_{\overline \Sigma} \ . \ee
However, all QFTs have ultraviolet (UV) obstructions to this factorization. These obstructions underlie the familiar divergence structure of entanglement entropy, with the leading term being proportional to the area of the entangling surface $\p \Sigma$ \cite{Sorkin:1984kjy, Bombelli:1986rw, Srednicki:1993im}. In gauge theories and perturbative gravity there are additional obstructions to Hilbert space factorization of a more infrared (IR) nature \cite{Donnelly:2011hn, Donnelly:2014gva}. These can variously be viewed as arising from constraints, edge charge matching conditions, or the existence of extended gauge invariant objects (e.g. Wilson loops) that cross the entangling surface. Sharply characterizing and resolving these IR issues, in both the classical and quantum theories, is our main interest in this paper. Edge modes are at the heart of this endeavor.

The UV obstructions to factorization have a well-known manifestation in the physics of a local observer with access only to the causal domain $\mc{D}(\Sigma)$ of $\Sigma$. The boundary of $\mc{D}(\Sigma)$ is a horizon where the local observer's sense of time breaks down, and the associated redshift leads to a continuous spectrum and certain pathologies.\footnote{Heuristically, the UV obstructions to the global factorization can be thought of as coming from having infinitely many modes localized near the horizon. These modes are UV with respect to global time, but because of the horizon redshift they have finite energy with respect to the local observer's time and are instead associated with a continuous spectrum.} One way forward is through algebraic QFT, eschewing Hilbert spaces altogether and working only with the local operator algebra on $\mc{D}(\Sigma)$. The algebra is Type III, which means that there are no pure states, no density matrices, no traces, and no notion of entanglement entropy. Instead one studies finite quantities like the relative entropy, which can be defined in terms of a modular operator that replaces the conventional notion of a density matrix.  However, it is not clear how the local operator approach can address IR issues related to the presence of extended operators that cross the entangling surface. 

A different approach to subregion physics was taken in \cite{Anninos:2020hfj} for the static patch of $D$-dimensional de Sitter space. They tamed the continuous subregion spectrum by devising a ``quasicanonical" thermal trace on $\Sigma$. They evaluated it for various free QFTs and compared the results with the Euclidean partition function on $S^D$, which is the Wick rotation of the static patch. They found agreement between the two methods for scalars and spinors, but for spins one and higher they encountered manifestly codimension-two discrepancies. For Maxwell theory the discrepancy took the form of the partition function of a codimension-two massless scalar. This suggests that their ``quasicanonical" partition functions did not capture the IR effects mentioned above. This presents an extremely sharp puzzle, begging for a careful treatment of these IR effects.
\begin{figure}[H]
    \centering
   \includegraphics[width=0.35\textwidth]{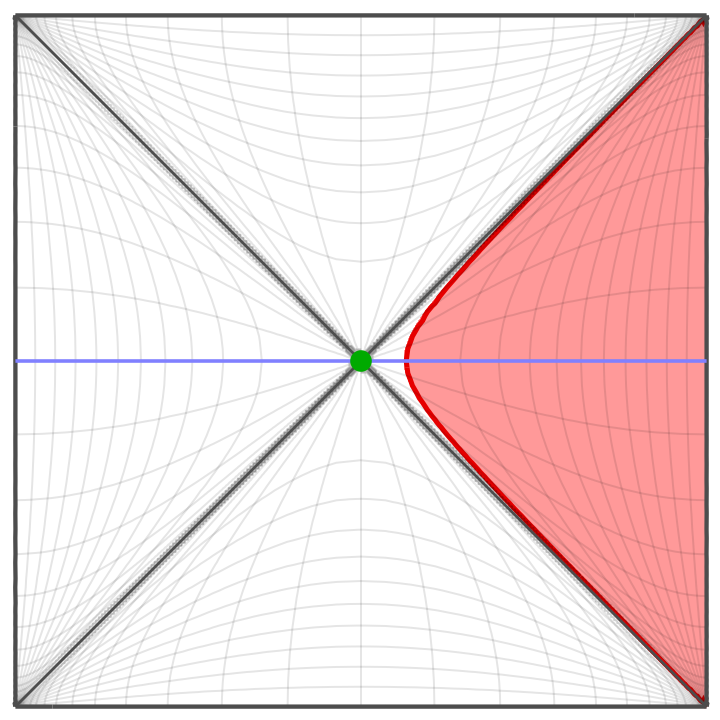}
        \caption{Penrose diagram for the global de Sitter geometry. The brick-wall-regulated subregion is shown in red. A global Cauchy slice is shown in blue.}
        \label{fig:dSintro}
\end{figure}
In this paper we have chosen to regulate $\mc{D}(\Sigma)$ by introducing a brick wall \cite{tHooft:1984kcu}, also known as a stretched horizon, so that the boundary of our subregion is timelike and the flow of time is nonsingular. See figure \ref{fig:dSintro} for the case of de Sitter space. This results in a discrete mode spectrum for the subregion theory and consequently an operator algebra of Type I. The standard tools of QFT then remain in our arsenal, including density matrices and a Hilbert space on $\Sigma$. The main challenge of the brick wall is that once the boundary is timelike, we must choose some boundary condition so that the subregion theory is well-defined. Consistency with the original global, unregulated theory in the limit of removing the brick wall imposes a stringent condition on the set of possible boundary conditions. The Euclidean version of this consistency condition is called shrinkability \cite{Donnelly:2016jet, Donnelly:2018ppr,Hung:2019bnq, Jafferis:2019wkd, Agia:2022srj}, and will be discussed at length in section \ref{sec:shrink}.

In Maxwell theory, neither of the standard perfectly electrically conducting (PEC) or perfectly magnetically conducting (PMC) boundary conditions is shrinkable. This failure can be traced to the fact that they admit no edge modes. This motivates us to introduce what we call the dynamical edge mode (DEM) boundary condition. It gives rise to dynamical edge modes and passes detailed checks on shrinkability. In particular we compute the edge mode contributions to the thermal partition function and entanglement entropy. In four dimensions Maxwell theory is conformal, and Donnelly and Wall proposed in their seminal paper(s) \cite{Donnelly:2014fua, Donnelly:2015hxa} that edge modes resolve a longstanding disagreement in the literature between the entanglement entropy as computed by the conformal anomaly method of \cite{Myers:2010tj, Casini:2011kv} as opposed to with other methods that incorporate only the bulk, propagating degrees of freedom \cite{Dowker:2010bu, Dowker:2010yj, Eling:2013aqa}. See also \cite{Huang:2014pfa, Soni:2016ogt}. However, as we will discuss, the calculation of the edge mode contribution to entanglement entropy in \cite{Donnelly:2014fua, Donnelly:2015hxa} neglects a zero mode effect that changes the universal log term in the entropy. One of our main results is that our edge mode treatment reproduces the entanglement entropy of 4D Maxwell theory even when accounting for this zero mode effect.

Another important distinction between our work and \cite{Donnelly:2015hxa} is the logic behind the introduction of the edge modes. In \cite{Donnelly:2015hxa} edge modes are introduced by hand via a path integral over non-dynamical superselection sectors: from the algebraic point of view this corresponds to introducing center variables that do not have a conjugate partner, and therefore commute with the local algebra. In contrast, we derive a conjugate pair of edge modes as a dynamical consequence of a local, shrinkable boundary condition.  For this reason, we obtain the entanglement entropy as a trace over a subregion Hilbert space, rather than a path integral prescription for integrating over non-dynamical center variables.\footnote{See also \cite{Blommaert:2018oue, Blommaert:2018rsf} for preliminary work in identifying the edge mode phase space in the special case of Rindler space.} Finally we also provide checks on the edge mode contribution to the thermal partition function of the static de Sitter patch in any dimension $D$ by comparing with results of \cite{Anninos:2020hfj}.

\subsubsection*{Overview of the paper}

In section \ref{sec:CovPhaseEdge} we study the covariant phase space formalism \cite{Lee:1990nz, Wald:1993nt, Iyer:1994ys, Iyer:1995kg} for Maxwell theory on a manifold $M$ whose boundary $\p M$ is timelike. We introduce the DEM boundary condition and show that its phase space naturally splits into bulk and edge parts. On a Cauchy slice $\Sigma$ of $M$ the spatial components of the gauge field, subject to a generalized Coulomb gauge, and the electric field, subject to the Gauss constraint, constitute free data for the theory and we split them into bulk and edge parts as
\be A_i = \tilde A_i + \p_i \alpha, \qquad E_i = \tilde E_i + \p_i \beta \; . \ee
We impose certain conditions to make the split unique. The edge modes $\alpha, \beta$ are uniquely determined by their boundary data, and so can be thought of as living on the boundary. The field $\alpha$ transforms by a shift under large gauge transformations, which are gauge transformations with boundary support that are physical under the DEM boundary condition. $\alpha$ is the finite region analogue of the Goldstone modes for large gauge transformations at asymptotic infinity described in \cite{Strominger:2017zoo}. We trade $\beta|_{\p\Sigma}$ for its boundary normal derivative, which has the same information but is more physically relevant; it is the electric flux through the boundary. We denote this by $E_\perp$. The symplectic form reads
\be \ba \Omega & = \int_\Sigma \delta \tilde A^i \delta \tilde E_i + \int_{\p\Sigma} \delta\alpha \, \delta E_\perp  = \Omega_{\rm bulk} + \Omega_{\rm edge} \; . \ea \ee
The bulk phase space precisely corresponds to that of the PMC boundary condition, so that the phase space factorizes as
\be \Gamma_{\rm DEM} = \Gamma_{\rm PMC} \times \Gamma_{\rm edge} \; . \ee
This is one of our main results, and it substantially extends the results of \cite{Donnelly:2016auv} in the case of abelian gauge theory. The Hamiltonian splits as well,
\be H = H_{\rm bulk} + H_{\rm edge} \; , \ee
so that the bulk and edge parts are entirely decoupled and the edge modes can be analyzed as a system in their own right. The edge Hamiltonian depends only on $E_\perp$. After specializing to static backgrounds, we go on to show that when $\p M$ is nearly a bifurcate horizon the kinetic operator in $H_{\rm edge}$ simplifies to a multiple of the inverse Laplacian on the horizon's bifurcation surface. This is a key step in characterizing the edge mode contribution to entanglement entropy as that of a (ghost) scalar on a codimension-two entangling sphere.

In section \ref{sec:EdgeP} we quantize and consider the thermal partition function as a canonical trace. It factorizes into bulk and edge parts,
\be \ba Z_{\rm DEM}(\beta) & = \Tr e^{-\beta H} \\
& = \Tr e^{-\beta H_{\rm bulk}} \Tr e^{-\beta H_{\rm edge}} \\
& = Z_{\rm PMC}(\beta) Z_{\rm edge}(\beta) \; . \ea \ee
The PMC piece captures the bulk, propagating degrees of freedom of the theory. We compute $Z_{\rm edge}(\beta)$ and find that it formally contains a factor of the volume of the group of large gauge transformations. Specifically, with $\Sigma$ as our Cauchy slice we define the group $\mc{G}'_{\p\Sigma}$ as the space of functions from $\p\Sigma$ to $U(1)$ with the constant mode omitted. This group is infinite, and so $Z_{\rm edge}(\beta)$ is as well. This motivates us to define a renormalized trace including a division by $\mc{G}_{\p\Sigma}$,
\be \label{edgetr} \bar Z_{\rm edge}(\beta) \equiv \frac{1}{|\mc{G}_{\p\Sigma}|} \Tr e^{-\beta H_{\rm edge}}\; .  \ee
Note that we are dividing by $\mc{G}_{\p\Sigma}$, which includes the normalizable constant mode, as opposed to $\mc{G}'_{\p\Sigma}$. The quickest argument for this choice is that dividing by $\mc{G}'_{\p\Sigma}$ would be nonlocal, but we offer further justification in sections \ref{sec:shrink} and \ref{sec:further}. We find that in the horizon limit $\bar Z_{\rm edge}(\beta)$ reduces to the reciprocal of the partition function of a codimension-two compact scalar, with the target radius set by the fundamental charge of the $U(1)$ Maxwell theory. This both generalizes and elevates the precision of results from \cite{Huang:2014pfa,Donnelly:2014fua,Donnelly:2015hxa,Soni:2015yga,Blommaert:2018oue,Blommaert:2018rsf,Chen:2023tvj}.

\begin{figure}[h]
    \centering
    \includegraphics[scale=.2]{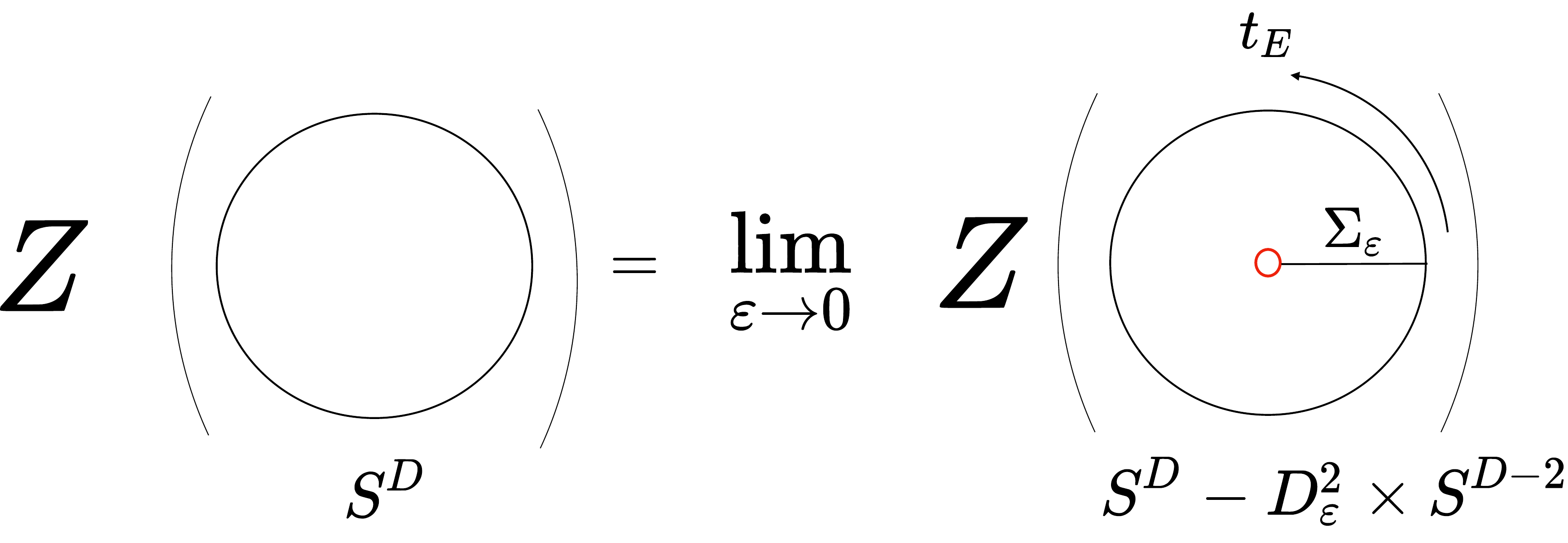}
    \caption{In Euclidean signature, the stretched horizon consists of a non-contractible thermal circle (in red) times a $(D-2)$-dimensional transverse manifold. The boundary condition is defined so that in the shrinking limit, the original path integral without the brick wall is recovered. }
    \label{fig:shrink}
\end{figure}

In section \ref{sec:shrink} we elaborate on the idea of shrinkability of a boundary condition. This is the Euclidean version of recovering the global physics as you remove the brick wall. See figure \ref{fig:shrink}.
We conjecture that our DEM boundary condition for Maxwell theory is shrinkable when one uses the renormalized trace. In section \ref{sec:further} we provide checks in all dimensions by comparing with results from \cite{Anninos:2020hfj} for the partition function of Maxwell theory on the sphere $S^D$. This is another main result, and it underlies the match between our computation of entanglement entropy in 4D Maxwell and the conformal anomaly. The renormalization of the subregion trace is highly constrained, because the same trace must be shrinkable for arbitrary subregions. We give a detailed discussion of how this works in 2D gauge theory, which is completely solvable.

In section \ref{sec:Proca} we address the theory of a massive vector, also known as Proca theory, whose Euclidean black hole partition functions exhibit a bulk-edge split in explicit examples \cite{Anninos:2020hfj,Grewal:2022hlo}. We find that some but not all of the insights for Maxwell carry over. In particular, neither the phase space nor the Hamiltonian splits into bulk and edge parts. In section \ref{sec:global} we come full circle and discuss the global point of view on our results, in particular the consequences for Hilbert space factorization and entanglement entropy. Finally in section \ref{sec:Discussion} we conclude and discuss possible applications and generalizations of our results.

We also include several appendices. Appendix \ref{appendix:nophase} presents a (configuration space) path integral derivation of the edge partition function that makes no reference to phase space. Appendix \ref{appendix:constgauge} clarifies a subtlety about how the covariant phase space formalism ``knows" about the constant gauge transformation. Appendix \ref{app:PIZedge} shows how to compute $Z_{\rm edge}$ with the phase space path integral, complementing the thermal trace calculation in the main text. Appendix \ref{appendix:zmode} carefully reviews the partition function of the 2D massless scalar, focusing in particular on the contribution from the zero mode. Finally appendix \ref{app:sphere_PI} collects relevant results from \cite{Anninos:2020hfj} and adapts them to our notation.

\section{Bulk and edge modes in Maxwell}
\label{sec:CovPhaseEdge}

\subsection{Covariant phase space review}

We begin with a review of the covariant phase space formalism for Maxwell theory on a spatially finite $D$-dimensional Lorentzian manifold $M$ whose boundary $\partial M$ is timelike. For the gauge group we use $G = U(1)$. The fundamental charge $q$ sets the size of $G$. The formalism is ignorant of gauge transformations not in the identity component, so one must augment the formalism and quotient them by hand when they exist. We will avoid this subtlety by assuming for simplicity that the first homology group of $M$ vanishes, i.e. that all loops are homologous to zero, and that $\p M$ has a single connected component. We further choose to work with the trivial gauge bundle; the effect of a nontrivial bundle is simple to incorporate later. Our starting point is the standard Maxwell Lagrangian,\footnote{Changing the Lagrangian by a total derivative, or equivalently changing the action by a boundary term, can affect which boundary conditions are variationally well-defined but does not affect the symplectic form. (At least, not beyond the total-derivative ambiguity that is always inherent in the symplectic potential. See \cite{Harlow:2019yfa, Geiller:2019bti} for a different perspective.) Our boundary conditions of interest will be variationally well-defined for the standard Maxwell Lagrangian.}
\be \label{eq:lag} L = \half F \wedge *F \;, \ee
whose variation is
\be \label{eq:L_var} \ba \delta L & = d\delta A \wedge *F  = \delta A \wedge d\,{*F} + d(\delta A \wedge *F) \; . \ea \ee
The first term defines the bulk equation of motion $d\,{*F}=0$, but it is not sufficient alone to make the action variationally well-defined. Even when the bulk equation of motion is satisfied, we still have\footnote{In a variational context one should technically consider some initial and final times, giving spacelike components to $\p M$, but stationarity of the action is defined only up to such terms \cite{Harlow:2019yfa}, so we ignore them.}
\be \label{eq:S_var} \ba \delta S & \weq \int_M d(\delta A \wedge *F) = \int_{\p M} \delta A \wedge *F \;,\ea \ee
where $\weq$ indicates on-shell equality. To make this vanish we choose a boundary condition.\footnote{Alternatively we could interpret this as a boundary equation of motion setting $i_{\p M} {*F} = 0$.} There are two standard choices. The perfectly electrically conducting (PEC) boundary condition is defined by\footnote{One might also consider the boundary condition where $i_{\p M}A$ is required to be exact rather than zero. We leave its study to future work.}
\be \label{eq:PEC} {\rm PEC:} \quad i_{\p M} A = 0\; , \ee
where $i_{\p M}$ indicates the pullback of a differential form to $\p M$. This is also known as the ``relative" boundary condition. Clearly it makes the action variationally well-defined. The other standard choice, called the perfectly magnetically conducting (PMC) boundary condition, is defined by
\be {\rm PMC:} \quad i_{\p M} {*F} = 0 \; .\ee
This is also known as the ``absolute" boundary condition. It is equivalent to setting $n^\mu F_{\mu\nu}|_{\p M} = 0$ where $n^\mu$ is the unit normal vector to $\p M$. This too clearly sets $\delta S \weq 0$. These are not the only valid boundary conditions though; one of our main results is a new boundary condition with many interesting consequences. We will discuss it in great detail later on. For now we only assume that some boundary condition has been chosen such that the action is variationally well-defined, i.e. such that \eqref{eq:S_var} vanishes. This defines an off-shell field space $\mathcal{F}$, the space of all allowed field configurations on $M$. We denote by $\widetilde{\mc{F}}$ the on-shell subspace of $\mc{F}$ satisfying the bulk equation of motion. We use $\delta$ to denote the exterior derivative on $\mc{F}$, but for one-forms on $\mc{F}$ there is no danger in lapsing into the perhaps more intuitive notion of field variations. The second term in \eqref{eq:L_var} defines the (pre-)symplectic potential density\footnote{Technically there is an ambiguity $\theta \to \theta + d\chi$ for $\chi$ any $(d-2)$-form locally constructed from fields, and different choices can lead to non-isomorphic phase spaces. However, most authors use the standard choice used here. An ``extended" (pre-)symplectic potential density related by such an ambiguity was used in \cite{Donnelly:2016auv}, but it turns out to be isomorphic to the standard choice as we will show.}
\be \theta \equiv \delta A \wedge *F \; .\ee
The (pre-)symplectic density is its field space exterior derivative,
\be \ba \omega & \equiv \delta \theta = -\delta A \wedge *\delta F \; .\ea \ee
We reserve the symbol $\wedge$ for the wedge product on spacetime; the wedge product on $\mc{F}$ is left implicit. Note that $\omega$ is conserved on-shell, 
\be d\omega \weq 0\; . \ee
The (pre-)symplectic form is defined by integrating $\omega$ over a Cauchy slice $\Sigma$, 
\be \Omega \equiv \int_\Sigma \omega = -\int_\Sigma \delta A \wedge *\delta F \; . \ee
The on-shell conservation of $\omega$, along with our variationally well-defined boundary condition, implies that $\Omega$ is independent of the choice of $\Sigma$. Henceforth we will always assume we are on-shell. We call $\Omega$ pre-symplectic, rather than symplectic, because even when restricted to the on-shell field space $\widetilde{\mc{F}}$ it is still degenerate due to gauge redundancy. First, there is a trivial gauge redundancy from the behavior of gauge transformations away from $\Sigma$: if $\lam|_\Sigma=0$ then $i_\Sigma \delta_\lam A = i_\Sigma d\lam = 0$ as well, and consequently $\Omega$ vanishes. We can use such gauge transformations to fix the time component $A_t$ to some chosen function on $M$, usually just zero, and then the residual gauge transformations are time-independent and so can be parametrized by functions on $\Sigma$. There is further redundancy on $\Sigma$ itself. Explicitly, plugging in a gauge transformation $\delta A = d\lambda$ for one of the variations gives\footnote{It is a moot point whether $\lam$ is real-valued or $U(1)$-valued. Our topological assumptions on $M$ and $\p M$ mean there are no paths for $\lam$ to wind around, and the constant mode of $\lam$ introduces no ambiguity since the Gauss law implies $\int_{\p\Sigma} *\delta F = 0$. So expressions like $\int_{\p\Sigma} \lam *\delta F$ are indeed well-defined.}
\be  \label{Om} \ba \Omega  = \int_\Sigma d\lam \wedge *\delta F = \int_{\p\Sigma} \lambda \,{*\delta F} \;.\ea \ee
This is a weighted integral of the electric flux through $\p\Sigma$. We see that if $\lambda$ has no support on $\p\Sigma$ then $\Omega$ vanishes, no matter what $i_{\p\Sigma} {*\delta F}$ is. We will refer to such gauge transformations as ``small". In the same vein, a gauge transformation with boundary support will be called ``large". This terminology is motivated by analogy with the study of asymptotic symmetries, as in e.g. \cite{Strominger:2017zoo}.\footnote{Some authors call a gauge transformation ``large" if it is not continuously connected to the identity transformation. This differs from our terminology here.} Large gauge transformations may or may not be allowed by our boundary condition. In general, phase space is defined as the quotient of the on-shell field space $\widetilde{\mc{F}}$ by the degenerate variations, which include small gauge transformations. Note also that a constant gauge transformation is unphysical in this context, since then $\delta A = d\lambda = 0$ identically. This goes hand-in-hand with the fact that $\int_{\p\Sigma} *F = 0$ due to the Gauss constraint $d\,{*F} = 0$. In a theory with charged matter the constant gauge transformation would be non-degenerate.\footnote{This may na\"ively appear to be in tension with the results of \cite{Donnelly:2013tia}, which showed that in the Maxwell path integral one must quotient by the full gauge group, not just the non-constant gauge transformations, but we show in appendix \ref{appendix:constgauge} that it is in fact consistent.}

\subsection{Dynamical edge mode boundary condition}

The physicality of large gauge transformations is a topic of heated and ongoing discussion, but in the present context there is a clear resolution. A given large gauge transformation is physical if it is both allowed and symplectically non-degenerate. This is where the choice of boundary condition comes into play. The PEC boundary condition \eqref{eq:PEC} disallows (non-constant) large gauge transformations altogether, since they would result in nonzero $i_{\p\Sigma} A$. The PMC boundary condition allows large gauge transformations, but makes them degenerate since ${i_{\p\Sigma} {*\delta F}}$ necessarily vanishes. We see that neither of these standard boundary conditions is compatible with physical large gauge transformations. This motivates us to introduce what we call the dynamical edge mode (DEM) boundary condition, which leads to physical large gauge transformations. We fix a choice of spatial Cauchy slice $\Sigma$, which for simplicity we assume meets $\p M$ orthogonally. Let $n^\mu$ be the outward unit normal vector to $\p M$. For convenience we will use Gaussian normal coordinates with respect to $\Sigma$, meaning that the metric in a neighborhood of $\Sigma$ takes the form (we focus on dimensions $D > 2$)
\be ds^2 = -dt^2 + g_{ij} dx^i dx^j\;, \ee
where $\Sigma$ is the $t=0$ surface. The spatial metric can still depend on time. The induced metric on $\Sigma$ is somewhat trivially $\hat g_{ij} = g_{ij}$, but we find that keeping the hat explicit improves the clarity of our discussion. For simplicity we assume $\p_t$ lies tangent to $\p M$. Then the DEM boundary condition states
\be {\rm DEM:} \quad A_t|_{\p M} = 0 = n^\mu F_{\mu i}|_{\p M}\;. \ee
This is essentially taking the PEC boundary condition for the $t$ direction and the PMC boundary condition for the other directions in $\p M$.\footnote{Setting the time component to zero on the boundary is reminiscent of treatments in 2D Yang-Mills \cite{Cordes:1994fc} and Chern-Simons theory \cite{Elitzur:1989nr}.} The key point is that it allows large gauge transformations along with arbitrary $i_{\p\Sigma} {*F}$, so that the large gauge transformations are non-degenerate. Henceforth we will assume the DEM boundary condition is imposed.

Let us explore the structure of the DEM phase space. As discussed above, our phase space is the quotient of the space of solutions $\widetilde{\mc F}$ by the symplectically degenerate variations. We start by using a degenerate gauge transformation with parameter
\be \lam(t, x) = -\int_0^t dt' \, A_t(t', x) \ee
to set $A_t=0$ in a neighborhood of $\Sigma$. Note that $\lam|_{t=0} = 0$, so it is indeed degenerate, and that $\p_t\lam|_{\p\Sigma} = 0$, so it respects the DEM boundary condition at $\p\Sigma$. Once this is done, solutions can be uniquely specified in terms of data $A_i, E_i$ on $\Sigma$, where we define
\be E_i \equiv F_{ti} = \p_t A_i \;.\ee
The electric field configurations must satisfy the Gauss constraint,
\be \ba 0 & = \nabla_\mu F^{\mu t} = \frac{1}{\sqrt{-g}} \p_\mu (\sqrt{-g} F^{\mu t}) = \frac{1}{\sqrt{\hat g}} \p_i (\sqrt{\hat g} F^{it}) = \hat\nabla_i E^i \; , \ea \ee
where $\hat\nabla_i$ is the covariant derivative on $\Sigma$ and $\hat g = \det \hat g_{ij}$. This constraint offers a convenient way to parametrize the boundary normal electric field $i_{\p\Sigma} \, {*F}$, which can also be written as
\be E_\perp \equiv n^i E_i|_{\p\Sigma} \;.\ee
We define the decomposition on $\Sigma$
\be \label{eq:Esplit} E_i = \tilde E_i + S \hat\nabla_i \beta \ee
where we fix some positive scalar $S$ and require
\be \label{eq:E_tilde_cond} \hat\nabla_i \tilde E^i = 0\; , \qquad n^i \tilde E_i|_{\p\Sigma} = 0\; . \ee
This implies
\be \label{eq:elliptic} \hat\nabla_i (S \hat\nabla^i \beta) = 0 \; , \ee
which is a second order elliptic PDE for $\beta$ on $\Sigma$. Solutions can be specified by either Dirichlet or Neumann data on $\p\Sigma$, and in fact the boundary normal electric field amounts to Neumann data,
\be S \hat\nabla_n \beta|_{\p\Sigma} = E_\perp \;. \ee
We have introduced the notation $\hat\nabla_n \equiv n^i \hat\nabla_i$. This ensures that the split \eqref{eq:Esplit} is unique.\footnote{Although $\hat\nabla_i\beta$ is unambiguous, $\beta$ is defined only up to a constant shift. We choose to eliminate this ambiguity by requiring $\int_{\p\Sigma} \beta = 0$.} In more detail, it defines a linear map $E_i \mapsto \tilde E_i$. The particular map depends on the choice of $S$, but the image of the map, i.e. the space of all $\tilde E_i$, is independent of $S$. A related decomposition was used in an off-shell context in \cite{Riello:2021lfl}. Plugging \eqref{eq:Esplit} into the symplectic form gives
\be \ba \Omega & = \int_\Sigma \delta A^i \delta \tilde E_i - \int_\Sigma \hat\nabla_i (S \delta A^i) \delta\beta + \int_{\p\Sigma} n^i \delta A_i S \delta \beta \;. \ea  \label{eq:syminter}\ee
We are using the shorthand notation $\int_\Sigma\equiv \int_\Sigma \sqrt{\hat g} \, d^{D-1}x$. Note that the first integral is gauge invariant thanks to the vanishing boundary normal of $\tilde E_i$. The second integral is pure gauge, and disappears if we gauge-fix
\be \hat\nabla_i (S A^i) = 0 \; . \label{eq:genguge}\ee
This can be imposed using a small gauge transformation $\lam$ satisfying
\be \hat\nabla_i (S \hat\nabla^i \lam) = -\hat\nabla_i (S A^i) \; . \ee
This is a sourced elliptic PDE for $\lam$, and it has a unique solution satisfying the Dirichlet boundary condition $\lam|_{\p\Sigma} = 0$. The uniqueness guarantees that there are no residual small gauge transformations. Once we impose the gauge \eqref{eq:genguge}, it is natural to parametrize the degrees of freedom as
\be \label{eq:Asplit} A_i = \tilde A_i + \hat\nabla_i \alpha \ee
where
\be \label{eq:A_tilde_cond} \hat\nabla_i (S \tilde A^i) = 0 \; , \qquad n^i \tilde A_i|_{\p\Sigma} = 0 \; . \ee
Note that $\alpha$ satisfies the elliptic PDE
\be \hat\nabla_i (S \hat\nabla^i \alpha) = 0 \ee
and is uniquely determined (up to a constant shift) by the Neumann data
\be \hat\nabla_n \alpha|_{\p\Sigma} = n^i A_i|_{\p\Sigma} \; . \ee
We fix the shift ambiguity by further stipulating that $\int_{\p\Sigma} \alpha = 0$. One might say that $\alpha$ is the ``Goldstone boson" of the large gauge symmetry. As a mathematical aside, we note that our decomposition of $E_i, A_i$ as one-forms on $\Sigma$ is essentially that of Hodge-Morrey-Friedrichs \cite{Schwarz}.\footnote{Yet another perspective is the following. Rather than committing to the gauge \eqref{eq:genguge}, one could allow an arbitrary $A_i$ and define the edge part $\alpha'$ as the unique (up to a constant shift) solution to $\hat\nabla^i(S\hat\nabla_i\alpha') = \hat\nabla^i(SA_i)$ with $\hat\nabla_n\alpha'|_{\p\Sigma} = n^i A_i|_{\p\Sigma}$. Then the bulk part $\tilde A_i \equiv A_i - \hat\nabla_i \alpha'$ is completely gauge invariant, and the boundary restriction $\alpha'|_{\p\Sigma}$ is invariant under small gauge transformations.} Plugging \eqref{eq:Asplit} into \eqref{eq:syminter}, the symplectic form now reads
\be \label{Omega} \ba \Omega & = \int_\Sigma \delta\tilde A^i \delta \tilde E_i + \int_{\p\Sigma}  \delta\alpha \, S \hat\nabla_n \delta\beta  \\
& = \int_\Sigma \delta\tilde A^i \delta \tilde E_i + \int_{\p\Sigma} \delta\alpha \, \delta E_\perp \\
&\equiv \Omega_{\rm bulk} + \Omega_{\rm edge} \; . \ea \ee
The DEM phase space is parametrized by bulk degrees of freedom $\tilde A_i, \tilde E_i$ satisfying respectively \eqref{eq:A_tilde_cond} and \eqref{eq:E_tilde_cond}, and boundary degrees of freedom $\alpha, E_\perp$ which are arbitrary functions on $\p\Sigma$ aside from integrating to zero. We refer to $\alpha, E_\perp$ as edge modes. We see that the phase space splits cleanly, and in the last line we have defined the bulk and edge symplectic forms. The function space of $\tilde E_i$ is independent of $S$, and in fact it is identical to that of the PMC boundary condition. Furthermore in the PMC case the vanishing of $E_\perp$ means that the $\alpha$ degrees of freedom are symplectically trivial and must be quotiented out, so the full set of PMC degrees of freedom is simply $\tilde A_i, \tilde E_i$. Therefore the bulk part of the DEM phase space is isomorphic to the PMC phase space. The edge parts of the gauge and electric fields, i.e. $\hat\nabla_i\alpha \subset A_i$ and $S \hat\nabla_i \beta \subset E_i$, do depend on $S$, but the edge phase space involves only the boundary values $\alpha|_{\p\Sigma}$ and $E_\perp$, and is therefore independent of $S$. Thus the main result of this section is the following decomposition of the DEM phase space:
\be \label{eq:phase_fac} \Gamma_{\rm DEM} \cong \Gamma_{\rm PMC} \times \Gamma_{\rm edge} \;. \ee
The novel object $\Gamma_{\rm DEM}$ has factorized into more familiar pieces. $\Gamma_{\rm edge}$ is closely related to the standard algebra of large gauge transformations and their conjugate charges, described for example in \cite{Donnelly:2016auv}, but in general an algebra of observables need not constitute a phase space. Although the boundary normal electric field at a point $x\in\p\Sigma$ is a valid local observable, it is not an independent degree of freedom: the Gauss law forbids electric field configurations supported only at a single point. So the electric field must extend into the bulk somehow, and this can affect the symplectic form. We have shown how to extend $E_\perp$ such that it remains orthogonal to the bulk degrees of freedom, and this allowed us to describe the phase space, not just the algebra, of large gauge transformations and their conjugate degrees of freedom. Furthermore this phase space $\Gamma_{\rm edge}$ is not merely a subspace of $\Gamma_{\rm DEM}$, but rather an independent factor. We emphasize that this all would have gone through just the same on a causal diamond with no boundary condition like the one studied in \cite{Donnelly:2016auv}, but we worked in the more stringent context of a timelike boundary with a boundary condition, making the existence of $\Gamma_{\rm edge}$ all the more remarkable. As we will see later, the simple structure of $\Gamma_{\rm DEM}$ just shown provides great control over the quantum dynamics and entanglement entropy of edge modes.

\subsection{Edge Hamiltonian(s)}

The characterization of phase space was a matter of kinematics, completely indifferent to the choice of time evolution, or equivalently the choice of Hamiltonian. We turn now to dynamics. In terms of the data $A_i, E_i$ on $\Sigma$, the Lagrangian density \eqref{eq:lag} reads
\be \ba L|_\Sigma & = \half E_i E^i - \hat\nabla_{[i} A_{j]} \hat\nabla^{[i} A^{j]} \;. \ea \ee
The corresponding Hamiltonian density is
\be \ba H_{\rm density} & = \frac{\p L}{\p E_i} E_i - L  = \half E_i E^i + \hat\nabla_{[i} A_{j]} \hat\nabla^{[i} A^{j]} \; .\ea \ee
To get the Hamiltonian generating the time evolution $S' \p_t$ we multiply by $S'$ and integrate over $\Sigma$, yielding
\be \ba H & = \int_\Sigma S' H_{\rm density} \\
& = \int_\Sigma S' \left( \half E_i E^i + \hat\nabla_{[i} A_{j]} \hat\nabla^{[i} A^{j]} \right) \\
& = \int_\Sigma S' \left( \half (\tilde E_i + S \hat\nabla_i \beta) (\tilde E^i + S \hat\nabla^i \beta) + \hat\nabla_{[i} \tilde A_{j]} \hat\nabla^{[i} \tilde A^{j]} \right) \\
& = \int_\Sigma S' \left( \half \tilde E_i \tilde E^i + \hat\nabla_{[i} \tilde A_{j]} \hat\nabla^{[i} \tilde A^{j]} \right) - \int_\Sigma \left(\beta \hat\nabla_i (S S' \tilde E^i) + \half \beta \hat\nabla_i (S^2 S' \hat\nabla^i \beta) \right) + \half \int_{\p\Sigma} S^2 S' \beta \, \hat\nabla_n \beta \; . \ea \ee
The cross term only vanishes if $S S' = 1$. With this choice we have
\be \ba H & = \int_\Sigma \frac{1}{S} \left( \half \tilde E_i \tilde E^i + \hat\nabla_{[i} \tilde A_{j]} \hat\nabla^{[i} \tilde A^{j]} \right) + \half \int_{\p\Sigma} S \beta \, \hat\nabla_n \beta \\
& \equiv H_{\rm bulk} + H_{\rm edge} \; .\ea \ee
This tells us that the bulk-edge symplectic split using $S$ is naturally associated with the time evolution $\frac{1}{S} \p_t$.

The edge Hamiltonian takes the form of an integral over the boundary $\p\Sigma$, but it involves the normal derivative of $\beta$, whose evaluation requires knowledge of $\beta$ in a neighborhood of the boundary, stretching infinitesimally into the bulk. However, because $\beta$ satisfies \eqref{eq:elliptic}, we can actually write $\hat\nabla_n\beta|_{\p\Sigma}$ directly in terms of $\beta|_{\p\Sigma}$ or vice versa.\footnote{Recall constant Dirichlet and Neumann data are disallowed for $\beta$.} More abstractly, we can use the Dirichlet-to-Neumann operator that maps Dirichlet data for \eqref{eq:elliptic} to Neumann data. See \cite{Girouard} for a review of its key properties, in particular that it is a positive-semidefinite self-adjoint pseudodifferential operator on $\p\Sigma$. It can be constructed fairly explicitly in terms of the Green's function $G(x, x') = G(x', x)$ satisfying
\be \hat\nabla_i (S \hat\nabla^i G(x,x')) = \frac{1}{\sqrt{\hat g}} \delta^{(D-1)}(x-x') \label{eq:Greeneom}\ee
and
\be 0 = G(x,x')|_{x\in\p\Sigma}\;. \label{eq:Greenbdyzero}\ee
We can use $S(x) \hat\nabla_n G(x, x')$ as a spatial boundary-to-bulk propagator, noting
\be \ba \int_{\p\Sigma} S(x) \beta(x) \hat\nabla_n G(x,x') & = \int_\Sigma \hat\nabla_i (S(x) \beta(x) \hat\nabla^i G(x,x')) \\
& = \int_\Sigma \frac{1}{\sqrt{\hat g}} \beta(x) \delta^{(d-1)}(x-x') + \int_\Sigma S(x) \hat\nabla_i \beta(x) \hat\nabla^i G(x,x') \\
& = \beta(x') - \int_\Sigma G(x,x') \hat\nabla^i (S(x) \hat\nabla_i \beta(x)) + \int_{\p\Sigma} S(x) \hat\nabla_n\beta(x) G(x,x') \\
& = \beta(x') \; . \ea \ee
Therefore,
\be \ba S(x') \hat\nabla_{n'} \beta(x')\big|_{\p\Sigma} & = \int_{\p\Sigma} \beta(x) K(x,x') \ea \ee
where we have defined the integral kernel
\be K(x,x') \equiv S(x) S(x') \hat\nabla_n \hat\nabla_{n'} G(x,x') \; . \ee
The edge Hamiltonian then takes the form
\be H_{\rm edge} = \half \int_{\p\Sigma} \int_{\p\Sigma} \beta(x) K(x,x') \beta(x') \; .\ee
We will ultimately prefer to rewrite this in terms of $E_\perp = S \hat\nabla_n \beta|_{\p\Sigma}$ using the inverse integral kernel $K^{-1}(x,x')$ defined by
\be \frac{1}{\sqrt{\hat g}} \delta^{(D-1)}(x-x') = \int_{\p\Sigma} K(x,x'') K^{-1}(x'',x') \ee
which is essentially the Neumann-to-Dirichlet operator. We have
\be H_{\rm edge} = \half \int_{\p\Sigma} \int_{\p\Sigma} E_\perp(x) K^{-1}(x,x') E_\perp(x') \; . \ee
We will sometimes use the notations
\be \label{eq:K_abbrev} (K \beta)(x') \equiv \int_{\p\Sigma} \beta(x) K(x,x') \; , \qquad \left( \frac{1}{K} E_\perp \right)(x') \equiv \int_{\p\Sigma} E_\perp(x) K^{-1}(x,x') \; , \ee
in which case the edge Hamiltonian takes the form
\be H_{\rm edge} = \half \int_{\p\Sigma} \beta K \beta = \half \int_{\p\Sigma} E_\perp \frac{1}{K} E_\perp \; ,\ee
but one should keep in mind that generically the Dirichlet-to-Neumann operator and its inverse are nonlocal on $\p\Sigma$.

\subsection{Static spacetimes}
\label{subsec:static}

So far we have focused on some particular Cauchy surface $\Sigma$ and allowed for arbitrary (orthogonal) time evolution off of it, but in many cases of interest there is a distinguished global sense of time. In this subsection we explore the case where our manifold $M$ (and $\p M$) is static. Spacetime is naturally foliated by surfaces of constant static time $t$, all of which are isometric. We take $\Sigma$ to be any such surface. The metric is
\be ds^2 = g_{tt} dt^2 + g_{ij} dx^i dx^j \ee
with all components independent of $t$. Note that $g_{tt} \ne -1$ as we are no longer using Gaussian normal coordinates. Let us briefly reestablish our notation in these new coordinates. We define the electric field as
\be \label{eq:Edef} E_i = \sqrt{-g^{tt}} F_{ti} \; .\ee
The DEM boundary condition on $\p M$ is
\be A_t|_{\p M} = 0 \; , \qquad n^\mu F_{\mu i}|_{\p M} = 0 \; , \ee
where $n^\mu$ is the unit normal to $\p M$. Note $n^t=0$. We parametrize our fields as
\be E_i = \tilde E_i + \sqrt{-g^{tt}} \hat\nabla_i \beta \; , \qquad A_i = \tilde A_i + \hat\nabla_i \alpha \; , \label{eq:staticsplit} \ee
respectively satisfying the Gauss constraint 
\be \hat\nabla_i E^i = 0 = \hat\nabla_i \tilde E^i \ee
and the gauge condition
\be \hat\nabla_i (\sqrt{-g^{tt}} A^i) = 0 = \hat\nabla_i (\sqrt{-g^{tt}} \tilde A^i) \;,  \ee
along with the requirements
\be n^i \tilde E_i|_{\p M} = 0 = n^i \tilde A_i|_{\p M} \;.\ee
With $\Sigma$ as the $t=0$ surface, the solutions on $M$ parametrized by $\alpha(x), \beta(x)$ on $\Sigma$ are
\be A_t(t,x) = 0\; , \qquad A_i(t, x) = \hat\nabla_i \alpha(x) + t \hat\nabla_i \beta(x) \;.\ee
Once again we define the boundary normal electric field on $\p\Sigma$,
\be E_\perp = n^i E_i|_{\p\Sigma} \; .\ee
With these choices the symplectic form
\be \ba \Omega & = \int_\Sigma \delta \tilde A^i \delta \tilde E_i + \int_{\p\Sigma} \delta \alpha \, \delta E_\perp \\
& = \Omega_{\rm bulk} + \Omega_{\rm edge} \ea \ee
and Hamiltonian
\be \ba H & = \int_\Sigma \sqrt{-g_{tt}} \left( \half \tilde E_i \tilde E^i + \hat\nabla_{[i} \tilde A_{j]} \hat\nabla^{[i} \tilde A^{j]} \right) + \half \int_{\p\Sigma} \sqrt{-g^{tt}} \beta \, \hat\nabla_n \beta \\
& = H_{\rm bulk} + H_{\rm edge} \ea \ee
both split. In terms of the Dirichlet Green's function
\be \hat\nabla_i (\sqrt{-g^{tt}} \hat\nabla^i G(x,x')) = \frac{1}{\sqrt{\hat g}} \delta^{(D-1)}(x-x') \; ,\ee
we can define the Dirichlet-to-Neumann kernel
\be K(x,x') \equiv \sqrt{-g^{tt}(x)} \sqrt{-g^{tt}(x')} \hat\nabla_n \hat\nabla_{n'} G(x,x') \;. \ee
We have
\be \ba E_\perp(x') & = \int_{\p\Sigma} \beta(x) K(x,x') \ea \ee
and the inverse relation
\be \beta(x') = \int_{\p\Sigma} E_\perp(x) K^{-1}(x,x') \;.\ee
Using the more compact notation \eqref{eq:K_abbrev}, the edge Hamiltonian then takes the form
\be H_{\rm edge} = \half \int_{\p\Sigma} \beta K \beta = \half \int_{\p\Sigma} E_\perp \frac{1}{K} E_\perp \; .\ee

\subsection{When $\partial M$ is a stretched horizon}
\label{sec:stretch_horiz}

\begin{figure}[H]
    \centering
            \includegraphics[width=0.2\textwidth]{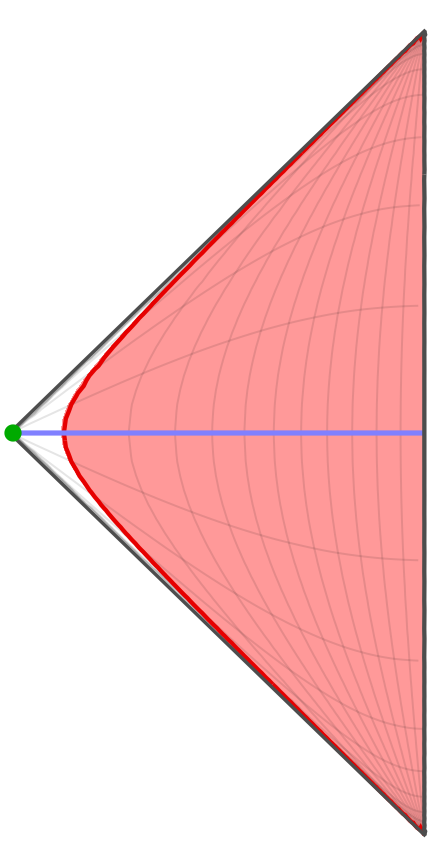}
        \caption{The Penrose diagram of a $dS_D$ static patch. The brick wall is shown in bold red. A Cauchy surface for the patch is shown in blue.}
        \label{fig:staticpatch}
\end{figure}

In general we do not expect the Dirichlet-to-Neumann operator to admit any simple description beyond the one given above, but in the case where $\p M$ approaches a bifurcate Killing horizon it simplifies and in particular becomes local on $\p\Sigma$. Let us embed $M$ in a static spacetime with a bifurcate Killing horizon such that the small proper spatial distance from $\p M$ to the horizon is $\eps$. For example, the case of the static patch in $dS_D$ is shown in figure \ref{fig:staticpatch}. 
We can use Gaussian normal coordinates for a neighborhood of $\p\Sigma$ within $\Sigma$, giving the metric
\be ds^2 = g_{tt} dt^2 + dr^2 + g_{ab} dx^a dx^b \ee
where $g_{ab}|_{r=\eps}$ is the metric on $\p\Sigma$. We assume $\eps$ is small enough so that these coordinates are valid all the way up to the bifurcation surface at $r=0$. The existence of a static bifurcation surface at $r=0$ implies \cite{Solodukhin:2011gn}
\be g_{tt} = -\kappa^2 r^2 + \mc{O}(r^4) \ee
where $\kappa$ is the horizon's surface gravity with respect to $\p_t$. In these coordinates the elliptic PDE satisfied by $\beta$ is
\be 0 = \frac{1}{\sqrt{-g^{tt}}} \p_r (\sqrt{-g^{tt}} \p_r \beta) + \frac{\p_r \sqrt{\det g_{ab}}}{\sqrt{\det g_{ab}}} \p_r \beta + \frac{\p_a \sqrt{-g^{tt}}}{\sqrt{-g^{tt}}} g^{ab} \p_b \beta - \Delta_{\p\Sigma} \beta \ee
where $\Delta_{\p\Sigma} = -\nabla_a \nabla^a$ is the transverse (scalar) Laplacian. The first three terms na\"ively obstruct separation of variables, but we will see that they are actually amenable in the horizon limit, with the middle two terms simply vanishing at leading order in $r$.

First recall the zeroth law of black hole mechanics, which states that the surface gravity is constant along a stationary horizon \cite{Bardeen:1973gs}. This means that $\kappa$ is independent of the transverse coordinates $x^a$, and consequently
\be \frac{\p_a \sqrt{-g^{tt}}}{\sqrt{-g^{tt}}} = \mc{O}(r^2) \; . \ee
The other property we need is that the bifurcation surface of a stationary horizon has vanishing extrinsic curvature \cite{Solodukhin:2011gn}, i.e. the normal Lie derivative vanishes,
\be 0 = \mc{L}_{\p_r} g_{ab}|_{r=0} = \p_r g_{ab}|_{r=0} \; .\ee
Then also the normal derivative of the determinant vanishes at $r=0$, so
\be \p_r \sqrt{\det g_{ab}} = \mc{O}(r) \;.\ee
We can now write the PDE for $\beta$ as
\be \label{eq:betaODE}
\ba 0 & = \frac{1}{\sqrt{-g^{tt}}} \p_r (\sqrt{-g^{tt}} \p_r \beta) - \Delta_{\p\Sigma} \beta + \p_r \beta \times \mc{O}(r) + \beta \times \mc{O}(r^2) \\
& = r \p_r \left(\frac{1}{r} \p_r \beta\right) - \Delta_{\p\Sigma} \beta + \p_r \beta \times \mc{O}(r) + \beta \times \mc{O}(r^2) \\
& = \p_r^2 \beta - \frac{1}{r} \left( 1 + \mc{O}(r^2) \right) \p_r \beta - \left( \Delta_{\p\Sigma} + \mc{O}(r^2) \right) \beta \;. \ea \ee
If we expand $\beta$ in eigenfunctions of the transverse Laplacian $\Delta_{\p\Sigma}$ then the leading terms take the form of a second order ODE in $r$. Explicitly we write
\be \beta(r, x^a) = \sum_{k\ne 0} \beta_k(r) f_k(x^a) \ee
where
\be \Delta_{\p\Sigma} f_k(x^a) = \lam_k f_k(x^a) \ee
with $\lam_k > 0$. We have set the constant mode of $\beta$ to zero since it does not contribute to the gradient of $\beta$, which is the real physical quantity. The approximate ODE in question is then
\be 0 = \p_r^2 \beta_k - \frac{1}{r} \left( 1 + \mc{O}(r^2) \right) \p_r \beta_k - \left( \lam_k + \mc{O}(r^2) \right) \beta_k \; . \ee
One can show using the Frobenius method that the leading behavior of the general solution is
\be \label{eq:beta_asymp} \beta_k(r) \propto 1 + \half \lam_k r^2 \log \sqrt{\lam_k} r + C r^2 + \mc{O}(r^3) \; . \ee
Demanding regularity in the bulk will fix the coefficient $C$, leaving freedom only in the overall normalization of the solution. From this we can explicitly read off the action of the Dirichlet-to-Neumann operator $K$ on the $k$ mode as
\be \frac{E_{\perp,k}}{\beta_k|_{r=\eps}} = \frac{-\sqrt{-g^{tt}} \p_r \beta_k(r)}{\beta_k(r)}\Big|_{r=\eps} = \frac{\lam_k}{\kappa} \log \frac{1}{\eps} + {\cal O}(\eps^0)\;. \ee
Based on this we identify in the $\eps\to 0$ limit
\be K \quad \longleftrightarrow \quad \frac{\log\eps^{-1}}{\kappa} \Delta_{\p\Sigma} \; .\ee
This simplification of $K$ is remarkable. It is generically a highly nonlocal operator on $\p\Sigma$, but in the horizon limit it becomes local. A similar conclusion was reached in \cite{Donnelly:2015hxa} for the special case when the static manifold $M$ factorizes into ${\cal B} \times \p\Sigma$ with ${\cal B}$ a 2D static manifold with bifurcate Killing horizon. Note that the edge Hamiltonian $H_{\rm edge} = \half \int_{\p\Sigma} E_\perp \frac{1}{K} E_\perp$ contains an overall factor of $\frac{1}{\log\eps^{-1}}$, sending the energy of the edge modes to zero and thus strengthening the analogy with soft and Goldstone modes in asymptotically flat spacetimes as in \cite{Strominger:2017zoo}.

Besides allowing us to compute the Dirichlet-to-Neumann operator $K$, the asymptotic form \eqref{eq:beta_asymp} for $\beta_k(r)$ also provides insight into the physical localization of edge modes to the boundary in the horizon limit. We assume some finite electric flux $E_\perp(x^a)$ through the boundary $\p\Sigma$, and we find that the electric field falls off rapidly away from the boundary. The asymptotic solution corresponding to $E_\perp$ is
\be \beta_k(r) = \frac{\kappa E_{\perp,k}}{\lam_k \log\eps^{-1}} \left( 1 + \half \lam_k r^2 \log \sqrt{\lam_k} r + C r^2 + \mc{O}(r^3) \right) \; . \ee
The corresponding radial electric field is
\be \ba E_{r,k}(r) & = -\sqrt{-g^{tt}} \p_r \beta_k(r)  = E_{\perp,k} \frac{\log\sqrt{\lam_k}r }{\log\sqrt{\lam_k}\eps}+ \mc{O}(r)\; . \ea \ee
Its peak value of $E_{\perp,k}$ is reached at the boundary $r=\eps$, and from there it decreases with increasing $r$. When it reaches the scale $r \sim \lam_k^{-1/2}$ it is parametrically small, suppressed by $\frac{1}{\log\eps^{-1}}$. The corresponding transverse electric field is
\be \ba E_a\big( r,x^b\big) & = \sqrt{-g^{tt}} \p_a \beta\big(r,x^b\big)   = \sum_k \frac{E_{\perp,k}}{\lam_k r \log\eps^{-1}} \p_a f_k\big(x^b\big)  \left( 1 + \mc{O}(r^2\log r) \right) \; . \ea \ee
We see that it is even bigger than the radial electric field, achieving its maximum $\mc{O}\big(\frac{1}{\eps\log\eps^{-1}}\big)$ value at the boundary $r=\eps$ and decreasing from there, once again becoming small like $\mc{O}\big(\frac{1}{\log\eps^{-1}}\big)$ at the scale $r \sim \lam_k^{-1/2}$. Similar observations about localization near the boundary were made in \cite{Blommaert:2018rsf} for the special case of Rindler.

\section{Edge partition function}
\label{sec:EdgeP}

One of the advantages of the phase space analysis carried out above is that some aspects of quantization are now nearly trivial. The Hilbert space consists of wavefunctionals on the configuration space part of phase space, i.e. the $\tilde A_i$ and $\alpha|_{\p\Sigma}$ variables. We can mode expand $\tilde A_i$ and build up a Fock space, and $\alpha|_{\p\Sigma}$ admits a similar analysis. For the operator formalism we can promote the Poisson bracket on phase space to the commutator, and for the phase space version of the path integral we can read off the measure from the symplectic form. Note also that since the phase space factorizes into bulk and edge parts, the corresponding Hilbert space must tensor factorize as
\be \mc{H}_{\rm DEM} \cong \mc{H}_{\rm bulk} \otimes \mc{H}_{\rm edge} \ee
with $\mc{H}_{\rm bulk}$ isomorphic to the PMC Hilbert space. Since the Hamiltonian splits as well, the thermal partition function\footnote{We wrote the thermal partition function as a trace here, but it can also be evaluated with the path integral. See appendix \ref{app:PIZedge}.} must factorize:\footnote{Here $\beta$ is the inverse temperature, and is unrelated to the edge mode degree of freedom discussed previously. In the following we hope it is clear from context which is which.}
\be \ba Z_{\rm DEM}(\beta) & \equiv \Tr e^{-\beta H} \\
& = \Tr_{\rm bulk} e^{-\beta H_{\rm bulk}} \Tr_{\rm edge} e^{-\beta H_{\rm edge}} \\
& = Z_{\rm bulk}(\beta) Z_{\rm edge}(\beta) \; . \ea \ee
The $Z_{\rm bulk}(\beta)$ factor is simply the partition function with the PMC boundary condition, and it has been computed for several geometries \cite{Dowker:2010bu, Dowker:2010yj, Huang:2014pfa, Donnelly:2015hxa}. Our focus here is on $Z_{\rm edge}(\beta)$. The large gauge transformations are physical symmetries, and so the partition function will contain a factor of their (infinite) group volume, rendering it ill-defined. An object to which we can actually assign a number is
\be \label{eq:Zbar_def} \bar Z_{\rm edge}(\beta) \equiv \frac{1}{|\mc{G}_{\p\Sigma}|} Z_{\rm edge}(\beta) \ee
where we define the group of large gauge transformations $\mc{G}_{\p\Sigma}$ as the group of $G$-valued functions on $\p\Sigma$. Our assumption that $\p\Sigma$ has a single connected component means that $\mc{G}_{\p\Sigma}$'s only multi-valued mode is the constant gauge transformation.  Note $\mc{G}_{\p\Sigma}$ is slightly larger (precisely by this zero mode) than the group parametrized by $\alpha$, since we require $\int_{\p\Sigma} \alpha = 0$. Let $\mc{G}'_{\p\Sigma}$ be this smaller group with the zero mode removed. A quick motivation for using $\mc{G}_{\p\Sigma}$ rather than $\mc{G}_{\p\Sigma}'$ is that excluding a zero mode is nonlocal, but we will offer more compelling reasons for the form of $\bar Z_{\rm edge}(\beta)$ in section \ref{sec:shrink}.

For concreteness in the following, expand in orthonormal eigenmodes on $\p\Sigma$ as
\be E_\perp(x^a) = \sum_{k\ne 0} E_{\perp,k} f_k(x^a) \; , \qquad \alpha(x^a) = \sum_{k\ne 0} \alpha_k f_k(x^a) \; . \ee
The symplectic form reads
\be \Omega_{\rm edge} = \sum_{k\ne 0} \delta \alpha_k \, \delta E_{\perp,k} \ee
implying the quantum commutator $[\alpha_k, E_{\perp,k}] = i$. Our edge Hilbert space admits a basis of states with definite $E_\perp$,
\be \ket{E_\perp} \equiv \prod_{k\ne 0} \ket{E_{\perp,k}}\;.\label{eq:Eperpbasis}\ee
Each $\ket{E_{\perp,k}}$ is like a momentum ket in 1D quantum mechanics, normalized as
\be \inner{E_{\perp,k}}{E'_{\perp,k}} = 2\pi\,\delta\big(E_{\perp,k} - E'_{\perp,k}\big) \; . \ee
Identifying the infinite quantity $2\pi\,\delta(0)$ as the volume of the range of the conjugate variable $\alpha_k$,\footnote{The relation $2\pi\,\delta(p) = \int_{-\infty}^\infty dx \, e^{ipx}$ is familiar from elementary quantum mechanics, or really just Fourier analysis. One can even regulate the infinity by taking the quantum mechanical particle to live on a circle, $x \sim x + 2\pi R$.} the product over all modes gives
\be \prod_{k\ne 0} \mu \inner{E_{\perp,k}}{E_{\perp,k}} = |\mc{G}_{\p\Sigma}'| \ee
where we have introduced an arbitrary mass scale $\mu$ to keep everything dimensionless. In this basis, we compute the edge partition function \eqref{eq:Zbar_def}:
\be \label{eq:ZedgeTr} \ba \bar Z_{\rm edge}(\beta) 
& = \frac{1}{|\mc{G}_{\p\Sigma}|} \int \mc{D} E_\perp \bra{E_\perp} e^{-\beta H_{\rm edge}} \ket{E_\perp} \\
& = \frac{1}{|\mc{G}_{\p\Sigma}|} \int \mc{D} E_\perp \inner{E_\perp}{E_\perp} \, e^{-\half\beta\int_{\p\Sigma} E_\perp \frac{1}{K} E_\perp} \\
& = \frac{1}{|\mc{G}_{\p\Sigma}|} \int \left( \prod_{k\ne 0} \mu \inner{E_{\perp,k}}{E_{\perp,k}} \frac{dE_{\perp,k}}{2\pi\mu} \right) e^{-\half\beta\int_{\p\Sigma} E_\perp \frac{1}{K} E_\perp} \\
& = \frac{|\mc{G}'_{\p\Sigma}|}{|\mc{G}_{\p\Sigma}|} \, {\rm det}'\left(\frac{K}{2\pi\mu^2\beta}\right)^{1/2}\; . \ea \ee
Here the prime in $\det'$ indicates that we omit the zero mode of $K$ in computing the determinant. The ratio of group volumes can be handled as follows. Mode expand $g(x^a) \in \mc{G}_{\p\Sigma}$ as
\be g(x^a) = \sum_k g_k f_k(x^a) \ee
where $g_k$ is real-valued for $k\ne 0$ but the zero mode coefficient $g_0$ is $G$-valued. We define the group volumes by integrating over all mode coefficients,
\be \ba \frac{|\mc{G}'_{\p\Sigma}|}{|\mc{G}_{\p\Sigma}|} & = \frac{\prod_{k\ne 0} \int \mu \, dg_k}{\prod_k \int \mu \, dg_k}  = \frac{1}{\int_0^{2\pi\sqrt{V_{\p\Sigma}}/q} \mu \, dg_0} = \frac{q}{2\pi\mu\sqrt{V_{\p\Sigma}}} \; . \ea\ee
Here we used the periodicity $g \sim g + \frac{2\pi}{q}$ and the fact that the normalized zero mode is simply $f_0(x^a) = V_{\p\Sigma}^{-1/2}$ where $V_{\p\Sigma}$ is the volume of $\p\Sigma$. Plugging back in, we get
\be \bar Z_{\rm edge}(\beta) = \frac{q}{2\pi \mu \sqrt{V_{\p\Sigma}}} \, {\rm det}'\left(\frac{K}{2\pi\mu^2\beta}\right)^{1/2} \; . \ee
In the horizon limit $\eps\to 0$ where $K=\frac{1}{\kappa}\log(\eps^{-1})\Delta_{\p\Sigma}$ this reduces to
\be \label{eq:bZedge} \boxed{\bar Z_{\rm edge}(\beta) = \frac{q}{2\pi \mu \sqrt{V_{\p\Sigma}}} \, {\rm det}'\left(\frac{\log\eps^{-1}}{2\pi\kappa\,\mu^2\beta}\Delta_{\p\Sigma}\right)^{1/2}} \; . \ee
We note that in general a factor inside a functional determinant can be pulled out with a power independent of the size of the manifold. For example on $S^2$ one has, for any factor $C$, that ${\rm det}'(C \Delta_{S^2}) = C^{-2/3} {\rm det}'(\Delta_{S^2})$. In odd dimensions the factor is simply $C^{-\dim\ker\Delta} = C^{-1}$. We can use this to pull out the factor of $\frac{\log \eps^{-1}}{2\pi\kappa\beta}$ with some power. Up to this factor, we see that $1/\bar Z_{\rm edge}(\beta)$ amounts to the partition function on $\p\Sigma$ of a compact scalar with target radius $\frac{1}{q}$. This is a key observation, and naturally sets up our discussion of shrinkability. Related observations in \cite{Donnelly:2015hxa, Blommaert:2018oue, Blommaert:2018rsf} are compared and contrasted in section \ref{sec:shrink}. Finally, we recombine with $Z_{\rm bulk}(\beta)$ to define
\be \ba \bar Z_{\rm DEM}(\beta) & \equiv \frac{1}{|\mc{G}_{\p\Sigma}|} Z_{\rm DEM}(\beta)  = Z_{\rm bulk}(\beta) \bar Z_{\rm edge}(\beta) \; . \ea \ee
Henceforth we will always normalize time such that $\kappa=1$ and accordingly a non-singular Euclidean section corresponds to $\beta=2\pi$.

\section{Shrinkability}
\label{sec:shrink}

We explained in the introduction that we want to choose a boundary condition on the brick wall such that the $\eps\to 0$ limit recovers the global physics with no brick wall. Shrinkability is the Euclidean version of this condition, and we will demand it of our subregion theory. In Euclidean QFT, a boundary condition is said to be shrinkable if when applied to a vanishingly small hole the result is as if there were no hole at all \cite{Hung:2019bnq, Jafferis:2019wkd, Agia:2022srj, Agia:2024wxx}. More concretely, consider the example of the Euclidean plane with a small disk of radius $\eps$ cut out, denoted $\mathbb{R}^2_\eps$. Denote (unnormalized) correlators on the full plane by $\langle \dots \rangle_{\mathbb{R}^2}$. Correlators on the excised plane require a boundary condition to be well-defined. Let $B_\eps$ be a one-parameter family of boundary conditions labeled by $\eps$, and denote correlators on the excised plane with this boundary condition by  $\langle \dots \rangle^{B_\eps}_{\mathbb{R}^2_\eps}$. The family of boundary conditions $B_\eps$ is said to be shrinkable if for all correlators (with insertions independent of $\eps$) we have
\be \lim_{\eps\to 0} \langle \dots \rangle^{B_\eps}_{\mathbb{R}^2_\eps} = \langle \dots \rangle_{\mathbb{R}^2} \; . \ee
Generically $B_\eps$ will involve an $\eps$-dependent normalization factor that is singular as $\eps\to 0$, which compensates for UV divergences from the physics in a small neighborhood of the excision. Hopefully this makes the definition clear in general. We are interested in particular in local, codimension-two shrinkable boundary conditions. The codimension-two requirement allows us to think of a periodic Euclidean time coordinate encircling the hole (for $\mathbb{R}^2_\eps$ this is the polar angle $\theta$), and the requirement of locality allows us to define Hilbert spaces on the slices of constant Euclidean time that end on our small hole. See \cite{Agia:2022srj, Agia:2024wxx} for in-depth discussions in the case of 2D CFT. Besides its inherent interest, shrinkability is important because it offers one of the most promising general definitions of entanglement entropy \cite{Jafferis:2019wkd, Hung:2019bnq}. We will return to and expand on this point in section \ref{sec:global}.

Before moving on we note that, to our knowledge, neither existence nor uniqueness of shrinkable boundary conditions is guaranteed in general. For example in the 2D compact boson both the Dirichlet and Neumann boundary conditions are shrinkable \cite{Agia:2022srj}. In fact, in 2D CFT any boundary condition with nontrivial identity overlap (and a nontrivial gap above the identity) will be shrinkable \cite{Hung:2019bnq}. Nonetheless, for Maxwell our DEM boundary condition is the only shrinkable one of which we are aware.\footnote{See however \cite{Blommaert:2018oue}, whose approach may amount to a distinct shrinkable boundary condition.}

\subsection{2D scalar example}

\begin{figure}[H]
    \centering
    \includegraphics[width=0.6\textwidth]{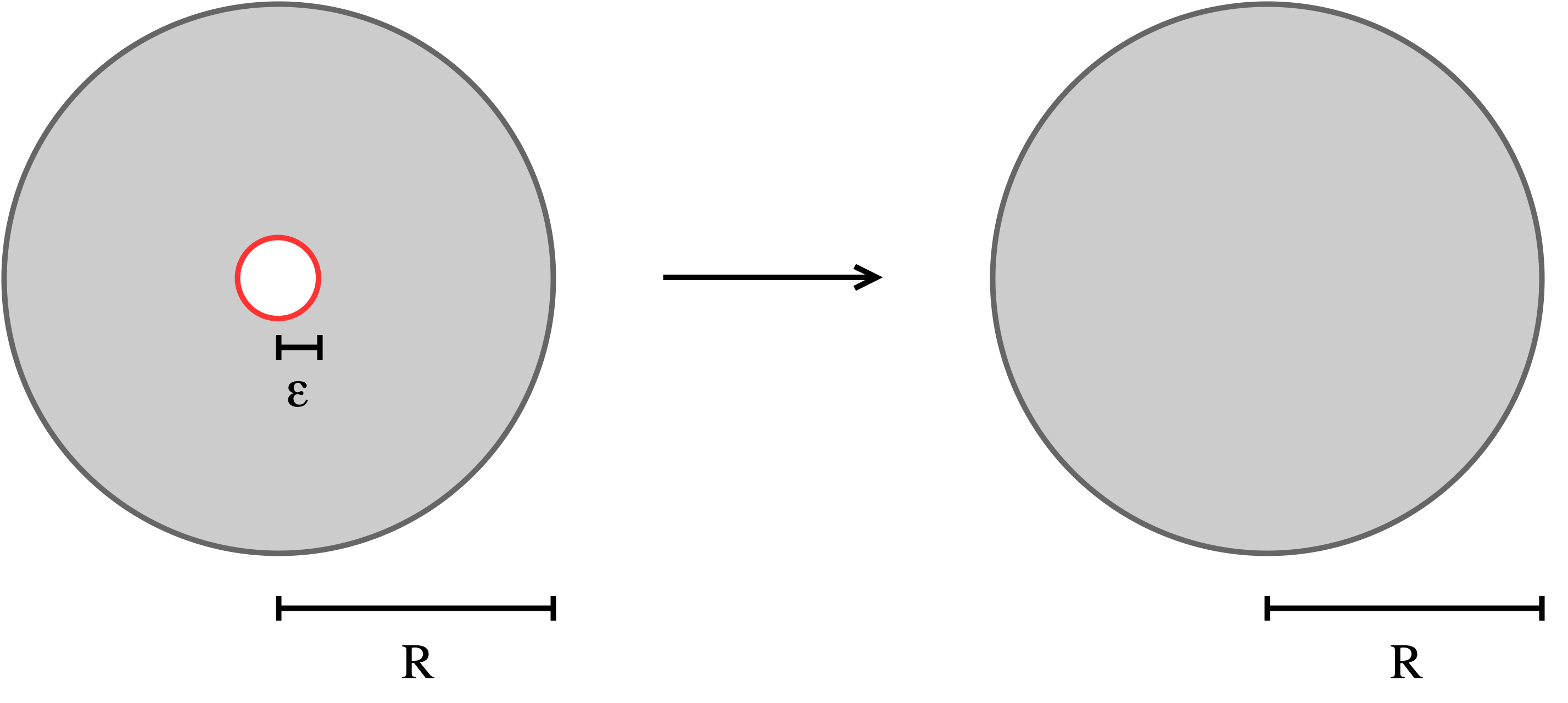}
        \caption{With an appropriate boundary condition, the annulus with inner radius $\eps$ can recover the disk as $\eps\to 0$.}
    \label{fig:ann-shrink}
\end{figure}

We find it instructive to first discuss shrinkability in the context of the 2D compact scalar, where things are relatively simple. We consider an annulus whose inner radius is shrinking to zero, and we attempt to recover the disk. Let the outer radius of the annulus be $R$ and the inner radius be $\eps$. See figure \ref{fig:ann-shrink}. Consider first the Neumann boundary condition. There is a zero mode contribution $C$ that we will not need to deal with explicitly because it is the same for the annulus and disk.\footnote{$C$ is proportional to the square root of the volume of the manifold, i.e. to $R$, and to the target radius of the scalar.} With $\mu$ as a renormalization scale, the Neumann partition functions are \cite{Weisberger:1987kh}\footnote{The Neumann annulus partition function also has contributions from sectors with nontrivial winding, but they are relatively suppressed by positive powers of $\eps$ and therefore do not affect the leading term of interest.}
\be \ba Z^N_{\rm ann}(\eps, R; \mu) & = C \, {\rm det}'_{\rm ann}(\mu^{-2}\Delta_N)^{-1/2} \\
& = \frac{C}{(\eps/R)^{1/6} \mu \, R \sqrt{2\pi}} \left( 1 + \mc{O}(\eps/R) \right) \\
Z^N_{\rm disk}(R; \mu) & = C \, {\rm det}'_{\rm disk}(\mu^{-2}\Delta_N)^{-1/2} \\
& = \frac{C}{(\mu \, R)^{5/6}} \, \exp[\zeta'(-1) - \frac{7}{24} + \frac{1}{12}\log 2 - \frac{1}{4}\log\pi] \ea \ee
where the prime in $\rm{det}'$ indicates that we omit the zero mode in computing the determinant. Both partition functions have the same leading $R$-dependence of $\,R^{-5/6}$, which is the only nontrivial takeaway for our purposes. We see that choosing
\be f(\mu\eps) = (\mu\eps)^{1/6} \sqrt{2\pi} \exp[\zeta'(-1) - \frac{7}{24} + \frac{1}{12}\log 2 - \frac{1}{4}\log\pi] \ee
leads to
\be \lim_{\eps\to 0} f(\mu\eps) Z^N_{\rm ann}(\eps, R; \mu) = Z^N_{\rm disk}(R; \mu) \ee
which suggests that the Neumann boundary condition, once rescaled by $f(\mu\eps)$, is shrinkable for the 2D compact scalar. Technically we have only checked shrinkability for the partition function but, barring any accidents, locality implies that it should also hold for generic correlators. Indeed, shrinkability of the Neumann boundary condition was shown explicitly in \cite{Agia:2022srj}. Now consider this same example but with the Dirichlet boundary condition instead. The partition functions in this case are \cite{Weisberger:1987kh}\footnote{There are no winding sectors this time thanks to the Dirichlet boundary condition.}
\be \ba Z^D_{\rm ann}(\eps, R; \mu) & = {\rm det}_{\rm ann}(\mu^{-2}\Delta_D)^{-1/2} \\
& = \frac{\pi^{1/2}}{(\eps/R)^{1/6} \sqrt{\log\frac{1}{\mu\eps}}} \left( 1 + \mc{O}\left(\frac{1}{\log\frac{1}{\mu\eps}}\right) \right) \\
Z^D_{\rm disk}(R; \mu) & = {\rm det}_{\rm disk}(\mu^{-2} \Delta_D)^{-1/2} \\
& = (\mu \, R)^{1/6} \exp[\zeta'(-1) + \frac{5}{24} + \frac{1}{12}\log 2 + \frac{1}{4}\log\pi] \;. \ea \ee
Choosing
\be \label{eq:Dir_f} f(\mu\eps) = (\mu\eps)^{1/6} \sqrt{\log\frac{1}{\mu\eps}} \, \exp[\zeta'(-1) + \frac{5}{24} + \frac{1}{12}\log 2 - \frac{1}{4}\log\pi] \ee
leads to
\be \lim_{\eps\to 0} f(\mu\eps) Z^D_{\rm ann}(\eps, R; \mu) = Z^D_{\rm disk}(R; \mu) \ee
which suggests that the Dirichlet boundary condition, once rescaled by $f(\mu\eps)$, is shrinkable for the 2D compact scalar. This was confirmed in \cite{Agia:2022srj}.

In both the Neumann and Dirichlet cases we can view the annulus partition function as the thermal trace for a portion of 2D Rindler space. For concreteness focus on the Dirichlet case. Describe the annulus using polar coordinates
\be ds^2 = dr^2 + r^2 d\tau^2 \ee
where we identify $\tau \sim \tau + 2\pi$. The interval $\eps < r < R$ with the Dirichlet boundary condition at the endpoints gives a Hilbert space. Let $H$ be the generator of $\tau$ translations. We then manifestly have
\be \Tr e^{-2\pi H} = Z^D_{\rm ann}(\eps, R; \mu) \ee
and shrinkability implies that, with $f(\mu\eps)$ as in \eqref{eq:Dir_f},
\be \lim_{\eps\to 0} f(\mu\eps) \Tr e^{-2\pi H} = Z^D_{\rm disk}(R; \mu) \;. \ee
In this way the disk partition function can be interpreted as a thermal trace. We can also consider a general periodicity $\tau \sim \tau + \beta$, in which case $\Tr e^{-\beta H}$ corresponds to the partition function on a cone with the tip cut off. We note however that shrinking to the disk at $\beta=2\pi$ is logically independent from shrinking to a cone at general $\beta$. This much is clear from the case of a non-minimally coupled scalar with curvature coupling $\xi \mc{R} \phi^2$. Here $\mc{R}$ is the Ricci scalar, which vanishes everywhere on the cone except at the tip, where it has delta function support. Since $\Tr e^{-\beta H}$ corresponds to a cone with no tip, it cannot depend on $\xi$ at all. In contrast, the partition function on the cone does depend on $\xi$ through a contact term at the tip. Therefore for generic $\xi$ the thermal trace at $\beta$ will not shrink to the cone partition function, despite it shrinking to the disk at $\beta=2\pi$. This is not a problem. In fact it is a feature of shrinkability, allowing us to vary $\beta$ without introducing new curvature. This is desirable since, for example, one expects the entanglement entropy in flat space to be independent of any curvature couplings \cite{Agon:2013iva}.

\subsection{Shrinkability in Maxwell on $S^4$}

In $D$ dimensions Maxwell theory has $D-2$ propagating degrees of freedom. The first clear hint that these are not sufficient to capture all of the physics came from Kabat \cite{Kabat:1995eq}, who showed that the partition function of Maxwell on a cone differed by a contact term from the answer suggested by the propagating degrees of freedom alone. In contrast, for scalars and spinors he found agreement between the two techniques. This already suggests that the PMC and PEC boundary conditions, which kill all non-propagating degrees of freedom, are not shrinkable. 

The puzzle was sharpened in \cite{Dowker:2010bu, Eling:2013aqa, Huang:2014pfa} in the context of computing entanglement entropy for 4D Maxwell across a sphere, and articulated most clearly in \cite{Donnelly:2014fua, Donnelly:2015hxa}, where it was proposed that edge modes resolve the discrepancy. The quantum boundary condition of \cite{Donnelly:2014fua, Donnelly:2015hxa} involved an ad hoc, by-hand path integral over a family of classical boundary conditions labeled by electric flux through the boundary, with a measure deduced from detailed lattice considerations. In this approach the edge modes were understood only as superselection sectors, with no dynamics. In $D=4$, Maxwell is conformal, and thus its entanglement entropy across a sphere of radius $R$ in flat space is equivalent to the thermal entropy on the static patch of $dS_4$ \cite{Casini:2011kv}; the universal log coefficient of the latter equals that of the sphere partition function $Z(S^4)$, which is determined by the conformal anomaly to be
\be S_{\rm EE} \sim -\frac{31}{45} \log \mu R \label{eq:Max4Dlog}\ee
where the renormalization scale $\mu$ serves as the UV regulator. Here $\sim$ indicates equivalence of log terms, which are independent of the UV regularization scheme. For the bulk contribution they cited \cite{Dowker:2010bu},
\be S^{\rm bulk}_{\rm EE} \sim -\frac{16}{45} \log \mu R \; .\label{eq:Sbulk}\ee
Consistency then required\footnote{Recall that in even-dimensionsal CFTs the entanglement entropy and partition function satisfy $S_{\rm EE} \sim \log Z$ \cite{Casini:2011kv}.}
\be S^{\rm edge}_{\rm EE} \sim \log Z^{\rm DW}_{\rm edge} \stackrel{?}{\sim} -\frac{1}{3} \log \mu R \; .\label{eq:Sedgerequired}\ee
The $Z^{\rm DW}_{\rm edge}$ of \cite{Donnelly:2014fua, Donnelly:2015hxa} amounted to (the reciprocal of) the partition function $Z(S^2)$ of a 2D massless scalar \textit{minus its zero mode}. A similar conclusion was reached in \cite{Soni:2016ogt}. For any 2D CFT, the log term in $Z(S^2)$ is determined by its central charge; for the 2D scalar it is $\log Z(S^2) \sim \frac{1}{3} \log \mu R$. But the zero mode contributes $+\log \mu R$ on its own (we review this explicitly in appendix \ref{appendix:zmode}), so without it one gets instead\footnote{We thank Javier Mag\'an for pointing this out to us.}
\be \log Z^{\rm DW}_{\rm edge} \stackrel{?}{\sim} \frac{2}{3} \log \mu R \;. \ee
In contrast, our $\bar Z_{\rm edge}(\beta=2\pi)$ amounts to (the reciprocal of) the partition function of a compact scalar \textit{including its zero mode}, and therefore we find
\be \log \bar Z_{\rm edge} \sim -\frac{1}{3} \log \mu R \label{eq:edgelog}\ee
as required for consistency with the conformal anomaly method of computing entanglement entropy in 4D Maxwell.

The authors of \cite{Blommaert:2018rsf} offered a Hilbert space description for the edge modes of \cite{Donnelly:2015hxa}, along with an edge Hamiltonian, in the special case of Rindler space. In \cite{Blommaert:2018oue} three of these same authors embarked on a path integral interpretation of edge modes, and their results suggest that an augmented version of the PEC boundary condition, in which the boundary pullback of $A_\mu$ is required to be exact rather than zero, is shrinkable. They isolated a boundary contribution to the partition function that was nonlocal in time, whereas ours is local in time.

Here we conjecture that in any dimension $D$ the quantum boundary condition defining $\bar Z_{\rm DEM}(\beta=2\pi)$, i.e. the DEM boundary condition along with the quotient by $\mc{G}_{\p\Sigma}$, is shrinkable after rescaling away a possible $\eps$-dependent factor. Explicitly, with $M$ the closed Euclidean manifold obtained by Wick rotation from a static Lorentzian manifold with horizon boundary, we propose that
\be \boxed{Z(M) = \lim_{\eps\to 0} f(\mu\eps) \bar Z_{\rm DEM}(\beta=2\pi)} \ee
for some appropriate $f(\mu\eps)$, independent of $M$, with $\mu$ an arbitrary scale to keep things dimensionless. This prescription is unprecedented in its detail and generality, passing checks for which the aforementioned edge mode treatments either fail explicitly or suffer from ambiguities, and furthermore our prescription is compatible with a complete description of the edge phase space and dynamics. We offer several further checks in section \ref{sec:further}.

\subsection{Perspective on dividing by $\mc{G}_{\p\Sigma}$}

We show through our checks that the division by $\mc{G}_{\p\Sigma}$ in our shrinkable boundary condition is necessary, but here we offer some intuition for it. Let $M$ be the closed Euclidean manifold obtained by Wick rotation of a static Lorentzian manifold with horizon. When we introduce a brick wall a distance $\eps$ from the horizon we can foliate the Lorentzian manifold as $\mathbb{R} \times \Sigma$, where $\Sigma$ is a slice of constant time. This Wick rotates to $S^1 \times \Sigma$, with boundary $S^1 \times \p\Sigma$. We can use a small disk $D^2_\eps$ to help patch up the hole,
\be M = (S^1 \times \Sigma) \cup (D^2_\eps \times \p\Sigma) \;. \ee
In the $\eps\to 0$ limit we think of the disk $D^2_\eps$ as shrinking to a point, in which case $D^2_\eps\times\p\Sigma$ becomes $\p\Sigma$. The picture is very concrete when the Lorentzian manifold is the static patch of $dS_2$. Then $M = S^2$, $\Sigma$ is an interval, and $S^1 \times \Sigma$ corresponds to excising small disks around the north and south poles of $S^2$. We identify these two poles of $S^2$ as a copy of $\p\Sigma$.

The path integral for $Z(M)$ uses the measure $\mc{D}A / \mc{G}_M$, where $\mc{G}_M$ is the group of all gauge transformations on $M$, including constant gauge transformations. Defining $\mc{G}^s_M$ as the subgroup restricting to the identity on $D^2_\eps \times \p\Sigma$, we can rewrite the measure as
\be Z(M): \qquad \frac{\mc{D}A / \mc{G}^s_M}{\mc{G}_M / \mc{G}^s_M} \;. \ee
The denominator is simply $\mc{G}_{D^2_\eps\times\p\Sigma}$, the group of $G$-valued functions on $D^2_\eps\times\p\Sigma$.

We show in appendix \ref{appendix:nophase} that the trace over the DEM Hilbert space corresponds to the path integral on $S^1 \times \Sigma$ with the DEM boundary condition and measure $\mc{D}A / \mc{G}^s_{S^1\times\Sigma}$ where $\mc{G}^s_{S^1\times\Sigma}$ is the group of small gauge transformations on $S^1\times\Sigma$. Our shrinkable boundary condition is defined with a further division by $\mc{G}_{\p\Sigma}$, so it can be written as a path integral on $S^1\times\Sigma$ with measure
\be \bar Z_{\rm DEM}(S^1\times\Sigma): \qquad \frac{\mc{D}A / \mc{G}^s_{S^1\times\Sigma}}{\mc{G}_{\p\Sigma}} \;. \ee
Now let us compare this measure with that of $Z(M)$ above. Given the shrinkability of $\bar Z_{\rm DEM}(S^1\times\Sigma)$, we expect them to be the same in some sense. Towards this, first note that the groups $\mc{G}^s_M$ and $\mc{G}^s_{S^1\times\Sigma}$ are naturally isomorphic. Then the main question is whether $\mc{G}_{D^2_\eps\times\p\Sigma}$ and $\mc{G}_{\p\Sigma}$ can be identified. The intuition of $D^2_\eps$ shrinking to a point suggests that this is the case, and our checks support this picture.

\section{Further checks on shrinkability}
\label{sec:further}

In this section we provide further checks on the conjectured shrinkability of our renormalized DEM boundary condition.

\subsection{Maxwell on $S^D$ ($D\geq 3$)}
\label{sec:MaxwellSD}

\begin{figure}[H]
    \centering
    \includegraphics[width=0.35\textwidth]{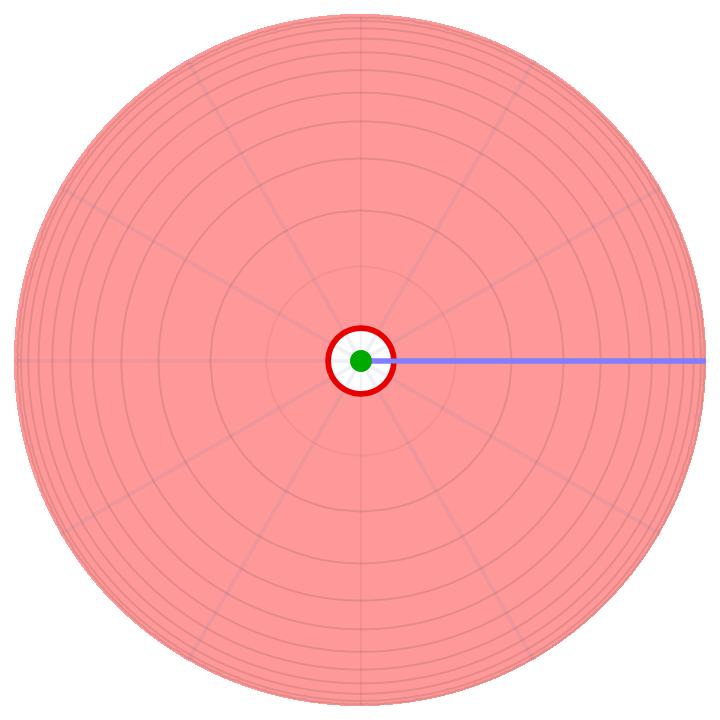}
        \caption{Upon a Wick rotation and periodically identifying the Euclidean time, a $dS_D$ static patch becomes the round $S^D$. The brick wall in figure \ref{fig:staticpatch} is mapped to the small red line encircling the origin.}
    \label{fig:brick-sphere}
\end{figure}

Our prime example is Maxwell theory on a $dS_{D}$ static patch for general $D\geq 3$, which Wick rotates to the round $S^{D}$. See figure \ref{fig:brick-sphere}.
The $S^{D}$ partition function for Maxwell is \cite{Giombi:2015haa} 
\begin{align}\label{eq:MaxPID}
	Z_\text{PI}(S^D)=\frac{ q}{ \sqrt{2\pi \mu^4 R^{D}\text{Vol}(S^{D})}} \frac{ \det'\left(\frac{-\nabla_{(0)}^2}{\mu^2}\right)^{1/2}}{\det(\frac{-\nabla_{(1)}^2 + \frac{D-1}{R^2}}{\mu^2})^{1/2}}\; .
\end{align}
Here $q$ is the fundamental charge, $R$ is the $dS$ length or radius of the sphere in the Euclidean signature, and $\text{Vol}(S^{D}) = \frac{2\pi^{\frac{D+1}{2}}}{\Gamma\left( \frac{D+1}{2}\right)}$ is the volume for a unit round $S^{D}$. 

Since Maxwell is not conformal for $D\neq 4$, $Z_\text{PI}(S^D)$ for $D\neq 4$ does not have the interpretation as (the universal part of) the entanglement entropy across a sphere with radius $R$ in flat space. However, it still has the interpretation of the thermal partition function for Maxwell in the $dS_D$ static patch at the de Sitter temperature $\beta = 2\pi$, and hence can be analyzed within our general framework. As pointed out in \cite{Anninos:2020hfj}, \eqref{eq:MaxPID} can be brought into the form\footnote{In the appendix G.3 of \cite{Anninos:2020hfj}, there is a factor $Z_{\rm KT}$ (where KT stands for Killing Tensor) that we have absorbed into $Z_{\rm G}$. In \eqref{eq:Gfactor} or \eqref{appeq:Gfactor}, $Z_{\rm KT}$ corresponds to the first term in the bracket.}
\begin{align}\label{eq:ADLSsplit}
    Z_\text{PI}(S^D) =  Z^{\rm ADLS}_\text{bulk} Z^{\rm ADLS}_\text{edge}Z_\text{G}\;.
\end{align}
Here the ``quasicanonical" partition function
\begin{align} \label{eq:Zbulkorigin}
    \log Z^{\rm ADLS}_\text{bulk} = \int_0^\infty \frac{dt}{2t}\frac{1+e^{-\frac{t}{R}}}{1-e^{-\frac{t}{R}}} \left[ (D-1) \cdot \frac{e^{-\frac{t}{R}}+e^{-(D-2)\frac{t}{R}}}{\left( 1-e^{-\frac{t}{R}}\right)^{D-1}}-\frac{1+e^{-(D-1) \frac{t}{R}}}{\left( 1-e^{-\frac{t}{R}}\right)^{D-1}}+1\right] ,
\end{align}
as we review in appendix \ref{app:quasicanonical}, can be independently defined in the Lorentzian static patch, and is sensitive only to the local propagating degrees of freedom. The quantity in the bracket is the Harish-Chandra character for a massless spin-1 field, which encodes its quasinormal modes (QNMs) in the static patch \cite{Sun:2020sgn}. The precise sense in which \eqref{eq:Zbulkorigin} is computing a thermal trace was clarified and extended to the cases of static BTZ and Nariai black holes in \cite{Law:2022zdq,Grewal:2022hlo}.

The edge partition function in \eqref{eq:ADLSsplit} is defined as the reciprocal of that in \cite{Anninos:2020hfj}, i.e. 
\begin{align}\label{eq:Zedgeorigin}
    \log Z^{\rm ADLS}_\text{edge} = -\int_0^\infty\frac{dt}{2t}\frac{1+e^{-\frac{t}{R}}}{1-e^{-\frac{t}{R}}} \left[ \frac{1+e^{-(D-3) \frac{t}{R}}}{\left( 1-e^{-\frac{t}{R}}\right)^{D-3}}-1\right] .
\end{align}
Note that there is an overall minus sign. The factor 
\begin{align}\label{eq:Gfactor}
    \log Z_\text{G} =& \log \frac{q}{ \sqrt{2\pi \mu^4 R^D\text{Vol}(S^D)}} + \int_0^\infty\frac{dt}{2t} \left[ -2\cdot \frac{1+e^{-\frac{t}{R}}}{1-e^{-\frac{t}{R}}}+2+e^{- (D-1)\frac{t}{R}}+e^{- (D-3)\frac{t}{R}} \right] 
\end{align}
in \eqref{eq:ADLSsplit} is associated with the constant gauge transformation. We explain in appendix \ref{app:sphere_PI} how to extract scheme-independent quantities from the UV-divergent integrals \eqref{eq:Zbulkorigin}, \eqref{eq:Zedgeorigin} and \eqref{eq:Gfactor}.

In the work of \cite{Anninos:2020hfj}, while \eqref{eq:Zbulkorigin} has a well-defined (quasi)canonical meaning, the Lorentzian interpretations for all other factors in  \eqref{eq:ADLSsplit}, i.e.
\begin{align}\label{eq:Zedgeredefine}
    {\tilde Z}^{\rm ADLS}_\text{edge} \equiv Z^{\rm ADLS}_\text{edge} Z_\text{G}\; ,
\end{align}
were lacking. In appendix \ref{app:ADLSmax}, we show that \eqref{eq:Zedgeredefine} equals exactly the partition function for a ghost compact scalar on $S^{D-2}$ with target space period $\frac{2\pi}{q}$ studied in appendix \ref{app:compactSD}, i.e. 
\begin{align}\label{eq:ADLSfinaledgecompact}
    {\tilde Z}^{\rm ADLS}_\text{edge} = \frac{1}{Z_{\rm PI}^{\rm compact}\left( S^{D-2}\right)} \;. 
\end{align}
with $Z_{\rm PI}^{\rm compact}\left( S^{D-2}\right)$ given in \eqref{appeq:compactPIresult}. In other words, we have
\begin{align}\label{eq:ADLSfinalsplit}
    Z_\text{PI}=  Z^{\rm ADLS}_\text{bulk} {\tilde Z}^{\rm ADLS}_\text{edge} \; . 
\end{align}
After rescaling away the $\eps$ dependence out front (which shrinkability allows), our edge partition function $\bar Z_{\rm edge}(\beta=2\pi)$ in \eqref{eq:bZedge} when applied to the $dS_D$ static patch precisely matches  ${\tilde Z}^{\rm ADLS}_\text{edge}$ \eqref{eq:Zedgeredefine} for all $D\geq 3$. This highly nontrivial quantitative agreement strongly suggests that ${\tilde Z}^{\rm ADLS}_\text{edge}$ captures the dynamical edge modes allowed by the DEM boundary condition.

As a demonstration of the generality of our results, we use the prescriptions \eqref{appeq:ZbulkIR} and \eqref{appeq:readoff} to compute the universal log coefficients of \eqref{eq:ADLSfinalsplit} in some low even $D\geq 4$:
\begin{center}
	\begin{tabular}{ |c|c|c|c|c|c|c|c|c| } 
		\hline
		$D$  & 4 & 6 & 8 &10 &12 &14 &16 \\
        \hline
        $\alpha_D^{\rm Max, bulk}$  &$-\frac{16}{45}$ & $-\frac{331}{945}$ & $-\frac{1592}{4725}$ & $-\frac{303601}{935550}$ & $-\frac{32062
   0081}{1021620600}$ & $-\frac{56598287}{185749200}$ & $-\frac{165704180437}{558242685000}$\\ 
   \hline
		$\alpha_D^{\rm Max, edge}$ & $-\frac13$ & $-\frac{29}{90}$ & $-\frac{1139}{3780}$ & $-\frac{32377}{113400}$& $-\frac{2046263}{7484400}$ & $-\frac{5389909963}{204324120
   00}$& $-\frac{31374554287}{12259447
   2000}$\\ 
		\hline
    $\alpha_D^{\rm Max, total}$& $-\frac{31}{45}$ & $-\frac{1271}{1890}$ & $-\frac{4021}{6300}$ & $-\frac{456569}{748440}$ & $-\frac{1199869961}{2043241200}$ & $-\frac{893517041}{1571724000}$ & $-\frac{17279945447657}{31261590360000}$
  \\ 
		\hline
	\end{tabular}
\end{center}
This contains in particular the $D=4$ results \eqref{eq:Sbulk}, \eqref{eq:edgelog} and \eqref{eq:Max4Dlog}.

\subsection{Comparison with Donnelly and Wall's Measure}
\label{subsec:measure}

The measure for $E_\perp$ used in \cite{Donnelly:2015hxa} differs from ours by an anomaly-dependent factor that leads to discrepancies between our partition functions in even dimensions. Explicitly, their measure (80) is\footnote{Strictly speaking, they wrote their measure in terms of the number $N$ of lattice sites on $\p\Sigma$. We have rewritten it in terms of the lattice spacing $a$, related by $N = V_{\p\Sigma} / a^{D-2}$.}
\be (\mc{D} E_\perp)_{\rm DW} = \frac{a^{\frac{D-2}{2}}}{\sqrt{V_{\p\Sigma}}} \prod_{k\ne 0} \frac{a^{\frac{D-2}{2}}}{q} dE_{\perp, k} \ee
where the small length scale $a$ is their lattice spacing on $\p\Sigma$. As explained for example in \cite{Donnelly:2015hxa}, we can pull factors out of a functional determinant at the cost of an anomaly-dependent factor. In this case the relevant anomaly is that of a $(D-2)$-dimensional scalar. The upshot is that for any $k$-independent number $C$ we have an equivalence of measures
\be \prod_{k\ne 0} C \, dE_{\perp, k} = C^{-1+A} \prod_{k\ne 0} dE_{\perp, k} \ee
where $A$ is determined by the Type A trace anomaly of the scalar. The quantity $A$ is dimension-dependent and vanishes when $D$ is odd. We can now rewrite their measure as
\be (\mc{D} E_\perp)_{\rm DW} = \left( \frac{a^{\frac{D-4}{2}}}{q} \right)^A \frac{q \, a}{\sqrt{V_{\p\Sigma}}} \prod_{k\ne 0} a \, dE_{\perp, k} \;. \ee
This is to be compared with our measure in \eqref{eq:ZedgeTr}
\be \frac{|\mc{G}'_{\p\Sigma}|}{|\mc{G}_{\p\Sigma}|} \mc{D} E_\perp = \frac{q}{2\pi\mu\sqrt{V_{\p\Sigma}}} \mc{D}E_\perp = \frac{q}{2\pi\mu\sqrt{V_{\p\Sigma}}} \prod_{k\ne 0} \frac{dE_{\perp,k}}{2\pi\mu} \;.\label{eq:ourmeasure}\ee
Supposing that the two regularization schemes are related by $2\pi\mu \leftrightarrow a^{-1}$, we see that the two measures agree up to the anomaly term $(a^{\frac{D-4}{2}} / q)^A$ present only in even dimensions. Donnelly and Wall state that this anomalous factor can be safely neglected. But for even $D \ne 4$ this affects the universal $\log a$ term in the log partition function. And although in $D=4$ the anomaly term reduces to $q^{-A}$, it still makes a physical difference since the coefficient of $\log q$ is scheme-independent. Considering that our boundary treatment correctly reproduces the Maxwell partition function on $S^D$ for all $D\geq 2$ as shown in sections \ref{sec:MaxwellSD} (for $D\geq 3$) and \ref{sec:2DMaxwell} (for $D=2$), and that the boundary treatment of \cite{Donnelly:2015hxa} differs from ours by universal terms, theirs must not reproduce $Z(S^D)$ in even dimensions.\footnote{The analysis in \cite{Donnelly:2015hxa} for a general dimension $D$ was separate from their analysis for the special case of $D=4$. Their zero mode omission in $D=4$ discussed in section \ref{sec:shrink} is independent from the issues discussed here.} If we modify their prescription for $Z_{\rm edge}^{\rm DW}$ by replacing their measure $(\mc{D}E_\perp)_{\rm DW}$ with our \eqref{eq:ourmeasure}, then we do find agreement with our $\bar Z_{\rm edge}$:
\be \bar Z_{\rm edge} = \frac{q}{2\pi\mu\sqrt{V_{\p\Sigma}}} \int \mc{D}E_\perp e^{-S_{\rm cl}[E_\perp]} \; . \ee
Here $S_{\rm cl}[E_\perp]$ is the Euclidean action of the unique classical solution with static boundary flux $E_\perp$.

\subsection{Shrinkability in 2D Maxwell}\label{sec:2DMaxwell}

Maxwell theory is solvable in two dimensions, allowing us to check our results in extreme detail. The DEM boundary condition in two dimensions reduces to the statement that the gauge field pulls back to zero on the boundary, which in particular means that the holonomy around a boundary circle is trivial. This is known to be shrinkable \cite{Donnelly:2014gva}, as long as one divides by $|G|$ for each hole. It seems plausible that the triviality of the holonomy may be the key in higher dimensions too, but that is beyond the scope of this paper. In any dimension the Maxwell partition function on a closed manifold $M$ is given by the path integral (see e.g. \cite{Donnelly:2013tia})
\be \label{gaugepath} Z(M) = \sum_{\rm bundles} \int \frac{\mc{D} A}{\mc{G}_M} e^{-S} \ee
where $\mc{G}_M$ is the group of $G$-valued functions on $M$. For $M=S^2$ there are no nontrivial loops and therefore no multi-valued elements of $\mc{G}_M$, but there are nontrivial magnetic bundles with background flux $F_{\theta\phi} = \frac{n}{2q R^2}$ where $R$ is the radius of the sphere. These add $\frac{\pi n^2}{2q^2 R^2}$ to the action. Denoting the trivial bundle contribution by $Z_0(S^2)$ we have
\be \label{eq:Z0def} \ba Z(S^2) & = Z_0(S^2) \sum_{n\in\mathbb{Z}} e^{-\frac{\pi n^2}{2q^2R^2}}  = Z_0(S^2) \vartheta_3(e^{-\frac{2\pi^2}{Vq^2}}) \ea \ee
where in the second equality we introduced the sphere volume $V = 4\pi R^2$. The full partition function is given in \cite{Witten:1991we} as\footnote{The convention in \cite{Witten:1991we} for the Haar measure is such that $|G|=1$. We have converted to our convention where $|U(1)| = \frac{2\pi\mu}{q}$.}
\be \label{eq:WittenZS2} \ba Z(S^2) & = \sum_{n\in\mathbb{Z}} \frac{q^2}{(2\pi\mu)^2} e^{-\half V q^2 n^2}  = \frac{q^2}{(2\pi\mu)^2} \vartheta_3(e^{-\half Vq^2}) \ea \ee
where $\mu$ is a renormalization scale. Using the modular properties of the theta function we can rewrite this as
\be Z(S^2) = \frac{q}{(2\pi)^{3/2} \mu^2 V^{1/2}} \vartheta_3(e^{-\frac{2\pi^2}{Vq^2}}) \; .\ee
By comparison with \eqref{eq:Z0def} we see that the trivial bundle contributes
\be Z_0(S^2) = \frac{q}{(2\pi)^{3/2} \mu^2 V^{1/2}} \ee
which is consistent with the 2D case of \eqref{eq:MaxPID} for arbitrary dimensions, up to an unphysical rescaling of $\mu$ by $\sqrt{2\pi}$. Now let us compare these sphere quantities with $\bar Z_{\rm DEM}(\beta=2\pi)$ for the static patch of de Sitter, with metric
\be ds^2 = R^2 (-\sin^2\theta \, dt^2 + d\theta^2) \; . \ee
In this context our Cauchy surface $\Sigma$ is the interval $\eps < \theta < \pi-\eps$. Its boundary consists of two points. Most of the covariant phase space analysis of section \ref{sec:CovPhaseEdge} carries over directly. Defining $E_\theta = \sqrt{-g^{tt}} F_{t\theta}$, the symplectic form is
\be \Omega = \int_\eps^{\pi-\eps} d\theta \sqrt{g_{\theta\theta}} \, \delta A_\theta \delta E^\theta \;. \ee
We can impose the gauge
\be 0 = \p_\theta (\sqrt{-g^{tt} g^{\theta\theta}} A_\theta) \ee
using the small, single-valued gauge transformation
\be \lam = -\int_\eps^\theta d\theta' A_\theta(\theta') + \frac{\int_\eps^\theta d\theta' \sqrt{-g_{tt} g_{\theta\theta}}}{\int_\eps^{\pi-\eps} d\theta' \sqrt{-g_{tt} g_{\theta\theta}}} \int_\eps^{\pi-\eps} d\theta' A_\theta(\theta') \;. \ee
Once this has been done we can use a function $\alpha:\Sigma\to\mathbb{R}$ to parametrize
\be A_\theta = \p_\theta \alpha \; . \ee
The gauge condition implies $\p_\theta (\sqrt{-g^{tt}g^{\theta\theta}} \p_\theta \alpha) = 0$, so the two boundary values of $\alpha$ determine $\alpha$, and therefore $A_\theta$, on all of $\Sigma$. In fact, only the difference $\Delta\alpha \equiv \alpha|_{\theta=\pi-\eps} - \alpha|_{\theta=\eps}$ is symplectically non-degenerate. Unlike the analysis in section \ref{sec:CovPhaseEdge}, here $\p\Sigma$ has more than one connected component and so the group of small gauge transformations has non-identity components. In particular a small gauge transformation can wind any integer number of times between $\theta=\eps$ and $\theta=\pi-\eps$. Quotienting by them further reduces our configuration space down to $\Delta\alpha$ modulo $\frac{2\pi}{q}$. This is just the value of the Wilson line along $\Sigma$. Turning now to the electric field, the Gauss constraint is
\be 0 = \p_\theta (\sqrt{g^{\theta\theta}} E_\theta) \;. \ee
This says that $\sqrt{g^{\theta\theta}} E_\theta$ is constant. We can use it to write $E_\theta$ in terms of the normalized boundary flux $E_\perp = \sqrt{g^{\theta\theta}} E_\theta|_{\theta=\eps}$,
\be E_\theta(\theta) = \sqrt{g_{\theta\theta}(\theta)} \, E_\perp \;. \ee
In terms of $\Delta\alpha$ and $E_\perp$ the symplectic form reads
\be \Omega = \delta \Delta\alpha \, \delta E_\perp \;.\ee
Since the configuration space is $G$-valued, in the quantum theory its conjugate variable will be discrete, $E_\perp \in q\mathbb{Z}$. One can show that the Hamiltonian is
\be H = \frac{V E_\perp^2}{4\pi} \ee
where once again $V = 4\pi R^2$. The canonical trace is then
\be \ba Z_{\rm DEM}(\beta=2\pi) & = \Tr e^{-2\pi H} = \sum_{n\in\mathbb{Z}} e^{-\half V q^2 n^2} = \vartheta_3(e^{-\half Vq^2})\;.\ea \ee
The final step is to divide by the volume of the group of large gauge transformations (including its zero mode), which in this case is $|\mc{G}_{\p\Sigma}| = |G|^2 = (2\pi\mu/q)^2$. We include $\mu$ in this definition of volume to make it dimensionless. The result is
\be \ba \bar Z_{\rm DEM}(\beta=2\pi) & = \frac{1}{|\mc{G}_{\p\Sigma}|} \Tr e^{-2\pi H}  = \frac{q^2}{(2\pi\mu)^2} \vartheta_3(e^{-\half Vq^2}) \;.\ea \ee
We have recovered the sphere partition function \eqref{eq:WittenZS2}, which demonstrates the shrinkability of the boundary condition defining $\bar Z_{\rm DEM}$ and confirms the correctness of including the constant gauge transformation in our quotient by large gauge transformations. For completeness we note that in the path integral evaluation of this trace, which is equivalent to that of a compact scalar on $S^1$, the winding sum plays the role of the sum over bundles on $S^2$ term by term.

\subsection{Shrinkability in 2D Yang-Mills} \label{subsection:2DYM}

Our shrinkable quantum boundary condition uses a renormalized trace on the Hilbert space that involves a division by $|\mc{G}_{\p\Sigma}|$, the volume of large gauge transformations. As noted above, the inclusion of the constant gauge transformation affects the radius dependence and is crucial for the matching to the conformal anomaly calculation of entanglement entropy in 4D Maxwell. We have already justified the division by $|\mc{G}_{\p\Sigma}|$ in several ways, and explicitly checked its necessity for shrinkability in the case of a sphere $S^D$ with one of its azimuthal angles viewed as static Euclidean time. Here we offer one more check of shrinkability, generalizing to allow more punctures on $S^2$ in the case of 2D Yang-Mills (2DYM). In addition to extending the 2D Maxwell checks to the non-abelian case, the topological approach taken here leads to very simple formulas which are particularly well-suited for the discussion of shrinkability.

\paragraph{The partition function and the gauge theory measure}

2DYM is quasi-topological, with the partition function on an arbitrary Euclidean manifold $M$ depending only on the topology and area of $M$. We can think of the coupling $g_{\rm YM}$ and the area as the two continuous physical parameters of the theory, along with a renormalization scale $\mu$. The partition function can be computed exactly following sewing rules analogous to those of a TQFT. The result is \cite{Witten:1991we}
\begin{align} \label{eq:2dYMP}
    Z(M) = \sum_{R} \left( \frac{\dim R}{|G|} \right)^{\chi(M)} e^{-V C_{2}(R)}
\end{align}
where $R$ labels irreducible representations of the gauge group $G$, $C_{2}(R)$ is the quadratic Casimir of $R$, and $\chi(M)$ is the Euler characteristic of $M$. Here $V$ is a dimensionless combination of the area and the coupling. The normalization of the Haar measure used to compute $|G|$ in this formula is determined in principle by $g_{\rm YM}/\mu$. Rather than working out the specific dependence, it is more convenient to just treat $|G|$ and $V$ as our two independent dimensionless parameters. We note that \eqref{eq:2dYMP} reduces to \eqref{eq:WittenZS2} in the abelian case, with the sum over $n$ corresponding to the sum over representations.
   
The partition function \eqref{eq:2dYMP}, including the factor of $|G|$, was carefully derived in  \cite{Witten:1991we} from the Yang-Mills path integral with the $\frac{1}{\mathcal{G}_{M}}$ quotient in the measure as in \eqref{gaugepath}. There is an alternative interpretation of this measure which becomes manifest when $G$ is discrete. In this case, the Casimir is zero, and \eqref{eq:2dYMP} becomes  the standard topological partition function of discrete gauge theory,\footnote{This is also known as Dijkgraaf-Witten gauge theory.} which is defined in terms of counting gauge bundles on $M$. For discrete groups, each gauge bundle corresponds to a choice of holonomy, i.e. a homomorphism $g: \pi_{1}(M) \to G$, modulo gauge transformations.
A simple case is the sphere, where we get
\begin{align}
    Z(S^2)= \frac{1}{|G|^2} \sum_{R} (\dim R)^2= \frac{1}{|G|} \; .
\end{align}
Since all loops on the sphere are contractible, it supports only the trivial bundle. Thus, we see that the gauge theory measure \eqref{gaugepath} actually counts each bundle with a symmetry factor given by  the reciprocal of the order of its automorphism group;\footnote{This is just like in Feynman diagrams where we weight each graph by a symmetry factor.} these are just the gauge transformations that fix the bundle, which in this case is the entire group $G$. This is called the orbifold measure in mathematics, and is commonly used as an a priori definition of discrete gauge theory.

\paragraph{Shrinkability for a single interval }

In the topological approach to 2DYM, the sphere partition function is computed from the norm of the Hartle-Hawking vacuum, defined by the path integral on a hemisphere. This is a state on the Hilbert space $\mathcal{H}_{S^1}$ of a circle, which consists of class functions on the group $G$. These depend on the conjugacy class of the spatial holonomy
\begin{align} 
   U = \mc{P} \exp \left[i\oint_{S^1} A\right].
\end{align}
$\mathcal{H}_{S^1}$ is spanned by the group characters, which are labeled by irreducible representations $R$ of the gauge group. In terms of the corresponding normalized basis states $\ket{R}$, the Hartle-Hawking wavefunction is
\be \mathtikz{\etaC{0}{0}} \hspace{-3mm} = 
    \ket{HH} = \sum_{R} \frac{\dim R}{|G|} e^{ -\frac{V}{2} C_{2}(R)} \ket{R} .
\ee
Gluing two hemispheres together gives $Z(S^2)=\braket{HH|HH}$. To obtain a trace interpretation for $Z(S^2)$, we remove two infinitesimal disks from the sphere and impose the DEM boundary condition, along with dividing by one factor of $|G|$ for each hole. In 2D, DEM just means setting the gauge component tangent to the boundary to zero. As explained in \cite{Donnelly:2014gva}, quantizing with this boundary condition gives an interval Hilbert space 
\begin{align}
    \mathcal{H}_{I} =L^{2}(G)
\end{align}
of square-integrable functions on $G$. These are functions of the Wilson line 
\begin{align}
U = \mc{P} \exp \left[i\int_{I} A\right]
\end{align} 
that goes between the two endpoints of the interval.  A normalized basis for $L^{2}(G)$ is given by the representation matrix elements $R_{ab}$ of the group,
\be
    \braket{U|R,a,b} = \sqrt{\frac{\dim R}{|G|}} \, R_{ab}(U),\qquad a,b=1,\cdots, \dim R \; .
\ee
Here we normalize as $\inner{U}{U'} = \delta(U,U')$, where $\int_G dU \, \delta(U,U') = 1$ and $\int_G dU = |G|$. The states $\ket{R,a,b}$ transform nontrivially under the gauge group restricted to the boundaries of the interval, which corresponds to the large gauge transformations:
\begin{align}
    R_{ab}(U) \to R_{ab}(g_{L}Ug_{R}^{-1} ) ,\qquad g_{L},g_{R} \in G \; .
\end{align}
After replacing the infinitesimal disks with the shrinkable boundary condition, the spacetime now has the topology of a cylinder. Our shrinkable boundary condition demands that the corresponding partition function is computed with a \textit{renormalized} trace over the interval Hilbert space, in which we divide by $|\mathcal{G}_{\p \Sigma}|$: 
\begin{align}\label{S2shrink}
    Z(S^2) &= \lim_{\eps \to 0}  \frac{1}{|G|^2} \Tr_{\mathcal{H}_{I}} e^{- V(\eps) \hat{C}_{2}} = \frac{1}{|G|^2} \sum_{R} \sum_{a,b=1}^{\dim R} e^{-V C_{2}(R)} \; .
\end{align} 
In the first line, we used the fact that the modular Hamiltonian on the interval is given by the area times the quadratic Casimir \cite{Cordes:1994fc}. The second expression reproduces the partition function on the sphere, where $(\dim R)^2$ is now interpreted as a degeneracy factor from an explicit sum over states labeled by $a,b$. Note that a group volume factor had to be inserted ``by hand", rather than coming from the sum over states: it is determined by matching the trace over states to the sphere partition function.

\paragraph{Shrinkability for multiple intervals}   
   
Having renormalized the trace over an interval using the sphere partition function as an input, we can now proceed to apply this shrinkable boundary condition to different spacetime manifolds and subregions. How this works is not obvious a priori. If we fix the sphere topology, the $(\dim R)^2$ factor is easily interpreted as a degeneracy factor for states at the two endpoints of an interval. But for two disconnected intervals we would have four edges, and tracing over these would na\"ively give $(\dim R)^4$. However we will see that shrinkability is preserved in general provided that we divide by the volume of large gauge transformations, which is given by one factor of $|G|$ for each endpoint.
   
To understand shrinkability of the renormalized trace in the more general setting, we consider a periodic time evolution of $n$ disconnected intervals that sweeps out a 2D Euclidean spacetime with holes cut out around the endpoints.   For a single connected spacetime, this is only possible if the intervals collide and interact, cutting and rejoining in order to continue the flow. When the spacetime is a sphere and the region is two intervals, the reduced density matrix implementing this evolution was computed in \cite{Donnelly:2018ppr}. 
\begin{figure}
  \centering
   \begin{subfigure}{0.2\textwidth}
            \centering
            \includegraphics[width=\textwidth]{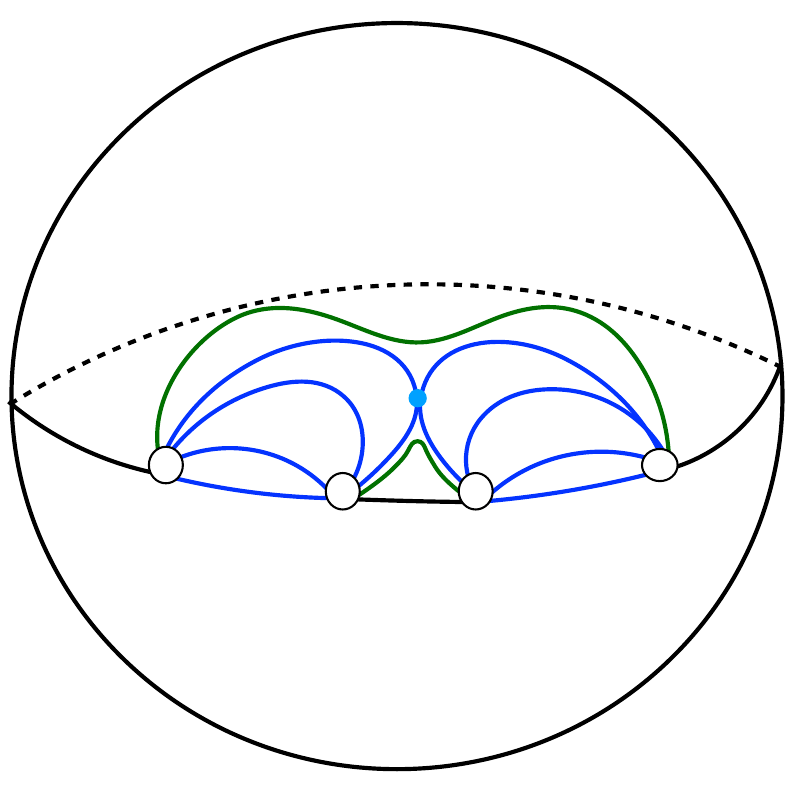}  
        \end{subfigure}
        \hspace{2cm}
        \begin{subfigure}{0.2\textwidth}  
            \centering 
            \includegraphics[width=\textwidth]{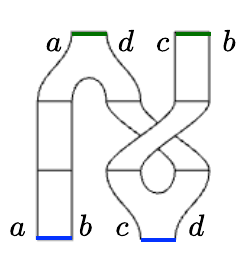}
        \end{subfigure}
        \caption{The left figure shows the time evolution (modular flow) that sweeps out the sphere with two intervals.  The initial time slice consists of two blue intervals, which break and rejoin at the interaction point (blue dot). The green intervals show the time slice after this interaction. On the right figure, we resolve the interaction point into basic operations that split, join and swap the intervals. These operations compose to give the operator $\rho^{1/2}$.}
        \label{fig:2int}
\end{figure}
In particular, evolution by half of the period is given by the operator
\begin{equation}\label{eq:two}
    \rho^{1/2} (\eps) = 
    \sum_{R,a,b,c,d} \frac{e^{-\frac{1}{2} V(\eps)  C_{2}(R)} }{\dim R/|G|} \ket{R,a,b} \ket{R,c,d} \bra{R,a,d} \bra{R,c,b} \, .
\end{equation}
The action of this operator can be understood by considering the flow depicted in figure \ref{fig:2int}, where an initial pair of (blue) intervals cut and reglue to form the final pair of (green) intervals. The contraction of the indices $a$, $b$, $c$, and $d$ in $\rho^{1/2}$ follows from the flow of the boundaries as indicated in the right of figure \ref{fig:2int}. Crucially, there is a $\frac{|G|}{\dim R}$ factor due to the interaction point where the two intervals collided. The complete evolution that sweeps out a sphere with four holes is obtained by squaring this operator and taking the renormalized trace:
\begin{align}
   \lim_{\eps\to 0} \frac{1}{|G|^4} \Tr \rho = \lim_{\eps\to 0} \frac{1}{|G|^4}\sum_{R} \frac{e^{- V(\eps) C_{2}(R)} }{(\dim R/|G|)^2} \sum_{a,b,c,d} =Z (S^2) \; .
\end{align}
We could repeat this exercise for an arbitrary number of intervals. Increasing the number of intervals increases the number of endpoints as well as the number of interaction points. These two effects give compensating factors that always lead to the same sphere partition function.  

\section{Proca on the edge}\label{sec:Proca}

Although the massive case has not garnered as much attention as its massless counterpart, edge mode contributions are naturally anticipated in quantities such as entanglement entropy. This stems partly from the fact that the Maxwell contact term's technical origin, as noted in \cite{Kabat:1995eq}, lies in the linear curvature coupling $\mc{R}$ in the kinetic term $S\sim\int A \left(-\nabla^2+\mc{R} \right)A+\dots$, which is also present for massive spinning fields \cite{Solodukhin:2015hma}. Moreover, although massive tensor theories lack gauge symmetry, they do have transversality constraints that follow from the equation of motion. From a global point of view (further elaborated in section \ref{sec:global}), one might then expect the appearance of edge modes. This latter point was discussed in \cite{Blommaert:2018rsf} for the Proca field on Rindler space.

In fact, it was found in the explicit examples of the $dS_D$ static patch \cite{Anninos:2020hfj} and the static BTZ and the Nariai black holes \cite{Grewal:2022hlo} that the Euclidean partition function for a massive vector field (and its higher spin generalizations) exhibits a bulk-edge split 
\begin{align}\label{eq:procabulkedge}
    Z_{\rm PI}=& Z_{\rm bulk}  Z_{\rm edge} \; .
\end{align}
Similar to the Maxwell case, $Z_{\rm bulk}$ has an independent definition as a quasicanonical partition function, while $Z_{\rm edge}$ takes the form of a path integral for a ghost scalar with mass $m^2$ on a codimension-two sphere. The explicit formulas for the case of the $dS_D$ static patch are given in appendix \ref{appsec:proca}. In \cite{Grewal:2022hlo}, the bulk-edge split \eqref{eq:procabulkedge} was understood from the perspective of the QNM method \cite{Denef:2009kn}, and it was argued that $Z_{\rm edge}$ is related to a special subset of QNMs.

In light of our comprehensive analysis of the Maxwell case in the preceding sections, it is natural to ask whether $Z_{\rm edge}$ in \eqref{eq:procabulkedge} admits a canonical interpretation as a thermal trace over edge modes living on the stretched horizon as well. Exploring this possibility is the primary focus of this section.

\subsection{Action, variations, and equation of motion}

The free massive vector action in the Proca formulation is 
\be \label{eq:proca_action} S = \int_M \left( -\frac{1}{4} F_{\mu\nu} F^{\mu\nu} - \half m^2 A_\mu A^\mu \right) \, , \ee
which is essentially the Maxwell action plus a mass term. Here we consider the theory on a static spacetime $M$ with metric 
\be	ds^2 = g_{tt} dt^2 + g_{ij} dx^i dx^j \ee
where the components are independent of $t$. To analyze the problem in our framework, we again put a brick wall at a proper spatial distance $\eps$ from the horizon so that the boundary $\partial M_\eps$ of the regulated spacetime $M_\eps$ is static and timelike.

The variation is
\be \label{eq:proca_variation} \delta S = \int_{M_\eps} \delta A^\nu (\nabla^\mu F_{\mu\nu} - m^2 A_\nu) + \int_{\p M_\eps} \delta A^\mu F_{\mu\nu} n^\nu \ee
from which we can read off the equation of motion
\be \label{eq:procaeom} \nabla^\mu F_{\mu\nu} = m^2 A_\nu \ee
and also see that the boundary term 
\begin{align}
    \int_{\p M_\eps} \delta A^\mu F_{\mu\nu} n^\nu 
\end{align}
is the same as for Maxwell. For this reason we will abuse language slightly and refer to PMC or PEC boundary conditions for Proca. 

Taking the divergence of \eqref{eq:procaeom}, one deduces that $A_\mu$ is transverse
\begin{align}\label{eq:procaAeom}
     \dot{A}^t +\frac{1}{\sqrt{-g_{tt}}}\hat\nabla_i \left( \sqrt{-g_{tt}}A^i\right)=0 \;. 
\end{align}
In terms of the ``electric" field
\be E_i = \sqrt{-g^{tt}} F_{ti}\; , \ee
the conjugate momentum for $A_i$, the $t$-component of \eqref{eq:procaeom} says
\begin{align}\label{eq:procaEeom}			
     \hat\nabla_i E^i = -m^2 \sqrt{-g_{tt}}A^t\;.
\end{align}
Unlike the Maxwell case ($m^2=0$), in this case the longitudinal component of $A_i$ is physical; similarly we see the longitudinal component of $E^i$ can take nonzero values.

\subsection{Extending the Donnelly-Wall prescription}

We consider the partition function on the Euclidean manifold $M_E$ given by Wick rotating $M$:
\begin{align}
    Z[M_E] = \int \mathcal{D}A \, e^{-S_E[A]}
\end{align}
where $S_E[A]$ is the Euclidean version of \eqref{eq:proca_action}. The Lorentzian horizon of $M$ is mapped to the Euclidean origin. As described in the beginning of this section, this path integral exhibits a bulk-edge split \eqref{eq:procabulkedge} in multiple examples. As a first step towards understanding the edge partition function, we examine the Euclidean path integral from the perspective of \cite{Donnelly:2015hxa}, whose extension to this case is straightforward. 

Following \cite{Donnelly:2015hxa}, we excise a hole of proper radius $\eps$ from the origin, and consider the family of PMC-like boundary conditions on $\p M_{E,\eps}$: 
\be n^\mu F_{\tau\mu}|_{\p M_{E,\eps}} = \sqrt{g_{\tau\tau}} E_\perp \; , \qquad n^\mu F_{\mu i}|_{\p M_{E,\eps}} = 0 \; , \ee
so that the modified path integral $Z[M_{E,\eps},E_\perp]$ contains an explicit dependence on the prescribed $E_\perp$. We split the field into an on-shell classical part incorporating $E_\perp$ and a fluctuation part satisfying the PMC boundary condition
\be n^\mu F_{\mu \nu}|_{\p M_{E,\eps}} = 0 \; , \qquad \nu = \tau, i \; .\ee
Since the Proca action is quadratic, the partition function factorizes as
\be Z[M_{E,\eps},E_\perp] = e^{-S_{\rm cl}[E_\perp]} Z_{\rm PMC}[M_{E,\eps}] \;. \ee
A proposal mimicking \cite{Donnelly:2015hxa} is that in the $\eps\to 0$ limit integrating over $E_\perp$ reproduces the partition function on the manifold with the hole filled in, i.e.
\begin{align}
    Z[M_E] = \lim_{\eps\to 0} f(\mu\epsilon) Z_{\rm PMC}[M_{E,\eps}] Z^{\rm DW}_{\rm edge}[\p M_{E,\eps}] \;,
\end{align}
for some appropriate $f(\mu\eps)$ with $\mu$ an arbitrary scale to keep things dimensionless. Here we have defined an edge partition function
\be Z^{\rm DW}_{\rm edge}[\p M_{E,\eps}] \equiv \int \mc{D} E_\perp \, e^{-S_{\rm cl}[E_\perp]} \;. \label{eq:procaDWedge}\ee
Following \cite{Donnelly:2015hxa}, we only include configurations of $E_\perp$ with $\p_\tau E_\perp = 0$. As pointed out in subsection \ref{subsec:measure}, this prescription of \cite{Donnelly:2015hxa} does reproduce the edge partition function \eqref{eq:procabulkedge} as long as a correct path integral measure is employed. In the Maxwell case, there is a zero mode subtlety, leading to the measure \eqref{eq:ourmeasure}. In the Proca case, since such a subtlety is absent, the measure is simply
\be  \mc{D} E_\perp =  \prod_{k=0}^\infty \frac{dE_{\perp,k}}{2\pi\mu} \; .\ee
To evaluate \eqref{eq:procaDWedge}, we make the ansatz that $A_\tau$ is the only nonzero component of the classical solution, and furthermore that $\p_\tau A_\tau=0$. The on-shell action then reduces to
\be \ba S_{\rm cl}[E_\perp] 
& = \half \int_{\p M_{E,\eps}} A^\mu F_{\mu\nu} n^\nu  = \half\beta\int_{\p\Sigma_\eps} \sqrt{g^{\tau\tau}} A_\tau E_\perp \ea \ee
where $\beta = \int_{S^1} d\tau$ is the coordinate length of the thermal circle. The next step is to explicitly solve for $A_\tau|_{\p\Sigma_\eps}$ in terms of $E_\perp$ using the $\tau$ component of the equation of motion,
\be \label{eq:procaEueom} \ba m^2 A^\tau & = \nabla_\mu F^{\mu\tau} = \frac{1}{\sqrt{g}} \p_i (\sqrt{\det g_{ij}} \sqrt{g^{\tau\tau}} \p^i A_\tau) \; . 
\ea \ee
Let us adopt Gaussian normal coordinates for $\p\Sigma_\eps$ within $\Sigma_\eps$, so that the metric reads
\be ds^2 = g_{\tau\tau} d\tau^2 + dr^2 + g_{ab} dx^a dx^b \; . \ee
Reapplying the arguments in subsection \ref{sec:stretch_horiz}, at leading order in $r$, \eqref{eq:procaEueom} simplifies to
\be m^2 A_\tau \approx \frac{1}{\sqrt{g^{\tau\tau}}} \p_r (\sqrt{g^{\tau\tau}} \p_r A_\tau) + \nabla_a \nabla^a A_\tau \; . \ee
Comparing with \eqref{eq:betaODE}, we see that the $m^2$ term's whole effect is to shift the eigenvalue of the transverse Laplacian $\Delta_{\p\Sigma_\eps} = -\nabla_a \nabla^a$. Therefore we can borrow our earlier results to get the partition function. We conclude that
\begin{align}\label{eq:ZedgeDWproca}
   Z^{\rm DW}_{\rm edge}[\p M_{E,\eps}] = {\rm det} \left(\frac{\log\eps^{-1}}{\kappa\beta} (\Delta_{\p\Sigma_\eps} + m^2) \right)^{1/2} \, . 
\end{align}
Note that except for the factor $\frac{\log(\eps^{-1})}{\kappa\beta}$, this takes the form of a ghost scalar of mass $m^2$ living on $\partial\Sigma_\eps$, in agreement with the cases studied in \cite{Anninos:2020hfj,Grewal:2022hlo}.

\subsection{Classical phase space analysis}

Since the Donnelly-Wall prescription works for Proca in the same manner as the Maxwell case, one might wonder if $Z^{\rm DW}_{\rm edge}[\p M_{E,\eps}]$ has a canonical interpretation as a trace over an edge Hilbert space as we have shown in the Maxwell case. Having the same boundary term in its action variation \eqref{eq:proca_variation} as Maxwell, Proca does admit the DEM boundary condition
\be A_t|_{\p M} = 0, \qquad n^\mu F_{\mu i}|_{\p M} = 0 \; . \label{eq:procaDEM}\ee
The expression for the symplectic form also coincides with Maxwell,
\begin{align}
    \Omega  = \int_\Sigma \delta A^i \delta E_i\;, 
\end{align}
although one should keep in mind that the constraints for the two theories differ. The time component $A_t$ and its time derivative are absent in the symplectic form, so it is natural to use the constraints to eliminate them, leaving the spatial components $A_i, E_i$ as the independent degrees of freedom. Note that aside from the boundary condition \eqref{eq:procaDEM}, $A_i, E_i$ are not subject to any constraints. 

Inspired by our analysis in the Maxwell case, one naturally expects that edge modes explaining $Z^{\rm DW}_{\rm edge}[\p M_{E,\eps}]$ should be scalars having the following properties:
\begin{enumerate}
    \item They are uniquely parametrized by the normals of $A_i, E_i$ on $\p M_\eps$.

    \item Their gradients give the edge parts of $A_i$, $E_i$.
\end{enumerate}
This motivates us to attempt splits similar to \eqref{eq:Esplit} and \eqref{eq:Asplit},
\be A_i = \tilde A_i + \hat\nabla_i \alpha, \qquad E_i = \tilde E_i + S \hat\nabla_i \beta \; . \label{eq:procasplit}\ee
Here $S$ is an arbitrary positive scalar function. We require $\alpha, \beta$ to capture the boundary normals of $A_i, E_i$, so that $\tilde A_i, \tilde E_i$ satisfy the PMC boundary condition\footnote{One can see that the PMC boundary condition implies the vanishing of $n^\mu A_\mu|_{\p M_\eps}$ by taking the divergence of the boundary condition and using the equation of motion: $0 = \nabla^\mu (F_{\mu\nu} n^\nu)|_{\p M_\eps} = m^2 A_\nu n^\nu|_{\p M_\eps}$. This also relies on $\nabla^{[\mu} n^{\nu]}=0$, which is familiar from the definition of extrinsic curvature.}
\be n^i \tilde A_i|_{\p \Sigma_\eps} = n^i \tilde E_i|_{\p\Sigma_\eps} = 0 \; . \ee
We can take $\tilde A_i, \alpha|_{\p\Sigma_\eps}, \tilde E_i, \beta|_{\p\Sigma_\eps}$ as the independent variables. Note that $\tilde A_i, \tilde E_i$ are arbitrary vectors on $\Sigma_\eps$ aside from satisfying the PMC boundary condition. This very general approach fails to yield a split in the symplectic form:
\be \ba \Omega & = \Omega_{\rm bulk} + \Omega_{\rm cross} + \Omega_{\rm edge} \; . \ea \ee
While the ``bulk" and ``edge" terms
\begin{align}
    \Omega_{\rm bulk}  =\int_{\Sigma_\eps} \delta \tilde A^i \delta\tilde E_i\; , \qquad \Omega_{\rm edge} =\int_{\p\Sigma_\eps} \delta\alpha \, S \, n^i \hat\nabla_i \delta\beta\;,
\end{align}
look familiar from the Maxwell case, this distinction is of questionable meaning here because the cross term is non-vanishing:
\begin{align}\label{eq:procacross}
    \Omega_{\rm cross} = \int_{\Sigma_\eps} \left[ \hat\nabla_i \left(S \delta\tilde A^i\right) \delta\beta-\delta\alpha \hat\nabla^i \left(\delta\tilde E_i + S \hat\nabla_i\delta\beta\right) \right] \, ,
\end{align}
In fact, $\hat\nabla^i \delta \tilde E_i$ and $\hat\nabla_i(S\delta\tilde A^i)$ are completely arbitrary scalar functions on $\Sigma$ and so their contributions to $\Omega_{\rm cross}$ can only vanish if $\delta\alpha = \delta\beta=0$. Similarly one can show that the Hamiltonian does not split under \eqref{eq:procasplit} either. Hence, we conclude that a split of the general form \eqref{eq:procasplit} does not describe {\it independent} bulk and edge degrees of freedom.

Since we do not have a factorized classical phase space, upon quantization the Hilbert space does not factorize into a product 
\begin{align}
    \mathcal{H} \neq \mathcal{H}_{\rm bulk} \otimes \mathcal{H}_{\rm edge} \; .
\end{align}
In particular, $Z^{\rm DW}_{\rm edge}[\p M_{E,\eps}]$ cannot be described as coming from edge degrees of freedom that are completely decoupled from the bulk ones, in sharp contrast to the Maxwell case. The question of whether an approximate bulk-edge split emerges in the $\eps\to 0$ limit remains open, but is beyond the scope of this paper.


\section{The global point of view}
\label{sec:global}

\begin{figure}[H]
    \centering
   \includegraphics[width=0.35\textwidth]{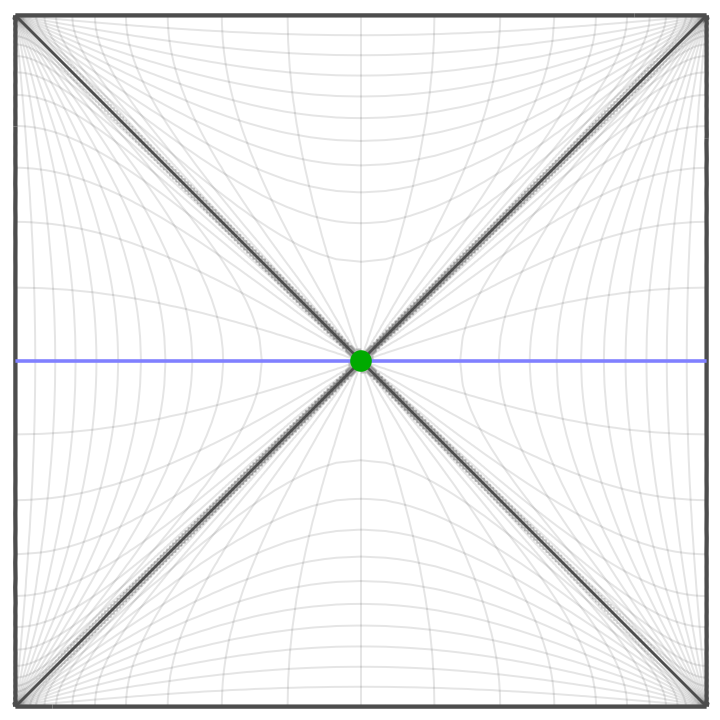}
        \caption{Penrose diagram for the global $dS$ geometry. The manifolds $M_L$, $M_R$ are the left and right static patches. The global Cauchy slice $Y$ is shown in blue. The interface $\p\Sigma$ is the green dot.}
        \label{fig:dSpenrose}
\end{figure}

Thus far we have considered the physics inherent to a fixed Lorentzian manifold $M$ with boundary. We now return to the perspective of the introduction and view $M$ as a subregion of some larger ambient spacetime. We partition a global Cauchy surface into two pieces as $Y = \Sigma_L \cup \Sigma_R$, and we define $M_L$, $M_R$ as the causal domains of $\Sigma_L$, $\Sigma_R$ respectively. We write $\p\Sigma \equiv \p\Sigma_L = \p\Sigma_R$ for the interface separating the left and right slices. A good example to have in mind is that of two complementary static patches in de Sitter space. See figure \ref{fig:dSpenrose}.
As mentioned in the introduction, even in scalar theories there are UV obstructions to the na\"ive Hilbert space factorization
\be \mc{H}_Y \stackrel{?}{=} \mc{H}_{\Sigma_L} \otimes \mc{H}_{\Sigma_R} \; . \ee
This manifests at the classical level as the fact that independently chosen field configurations on $\Sigma_L$ and $\Sigma_R$ will generically be discontinuous across $\p\Sigma$. The scalar's gradient will then have a delta function at $\p\Sigma$, and its Hamiltonian (whose density is the gradient squared) will diverge. This will dynamically damp out such discontinuous configurations, and spoil the independence of left and right. However, most QFT calculations require a UV regulator anyway, and most UV regulators solve this factorization issue, so it is often not emphasized. We will likewise ignore it in our discussion of classical phase space factorization. In gauge theory there are additional obstructions to factorization that can be thought of as coming from the IR \cite{Donnelly:2011hn, Donnelly:2014gva}. Independently chosen field configurations on $\Sigma_L$ and $\Sigma_R$ will generically violate the Gauss constraint at $\p\Sigma$. Specifically, the normal component in $\hat\nabla_i E^i$ will have a delta function at $\p\Sigma$. Violating a constraint is already a problem at the level of kinematics, and unlike the UV obstructions, it is associated with finite energy and therefore not damped out upon incorporating dynamics. The resolution at the level of the classical phase space was explained in \cite{Donnelly:2016auv}. When one includes edge modes in the left and right phase spaces $\Gam_L$, $\Gam_R$ then one can obtain the full phase space $\Gam_Y$ by taking the product space $\Gam_L \times \Gam_R$, restricting to the subspace where the constraint is satisfied, and quotienting by the newly symplectically trivial modes conjugate to the constraint-violating configurations. Our bulk-edge factorization in \eqref{eq:phase_fac} sharpens this procedure so that the left and right bulk modes factorize from the start and simply go along for the ride. After explaining this in detail we present the quantum version of the factorization story, in which our DEM boundary condition plays a prominent role, and we tie back to entanglement entropy.

\subsection{Phase space factorization along subregions}

In this subsection we ignore UV issues, so there is no need to introduce a brick wall. Accordingly there is no need for a boundary condition, although we still have the bulk-edge factorization $\Gam_L = \Gam_{L,{\rm bulk}} \times \Gam_{L,{\rm edge}}$ and similarly for $\Gam_R$. We work on a general manifold in Riemann normal coordinates
\be ds^2 = -dt^2 + g_{ij} dx^i dx^j \ee
where $g_{ij}$ can depend on time. We use a hat to emphasize spatial covariant derivatives $\hat\nabla_i$. As before, we can set $A_t=0$ and take the spatial components $A_i$ and $E_i = \dot A_i$ as the degrees of freedom on $Y$, modulo gauge transformations. The electric field obeys the Gauss constraint
\be \hat\nabla_i E^i = 0 \;. \ee
Let $n^i$ be the unit normal vector on $\p\Sigma$, oriented outward from $\Sigma_L$. Define the outward normal electric fluxes
\be E_\perp^L = n^i E_i|_{\p\Sigma} = -E_\perp^R \;. \ee
Similar to the DEM construction, we can parametrize part of the electric field in terms of $E_\perp^L$. We define $\beta^L, \beta^R$ as scalars on $\Sigma_L, \Sigma_R$ such that
\be \hat\nabla_i (S\hat\nabla^i \beta^L) = \hat\nabla_i(S\hat\nabla^i \beta^R) = 0 \ee
where $S$ is some positive scalar and
\be n^i S \hat\nabla_i \beta^L|_{\p\Sigma} = -n^i S \hat\nabla_i \beta^R|_{\p\Sigma} = E_\perp^L \;. \ee
Then on each side we have the decomposition\footnote{Note that $E_i - \tilde E_i$ is not globally a gradient. We expressed it as distinct gradients on $\Sigma_L$ and $\Sigma_R$, and its normal component on $\p\Sigma$ is defined through a limit from either side.}
\be \ba x \in \Sigma_L: \quad E_i & = \tilde E_i + S \hat\nabla_i \beta^L \; , \\
x \in \Sigma_R: \quad E_i & = \tilde E_i + S \hat\nabla_i \beta^R \; . \ea \ee
By construction $\tilde E_i$ satisfies the Gauss constraint and has vanishing normal component on $\p\Sigma$. Thus the restriction of $\tilde E_i$ to $\Sigma_L$ amounts to the bulk degrees of freedom for $\Sigma_L$, and likewise for the restriction of $\tilde E_i$ to $\Sigma_R$. Only the gradient part of $E_i$ resists a clean split. We can do the same thing for $A_i$, choosing the Coulomb-like gauge
\be \hat\nabla_i (S A^i) = 0 \ee
and defining
\be \ba x \in \Sigma_L: \quad A_i & = \tilde A_i + \hat\nabla_i \alpha^L \; , \\
x \in \Sigma_R: \quad A_i & = \tilde A_i + \hat\nabla_i \alpha^R \; , \ea \ee
where $\tilde A_i$ satisfies the same gauge and $n^i \tilde A_i|_{\p\Sigma} = 0$. The symplectic form is then
\be \label{eq:global_symp_edge} \ba \Omega & = \int_Y \delta A_i \delta E^i  = \int_{\Sigma_L} \delta \tilde A_i \delta \tilde E^i + \int_{\Sigma_R} \delta \tilde A_i \delta \tilde E^i + \int_{\p\Sigma} \delta(\alpha^L - \alpha^R) \delta E_\perp^L \;. \ea \ee
There is one conjugate pair for each point of $\p\Sigma$, with $\alpha^L - \alpha^R$ being determined by $n^i A_i|_{\p\Sigma}$.

Now consider starting with $\Sigma_L, \Sigma_R$ as two a priori separate regions and applying the decomposition of section \ref{sec:CovPhaseEdge}. Refer to $\Sigma_L$'s edge modes as $\alpha^L, E_\perp^L$ and $\Sigma_R$'s edge modes as $\alpha^R, E_\perp^R$. Defining
\begin{align}
    \alpha^\pm & \equiv \half (\alpha^L \pm \alpha^R) \;, \qquad E_\perp^\pm  \equiv \half (E_\perp^L \pm E_\perp^R) \; , 
\end{align}
which is local on the shared boundary $\p\Sigma$, the total symplectic form is
\be \ba \Omega & = \int_{\Sigma_L} \delta A_i \delta E^i + \int_{\Sigma_R} \delta A_i \delta E^i \\
& = \int_{\Sigma_L} \delta \tilde A_i \delta \tilde E^i + \int_{\Sigma_R} \delta \tilde A_i \delta \tilde E^i + \int_{\p\Sigma} (\delta\alpha^L \delta E_\perp^L + \delta\alpha^R \delta E_\perp^R) \\
& = \int_{\Sigma_L} \delta \tilde A_i \delta \tilde E^i + \int_{\Sigma_R} \delta \tilde A_i \delta \tilde E^i + 2 \int_{\p\Sigma} (\delta\alpha^+ \delta E_\perp^+ + \delta\alpha^- \delta E_\perp^-) \;. \ea \ee
We can recover the global phase space from this disjoint phase space by imposing the Gauss constraint across $\p\Sigma$, which amounts to continuity of the normal electric field, i.e. $E_\perp^L = -E_\perp^R$ or equivalently $E_\perp^+ = 0$. Then $\alpha^+$ is symplectically trivial and is quotiented out, and $E_\perp^- = E_\perp^L$ so we are left with
\be \Omega = \int_{\Sigma_L} \delta \tilde A_i \delta \tilde E^i + \int_{\Sigma_R} \delta \tilde A_i \delta \tilde E^i + \int_{\p\Sigma} \delta(\alpha^L - \alpha^R) \delta E_\perp^L \; . \ee
This is precisely the expression in \eqref{eq:global_symp_edge} for the global phase space, confirming explicitly that the fusion of the left and right phase spaces does indeed recover the original global phase space (up to UV obstructions). Furthermore we saw exactly which modes are lost upon imposing the Gauss law across $\p\Sigma$. In particular, no bulk modes were harmed. This gives a precise meaning to the familiar refrain that lifting the Gauss constraint liberates new degrees of freedom.

\subsection{Hilbert space factorization along subregions}

\begin{figure}[H]
    \centering
   \includegraphics[width=0.35\textwidth]{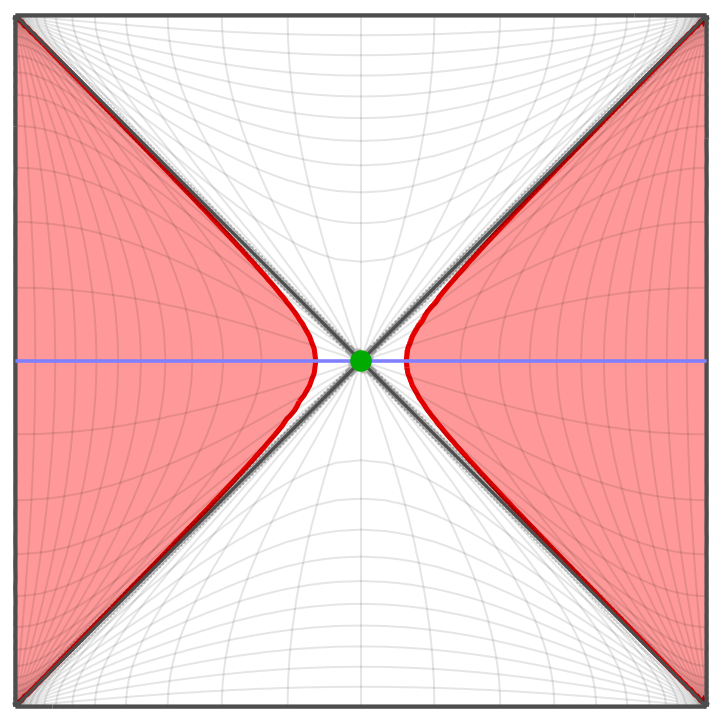}
        \caption{The brick-wall-regulated patches $M_L^\eps$, $M_R^\eps$ are shaded red, and bounded by the brick wall in bold red.}
        \label{fig:dbl_brick}
\end{figure}

In our discussion of classical factorization we ignored UV obstructions, but in our quantum discussion we will keep them front and center. They are one of the reasons we introduce a brick wall. Most of our comments in this subsection apply whenever $M_L$, $M_R$ are static, but for concreteness we will focus on de Sitter space. We place a brick wall at a proper spatial distance $\eps$ from the cosmological horizon. We write the left and right brick-wall-regulated patches as $M_L^\eps, M_R^\eps$, and their respective time slices as $\Sigma_L^\eps, \Sigma_R^\eps$. See figure \ref{fig:dbl_brick}.
Given this regularization and a choice of local boundary condition, we can quantize the theory within the usual QFT framework. The presence of the brick wall renders the (bulk) spectrum of the QFT discrete; this in turn implies that the algebra of operators is Type I, the same as ordinary quantum mechanics.\footnote{In free QFT, an individual mode is usually equivalent to a harmonic oscillator of the same frequency. The edge modes are an exception to this, being equivalent instead to non-relativistic free particles in quantum mechanics. The Hilbert space for an individual edge mode is $L^2(\mathbb{R})$. The bounded operators on this Hilbert space form an ordinary Type I algebra.} The question is now whether the regularized global Hilbert space $\mc{H}^\eps$ is a good approximation to the original global Hilbert space $\mc{H}$. We take shrinkability of the boundary condition on the Euclidean Wick-rotated manifold as strong evidence that the global Lorentzian physics is recovered in the $\eps\to 0$ limit. Therefore even though the operator algebra on $M_L^\eps$ is \textit{not} a subalgebra of the Type III algebra on $M_L$, it can be viewed as a Type I \textit{approximation} to this Type III algebra \cite{Soni:2023fke}. In the case of Maxwell we have argued in sections \ref{sec:shrink} and \ref{sec:further} that the DEM boundary condition is shrinkable, so let us specialize to this case in the following discussion.

\begin{figure}
    \hfill (a){
    \includegraphics[scale=0.08]{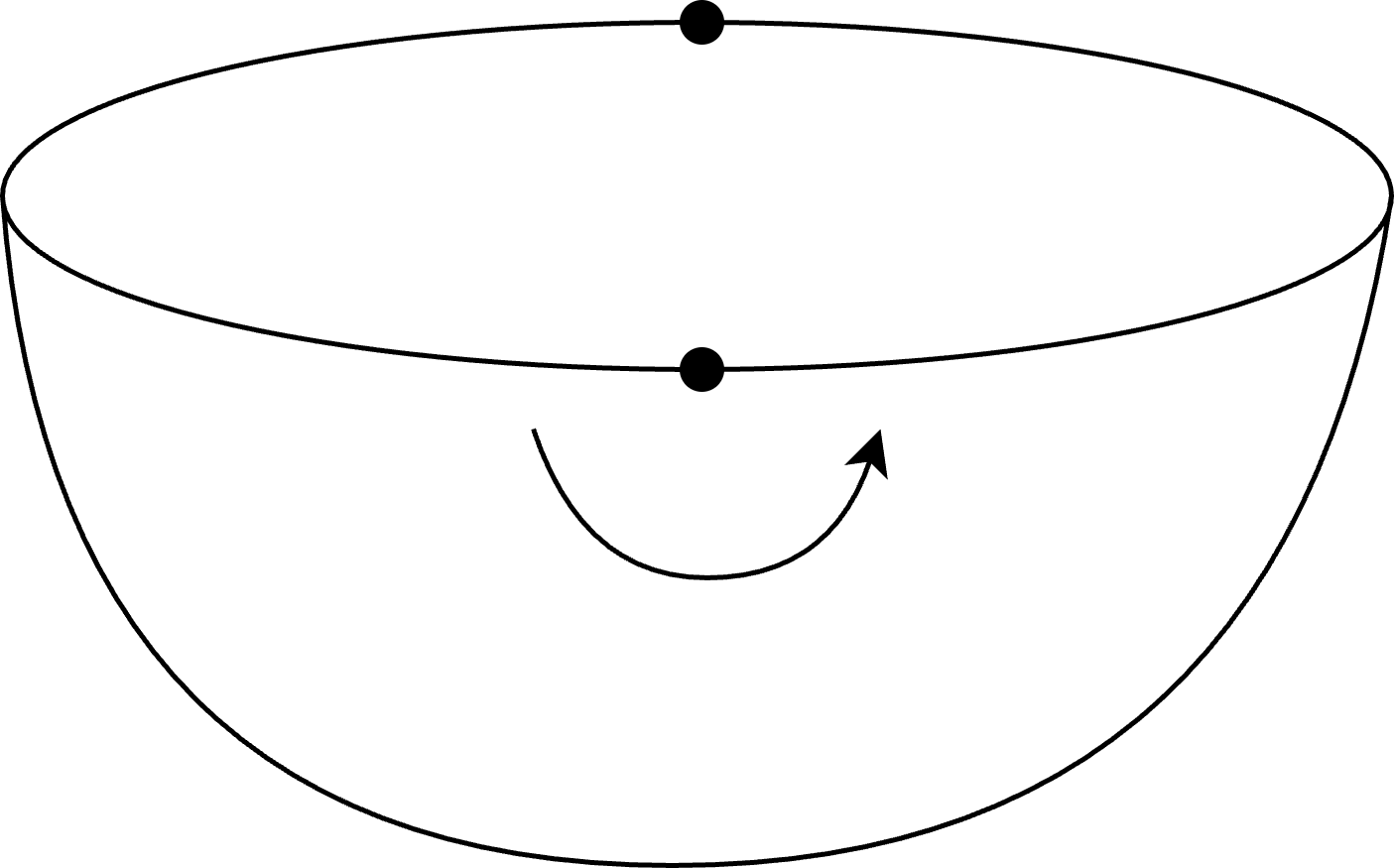}}
    \hfill (b){
    \includegraphics[scale=0.08]{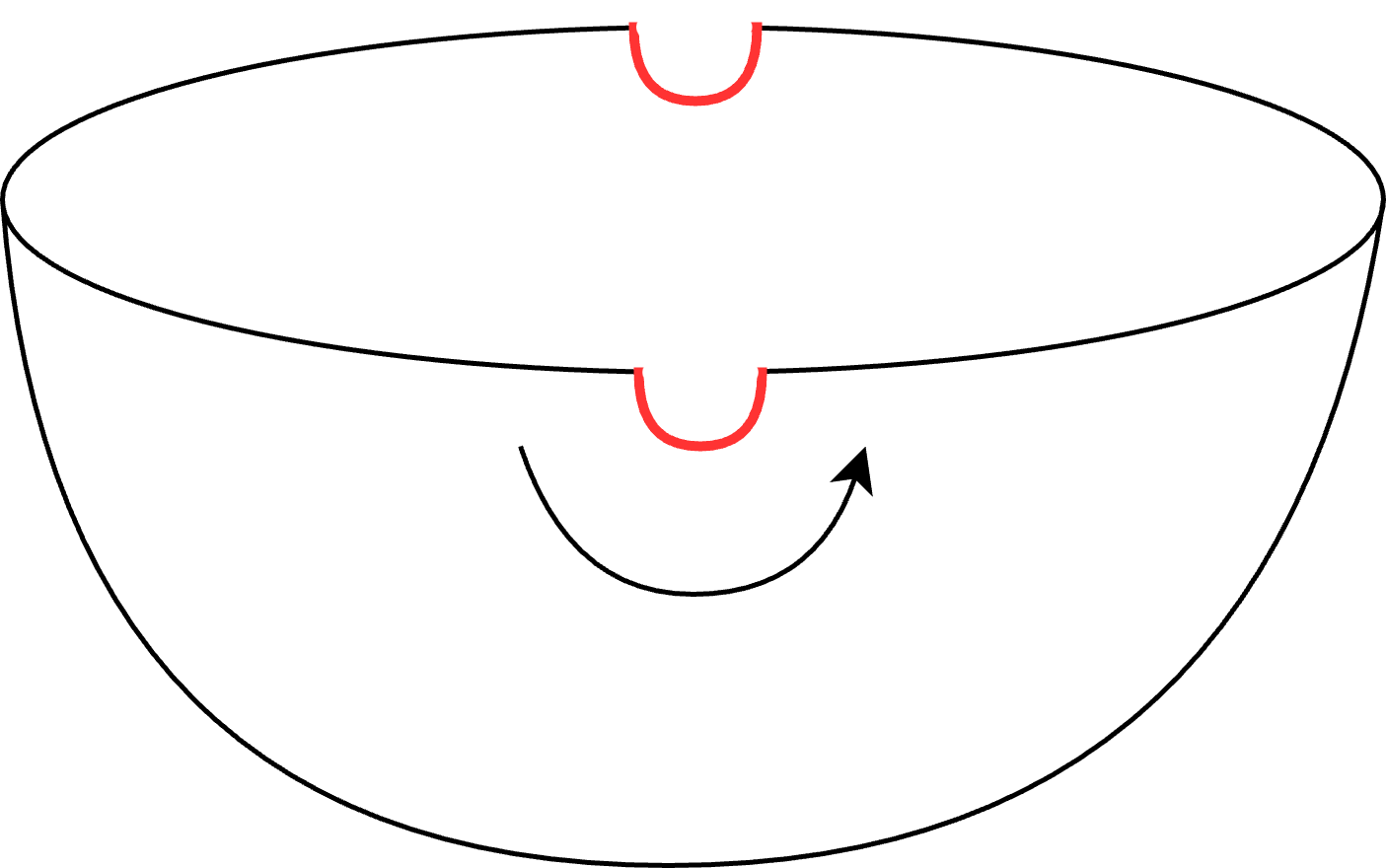}}
    \hfill\break
    \caption{(a) The Hartle-Hawking vacuum of global $dS_2$ is prepared with the Euclidean path integral on a hemisphere. The hemisphere and state are non-singular, but the flow of Wick-rotated static time (indicated by the arrow) has fixed points, which obstructs the factorization interpretation. (b) The regulated Hartle-Hawking vacuum uses a hemisphere with two small half-disks cut out, bounded by the brick wall (shown in red) on which we place the renormalized DEM boundary condition. The flow of Wick-rotated static time has no fixed points.}
    \label{fig:HH}
\end{figure}

The left and right regions are disjoint and the DEM boundary condition does not correlate them, so the regulated global Hilbert space is manifestly the tensor product of the left and right Hilbert spaces,
\be \mc{H}^\eps = \mc{H}_L^\eps \otimes \mc{H}_R^\eps \;. \label{eq:tensorHilbert}\ee
Shrinkability implies that $\mc{H}^\eps$ somehow approximates the original global Hilbert space $\mc{H}$ in the $\eps\to 0$ limit, but this does not guarantee any one-to-one correspondence between them. In fact we know this is not the case because, as in the classical case, generic independently chosen states from $\mc{H}_L^\eps$ and $\mc{H}_R^\eps$ have mismatched electric flux through $\p\Sigma_L$ and $\p\Sigma_R$. These states must somehow be tossed out in the $\eps\to 0$ limit. One practical advantage of the DEM boundary condition is that this happens almost automatically. Recall that the Hartle-Hawking vacuum\footnote{In the context of de Sitter space, this is also known as the Euclidean or Bunch-Davies vacuum \cite{Chernikov:1968zm,Bunch:1978yq}.} is prepared by the Euclidean path integral on a hemisphere of $S^D$. Likewise our regulated vacuum can be prepared by the Euclidean path integral on a hemisphere with a small half-ball cut out due to the brick wall. See figure \ref{fig:HH} for the case of $dS_2$. By a standard argument this path integral can be written as the exponential of the static Hamiltonian,\footnote{Technically $e^{-\pi H}$ is a linear operator on $\mc{H}_L^\eps$, as opposed to a state on $\mc{H}_L^\eps \otimes \mc{H}_R^\eps$. However they are naturally related by the adjoint on $\mc{H}_L^\eps$.}
\be \ket{0}_\eps = e^{-\pi H} = e^{-\pi H_{\rm bulk}} e^{-\pi H_{\rm edge}} \ee
where $H$ is the static time generator on $M_L^\eps$, normalized to unit surface gravity. Thanks to our brick wall regulator, $e^{-\pi H}$ actually exists, and thanks to our bulk-edge factorization we can isolate the edge part. The edge part can be evaluated explicitly as
\begin{align}
    e^{-\pi H_{\rm edge}} = \int \mathcal{D}E_{\perp} \, e^{-\frac{\pi}{2} \int  E_{\perp} \frac{1}{ K} E_{\perp}}\ket{E_\perp}  \bra{E_\perp} = \prod_{k\neq 0}  \int \frac{dE_{\perp,k}}{2\pi\mu} \, e^{\frac{-\pi E_{\perp,k}^2}{2\lam_k\log\frac{1}{\eps}}} \ket{E_{\perp,k}} \bra{E_{\perp,k}} \;.
\end{align}
This is diagonal in $\ket{E_{\perp,k}}$, which follows from the free evolution of the edge modes with respect to static time. This means that even at finite regulator the vacuum state sits in the diagonal subspace where the left and right normal electric fields match. Now, consider any state prepared by a path integral on a regulated hemisphere with insertions a finite distance from the boundary. Since the edge modes are physically localized near the brick wall as discussed in subsection \ref{sec:stretch_horiz}, we expect that these states will retain the same diagonal property as the vacuum in the $\eps\to 0$ limit. Therefore, for this general class of states, the unphysical constraint-violating states will not contribute to observables in the $\eps\to 0$ limit. It is in this sense that $\mc{H}^\eps$ approximates $\mc{H}$.

Finally let us return to the discussion of entanglement entropy. Because $\mc{H}^\eps$ factorizes, the process of partially tracing to get a reduced density matrix is well-defined. Starting from the regulated vacuum, partially tracing over $\mc{H}_R^\eps$ gives the (unnormalized) density matrix
\be \rho_L = e^{-2\pi H} = e^{-2\pi H_{\rm bulk}} e^{-2\pi H_{\rm edge}} \ee
on $\mc{H}_L^\eps$. Our shrinkability results above show that for some factor $f(\mu\eps)$ the sphere partition function is recovered by $\lim_{\eps\to 0} f(\mu\eps) \frac{1}{|\mc{G}_{\p\Sigma}|} \Tr \rho_L$. Inspired by \cite{Hung:2019bnq, Jafferis:2019wkd, Agia:2022srj}, this motivates us to \textit{define} the entanglement entropy of Maxwell theory as the $\eps\to 0$ limit of the von Neumann entropy of $\rho_L$ computed with the renormalized trace,
\be S_{\rm EE} \equiv \lim_{\eps\to 0} f(\mu\eps) \frac{-1}{|\mc{G}_{\p\Sigma}|} \Tr[\rho_L \log \rho_L] \; . \ee
As discussed above, for $D=4$ this agreees with the conformal anomaly method of \cite{Casini:2011kv}. Finally we note that $S_{\rm EE}$ can be rewritten as the thermal entropy on $M_L^\eps$, also computed with the renormalized trace,
\be S_{\rm EE} = \lim_{\eps\to 0} f(\mu\eps) \, (1-\beta\p_\beta) \log \left( \frac{1}{|\mc{G}_{\p\Sigma}|} \Tr e^{-\beta H} \right) \Big|_{\beta=2\pi} \; . \ee
This too can be split into bulk and edge parts.

\section{Outlook}
\label{sec:Discussion}
In this work we gave a canonical calculation of entanglement entropy in pure Maxwell theory, in which edge modes played an essential role. First, they allow a breaking of the Gauss constraint across the entangling surface, thereby allowing a factorization of the Hilbert space. Second, their contribution is required for shrinkability in general dimensions and for the entanglement entropy's log coefficient to agree with the conformal anomaly in four dimensions. An interesting alternative to the introduction of edge modes was proposed in \cite{Casini:2019nmu}, where the effects of heavy, \textit{physical} charges were incorporated into the entanglement entropy calculation in the IR Maxwell theory. More precisely, \cite{Casini:2019nmu} computed the mutual information between two regions separated by a physical corridor of size $\delta$,\footnote{In the presence of the corridor, the operator algebra factorizes into a tensor product, and this removes the subtleties associated to Type III algebras.} which is interpreted as a regularized form of entanglement entropy.\footnote{The mutual information can be defined without reference to density matrices and entanglement entropies, and is thus a well-defined observable to compute in the algebraic approach.} By including the effects of both electric and magnetic charges in the regime where their inverse masses are much smaller than $\delta$, \cite{Casini:2019nmu} found that the regulated entanglement entropy had a log coefficient equal to the trace anomaly.  Interestingly, a codimension-two scalar also appears in their calculation, due to integrating out heavy charges. This suggests there might be a deeper principle connecting these two perspectives.

Several interesting extensions of our work come to mind immediately. For example the generalization to $p$-form gauge theory, which appears in the study of higher-form symmetries and string theory, seems potentially straightforward. In particular, the Hodge-Morrey-Friedrichs decomposition mentioned above applies equally well to $p$-forms. Also, in the context of (anti-)de Sitter space, there are more exotic (partially) massless fields \cite{Deser:1983tm,DESER1984396,Higuchi:1986py,Brink:2000ag,Deser:2001pe,Deser:2001us,Deser:2001wx,Deser:2001xr,Zinoviev:2001dt,Hinterbichler:2016fgl}. The characterization of classical edge modes in \cite{Donnelly:2016auv} was for Yang-Mills theory and Einstein gravity, whereas we have restricted attention to Maxwell theory. Our results on the bulk-edge factorization of phase space, the edge Hamiltonian, and the DEM boundary condition all have plausible analogues in these even richer theories. In all cases one na\"ively expects the edge Hilbert space to be that of a gauge-parameter-valued field on the boundary $\p M$. Understanding them would potentially provide valuable insight into some of the pressing problems of our field, including the IR divergence structure of Yang-Mills theory, the microscopic entropy of horizons, and holography. We note that some potential stumbling blocks in the case of gravity are the ill-posedness of the initial boundary value problem with the standard Dirichlet boundary condition associated to the Gibbons-Hawking-York term \cite{Anderson_2008, An:2021fcq}, and the seeming inherent nonlocality of shrinkable boundary conditions in full gravity \cite{Jafferis:2019wkd}. To address the latter issue, the authors of \cite{Mertens:2022ujr, Wong:2023bhs} formulated a framework that incorporates such nonlocal boundary conditions in the context of 3D gravity. It is interesting to note that the computation in \cite{Mertens:2022ujr, Wong:2023bhs} of the bulk entanglement entropy due to gravitational edge modes also required a division by the total volume of the edge mode symmetry, including the zero mode.

There are enticing avenues to explore within Maxwell theory as well. It was argued in \cite{Hawking:2016msc} that black hole horizons carry electromagnetic soft hair containing a UV-divergent amount of information about the black hole, and that charge conservation constrains the end product of black hole evaporation. Our results provide the tools necessary to regulate and quantify this UV-divergent amount of information. Another intriguing connection is the resemblance of our edge action to the codimension-two action of \cite{Kapec:2021eug} that controls soft exchange in the Maxwell S-matrix. Finally, it would be interesting to study the infinite volume limit of our results and attempt to rigorously recover the soft and Goldstone modes described in \cite{Strominger:2017zoo}. This seems related to results in \cite{Chen:2023tvj} for the 4D conformal case.

\section*{Acknowledgments} 

It is a great pleasure to thank Nicholas Agia, Andreas Blommaert, Hong Zhe Chen, William Donnelly, Laurent Freidel, Daniel Jafferis, Javier Mag\'an, Thomas Mertens, Robert Myers, Ana-Maria Raclariu, Ronak Soni, and Aron Wall for useful discussions. AB is supported by the Celestial Holography Initiative at the Perimeter Institute for Theoretical Physics and the Simons Collaboration on Celestial Holography. Research at the Perimeter Institute is supported by the Government of Canada through the Department of Innovation, Science and Industry Canada and by the Province of Ontario through the Ministry of Colleges and Universities. AB was also supported by the US Department of Energy under contract DE-SC0010010 Task F and by Simons Investigator Award \#376208. AL was supported in part by the Croucher Foundation, the Black Hole Initiative at Harvard University and the Stanford Science Fellowship. GW is supported by the Science and Technology Facilities Council grant
ST/X000761/1.  GW was also supported by Harvard CMSA and the Oxford Mathematical Institute. 

\appendix

\section{Getting $Z_{\rm edge}$ without phase space}
\label{appendix:nophase}

In this appendix we show the factorization $Z_{\rm DEM}(\beta) = Z_{\rm bulk}(\beta) Z_{\rm edge}(\beta)$ and compute $Z_{\rm edge}(\beta)$ directly in terms of the configuration space path integral, with no reference to phase space. We use the gauge group $G=\mathbb{R}$ to avoid discussing discrete quotients and sums over bundles, and for further simplicity we use the ultrastatic Euclidean manifold $M = S^1 \times \Sigma$ with metric
\be ds^2 = d\tau^2 + g_{ij} dx^i dx^j \ee
where $g_{ij}$ is independent of $\tau$. As in the body of the paper, we will use a hat to emphasize derivatives on $\Sigma$. First we study the DEM partition function. We postulate that it takes the form
\be Z_{\rm DEM}(\beta) = \int \frac{\mc{D} A}{\mc{G}^s_M} \, e^{-S} \ee
where $\mc{G}^s_M$ is the group of small gauge transformations, i.e. those restricting to zero on the boundary. We start by splitting the gauge field into
\be A_\mu = A_\mu^C + \nabla_\mu \phi \ee
where $\phi|_{\p M}=0$, i.e. it is a small gauge parameter, and $A_\mu^C$ satisfies Coulomb gauge,
\be \hat\nabla^i A_i^C = 0 \; . \ee
Note this does not constrain $A_\tau^C$. Next we split the spatial components of $A_\mu^C$ as
\be A_i^C = \tilde A_i + \hat\nabla_i \alpha \ee
with
\be 0 = \hat\nabla^i \tilde A_i = \hat\nabla^i \hat\nabla_i \alpha \ee
and vanishing normal $n^i \tilde A_i|_{\p M} = 0$ so that the split is unique. We omit spatially constant $\alpha$ modes as degrees of freedom since they drop out. Note that $\alpha$ is spatially harmonic and therefore determined by its boundary values, so by $\mc{D}\alpha$ we implicitly mean only $\alpha|_{\p M}$ as the independent integration variables. We are interested in the change of variables
\be \mc{D}A = \mc{D}A^C_\tau \mc{D}\tilde A_i \mc{D}(\hat\nabla_i\alpha) \mc{D}(\nabla_\mu\phi) \; . \ee
Since $\mc{D}A_\tau = \mc{D}A_\tau^C + \mc{D}(\nabla_\tau\phi)$, and $\mc{D}(\nabla_\tau\phi)$ already appears elsewhere in the measure, we can simply replace $\mc{D}A_\tau^C \to \mc{D}A_\tau$.

The Jacobians can be deduced from the normalization of the path integral measure. We have by definition
\be 1 = \int \mc{D}A \exp[-\half\mu^2\int_M A_\mu A^\mu] \ee
where $\mu$ is an arbitrary mass scale needed to keep the exponent dimensionless. Note this implies that $\mc{D} A$ itself contains factors of $\mu$, e.g. $\mc{D} A_\tau = \prod_k \frac{\mu}{\sqrt{2\pi}} dA_{\tau, k}$ where $A_{\tau, k}$ are the mode coefficients of $A_\tau$. In the present ultrastatic case we assume that $\mc{D}A_\tau$ and $\mc{D}A_i$ are separately normalized, i.e.
\be 1 = \int \mc{D}A_\tau \exp[-\half\mu^2\int_M (A_\tau)^2] = \int \mc{D}A_i \exp[-\half\mu^2\int_M g^{ij} A_i A_j] \; . \ee
From the $\mc{D}A_i$ condition we see
\be \ba 1 & = \int \mc{D}A_i \exp[-\half\mu^2\int_M g^{ij} A_i A_j] \\
& = \int \mc{D} \tilde A_i \mc{D}(\hat\nabla_i\alpha) \mc{D}(\hat\nabla_i\phi) \, \exp[-\half\mu^2\int_M g^{ij} (\tilde A_i \tilde A_j + \hat\nabla_i \alpha \hat\nabla_j \alpha + \hat\nabla_i \phi \hat\nabla_j \phi)] \\
& = \int \mc{D}(\hat\nabla_i\alpha) \mc{D}(\hat\nabla_i\phi) \exp[-\half\mu^2\int_M (\hat\nabla_i \alpha \hat\nabla^i \alpha + \hat\nabla_i \phi \hat\nabla^i \phi)] \\
& = \int \mc{D}(\hat\nabla_i\alpha) \mc{D}(\hat\nabla_i\phi) \exp[-\half\mu^2\int_0^\beta d\tau \int_\Sigma (\hat\nabla_i \alpha \hat\nabla^i \alpha + \hat\nabla_i \phi \hat\nabla^i \phi)] \\
& = \int \mc{D}(\hat\nabla_i\alpha) \mc{D}(\hat\nabla_i\phi) \exp[-\half\mu^2\int_0^\beta d\tau \left( -\int_\Sigma \phi \,\hat\nabla_i \hat\nabla^i\phi + \int_{\p\Sigma} \alpha \, n^i \hat\nabla_i \alpha \right)] \; . \ea \ee
From here we can read off the Jacobians. Defining $\mc{D}\phi = \prod_k \frac{\mu^2}{\sqrt{2\pi}} d\phi_k$, the $\phi$ Jacobian has a factor of the Dirichlet Laplacian determinant $\det_{\rm D}(-\mu^{-2}\hat\nabla_i \hat\nabla^i)^{1/2}$ for each time slice.\footnote{Or more carefully, one for each Fourier mode on the $S^1$.} The boundary field $\alpha$'s Jacobian has a factor of $\det'(\mu^{-2} K)^{1/2}$ for each time slice, where $K$ is the Dirichlet-to-Neumann operator for the Laplacian $-\hat\nabla_i \hat\nabla^i$ on $\Sigma$ and the prime indicates omission of the constant $\alpha$ mode. The full measure is then
\be \frac{\mc{D} A}{\mc{G}^s_M} = \frac{\mc{D}A_\tau \mc{D}\tilde A_i \mc{D}(\hat\nabla_i\alpha) \mc{D}(\hat\nabla_i\phi)}{\mc{G}^s_M} = \mc{D} A_\tau \mc{D} \tilde A_i \mc{D}\alpha \, {\rm det}'(\mu^{-2} K)^{\infty/2} {\rm det}_{\rm D}(-\mu^{-2}\hat\nabla_i \hat\nabla^i)^{\infty/2} \; . \ee
The quotient by $\mc{G}^s_M$ precisely canceled $\mc{D}\phi$; recall both were defined to vanish on the boundary. The factor of infinity in the exponent $\infty/2$ is meant as shorthand for the fact that there is one square root determinant for each time slice. Now we turn to the action. Start by writing
\be \ba S & = \half\int_M \nabla_\mu A_\nu F^{\mu\nu} \\
& = \half\int_M A_\mu \nabla_\nu F^{\mu\nu} \\
& = \half\int_M A_\mu \frac{1}{\sqrt{g}} \p_\nu (\sqrt{g} F^{\mu\nu}) \\
& = \half\int_M A_\mu (\p_\tau F^{\mu\tau} + \frac{1}{\sqrt{g}} \p_i (\sqrt{g} F^{\mu i})) \\
& = \half\int_M \left[ A_i \p_\tau F^{i\tau} + A_\tau \frac{1}{\sqrt{g}} \p_i (\sqrt{g} F^{\tau i}) + A_j \frac{1}{\sqrt{g}} \p_i (\sqrt{g} F^{ji}) \right] . \ea \ee
We want to separate out the $A_\tau$ terms from the $A_i$ terms. The first step is to show that the cross terms vanish. Towards this, note
\be F^{\tau i} = g^{ij} (\p_\tau A_i - \p_i A_\tau) \; . \ee
The terms in the action containing both $A_\tau$ and $A_i$ are
\be S \supset \half\int_M \left[ -A^i \p_\tau \p_i A_\tau + A_\tau \frac{1}{\sqrt{g}} \p_i (\sqrt{g} \, g^{ij} \p_\tau A_j) \right] . \ee
Keeping in mind that our notation suppresses the measure factor $d^Dx \sqrt{g}$, we can integrate by parts to see that these terms cancel (with boundary terms vanishing thanks to the boundary condition). This leaves us with
\be \ba S & = \half\int_M \left[ -A_i \p_\tau (g^{ij} \p_\tau A_i) - A_\tau \frac{1}{\sqrt{g}} \p_i(\sqrt{g} \, g^{ij} \p_i A_\tau) + A_j \frac{1}{\sqrt{g}} \p_i(\sqrt{g} F^{ji}) \right] \\
& = -\half\int_0^\beta d\tau \int_\Sigma A_\tau \frac{1}{\sqrt{g}} \p_i (\sqrt{g} \hat\nabla^i A_\tau) + \half\int_M A_i \frac{1}{\sqrt{g}} \p_\mu (\sqrt{g} \bar F^{i\mu}) \\
& = -\half\int_0^\beta d\tau \int_\Sigma A_\tau \hat\nabla_i \hat\nabla^i A_\tau + \half\int_M A_i \nabla_\mu \bar F^{i\mu} \\
& = -\half\int_0^\beta d\tau \int_\Sigma A_\tau \hat\nabla_i \hat\nabla^i A_\tau + \frac{1}{4}\int_M \bar F_{\mu\nu} \bar F^{\mu\nu} \ea \ee
where $\bar F_{\mu\nu}$ is the field strength computed with $A_\tau$ set to zero, i.e. using only $A_i$. We can see now that the path integral over $A_\tau$ will give a factor of ${\rm det}_{\rm D}(-\mu^{-2}\hat\nabla_i\hat\nabla^i)^{-1/2}$ for each slice, with the Dirichlet condition coming from the fact that $A_\tau|_{\p M}=0$ in the DEM boundary condition. This perfectly cancels the $\phi$ Jacobian, leaving us with
\be \ba Z_{\rm DEM}(\beta) & = {\rm det}'(\mu^{-2}K)^{\infty/2} \int \mc{D}\tilde A_i \mc{D}\alpha \exp[-\half\int_M A_i\nabla_\mu \bar F^{i\mu}] \\
& = {\rm det}'(\mu^{-2}K)^{\infty/2} \int \mc{D}\tilde A_i \mc{D}\alpha \exp[-\half\int_M \left( -A^i \p_\tau^2 A_i + A_i \hat\nabla_j \bar F^{ij} \right)] \, . \ea \ee
Since $\tilde A_i, \hat\nabla_i\alpha$ are orthogonal the $\dot A^i \dot A_i$ term reduces to $\dot{\tilde A}^i \dot{\tilde A}_i + \hat\nabla^i\dot\alpha \hat\nabla_i\dot\alpha$. The $\dot{\tilde A}^i \dot{\tilde A}_i$ term combines with the $A_i\hat\nabla_j\bar F^{ij}$ term to give $A_i\hat\nabla_\mu\tilde F^{i\mu}$, where $\tilde F_{\mu\nu} = \nabla_\mu \tilde A_\nu - \nabla_\nu \tilde A_\mu$. The partition function now reads
\be \ba Z_{\rm DEM}(\beta) & = {\rm det}'(\mu^{-2}K)^{\infty/2} \int \mc{D}\tilde A_i \mc{D}\alpha \exp[-\half\int_M \hat\nabla^i\dot\alpha\hat\nabla_i\dot\alpha - \frac{1}{4} \int_M \tilde F_{\mu\nu} \tilde F^{\mu\nu}] \\
& = {\rm det}'(\mu^{-2}K)^{\infty/2} \int \mc{D}\alpha \exp[-\half\int_{\p M} \dot\alpha \, n^i \hat\nabla_i\dot\alpha] \int \mc{D}\tilde A_i \exp[-\frac{1}{4}\int_M \tilde F_{\mu\nu} \tilde F^{\mu\nu}] \\
& = {\rm det}'(\mu^{-2}K)^{\infty/2} \int \mc{D}\alpha \exp[-\half\int_{\p M} \dot\alpha K \dot\alpha] \int \mc{D}\tilde A_i \exp[-\frac{1}{4}\int_M \tilde F_{\mu\nu} \tilde F^{\mu\nu}] \, . \ea \ee
The partition function has factorized along bulk and edge degrees of freedom. The $\mc{D}\alpha$ integral will give a factor of $\det'(\mu^{-2}K)^{-1/2}$ for each time slice, except that it uses $\dot\alpha$ and not $\alpha$ so we must omit a single factor of the determinant, and that part of $\mc{D}\alpha$ is integrated over freely giving a factor of the volume of the group of large gauge transformations (with the constant mode omitted). This yields
\be Z_{\rm DEM}(\beta) = |\mc{G}'_{\p\Sigma}| \, {\rm det}'(\mu^{-2}K)^{1/2} \int \mc{D}\tilde A_i \exp[-\frac{1}{4}\int_M \tilde F_{\mu\nu} \tilde F^{\mu\nu}] \; . \ee
For the PMC boundary condition we postulate
\be Z_{\rm PMC}(\beta) = \int \frac{\mc{D}A}{\mc{G}_M} \, e^{-S} \ee
where $\mc{G}_M$ is the group of all gauge transformations on $M$. We decompose it as $\mc{G}_M = \mc{G}^s_M \mc{G}_{\p M}$, where $\mc{G}_{\p M}$ is the group of large gauge transformations on $M$, i.e. the space of functions on $\p M$. We can proceed similarly to the DEM case, splitting the field into $A_\tau, \tilde A_i, \hat\nabla_i\alpha, \nabla_\mu\phi$. We incur the same Jacobians as before. The action will still split along these lines, with the $\phi$ term vanishing. But now the $\alpha$ part vanishes too because PMC requires $n^\mu F_{\tau\mu}|_{\p M} = 0$. The $\mc{D}\alpha$ integral then cancels with $\mc{G}_{\p M}$, up to the spatially constant modes which we denote $\mc{G}^0_{\p M}$. This leaves us with
\be \ba Z_{\rm PMC}(\beta) & = {\rm det}'(\mu^{-2}K)^{\infty/2} {\rm det}_{\rm D}(-\mu^{-2}\hat\nabla_i\hat\nabla^i)^{\infty/2} \\
& \quad\times \int \frac{\mc{D}A_\tau}{\mc{G}^0_{\p M}} \exp[-\half\int_0^\beta d\tau \int_\Sigma A_\tau (-\hat\nabla_i\hat\nabla^i) A_\tau] \int \mc{D}\tilde A_i \exp[-\frac{1}{4}\int_M \tilde F_{\mu\nu} \tilde F^{\mu\nu}] \\
& = \left( \frac{{\rm det}'(\mu^{-2}K) {\rm det}_{\rm D}(-\mu^{-2}\hat\nabla_i\hat\nabla^i)}{{\rm det}'_{\rm N}(-\mu^{-2}\hat\nabla_i\hat\nabla^i)} \right)^{\infty/2} \int \mc{D}\tilde A_i \exp[-\frac{1}{4}\int_M \tilde F_{\mu\nu} \tilde F^{\mu\nu}] \, . \ea \ee
The $\mc{D}A_\tau$ integral gave a Neumann Laplacian determinant this time, rather than a Dirichlet determinant as in DEM, because PMC requires
\be 0 = n^i F_{\tau i}|_{\p M} = \p_\tau n^i A_i - n^i\p_i A_\tau \ee
and the gauge parameter $\phi$ drops out of the field strength, so the remaining spatial component is just $\tilde A_i$, whose boundary normal vanishes by definition. Therefore $n^i \p_i A_\tau|_{\p M}=0$, and so its path integral gives the Laplacian determinant with the Neumann boundary condition for each time slice. The spatial zero mode of $A_\tau$ dropped out of the action and was integrated freely, cancelling with $\mc{G}^0_{\p M}$. The ratio of determinants appearing in fact cancels by an identity due to Forman \cite{Forman}. This finally leaves
\be Z_{\rm PMC}(\beta) = \int \mc{D}\tilde A_i \exp[-\frac{1}{4}\int_M \tilde F_{\mu\nu} \tilde F^{\mu\nu}] \ee
and therefore also
\be Z_{\rm DEM}(\beta) = |\mc{G}'_{\p\Sigma}| \, {\rm det}'(\mu^{-2}K)^{1/2} Z_{\rm PMC}(\beta) \; , \ee
in agreement with the more rigorous phase space methods used in the body of the paper.

\section{The constant gauge transformation}
\label{appendix:constgauge}

One may wonder if our omission of the constant gauge transformation in our covariant phase space quotient is consistent with the fact that the correct measure to use in the configuration space path integral of $U(1)$ Maxwell theory on a manifold without boundary is $\frac{\mc{D}A}{\mc{G}_M}$, where $\mc{G}_M$ is the group of all gauge transformations, i.e. $G$-valued functions on $M$, including the constant gauge transformation. This seemingly superfluous quotient turns out to be necessary for locality and unitarity, among other properties \cite{Donnelly:2013tia}. In this appendix we show the consistency of the covariant phase space formalism and the usual configuration space path integral. We do so by reducing the configuration space path integral to a form that manifestly agrees with the covariant phase space formalism. We take the fundamental unit of charge to be $q$, so that our gauge parameter has periodicity $\lam \sim \lam + 2\pi/q$. For simplicity we consider a Euclidean product manifold $M = S^1 \times \Sigma$ where $\Sigma$ has no boundary nor first and second homology groups, $H_1(\Sigma, \mathbb{Z}) = H_2(\Sigma, \mathbb{Z}) = 0$. Then also $H_2(M, \mathbb{Z})=0$ and so there are no nontrivial magnetic bundles. We take the metric
\be ds^2 = d\tau^2 + g_{ij} dx^i dx^j \ee
where $\p_\tau g_{ij}=0$ and $\tau \sim \tau + \beta$. The partition function is
\be Z(\beta) = \int \frac{\mc{D} A}{\mc{G}_M} \, e^{-S} \; . \ee
We evaluate it in Coulomb gauge. That is, we use a single-valued scalar $\phi$ to split the spatial part of the gauge field as
\be A_i = \tilde A_i + \hat\nabla_i \phi \ee
where $\hat\nabla^i \tilde A_i = 0$ and the hat emphasizes that this is a spatial derivative. We omit spatially constant $\phi$ configurations since they drop out. To be fully explicit, we mode expand $\phi$ as
\be \phi(\tau,x) = \beta^{-1/2} \sum_{n\in\mathbb{Z}} e^{2\pi in\tau/\beta} \sum_{k\ne 0} \phi_{n,k} f_k(x) \ee
where $f_k$ are the orthonormal eigenfunctions of the Laplacian on $\Sigma$. Reality of $\phi$ requires $\phi_{-n,k} = \phi_{n,k}^*$. Note the sum is over $k\ne 0$. We also expand elements of $\mc{G}_M$ as
\be \lam(\tau,x) = \frac{2\pi m\tau}{q \beta} + \beta^{-1/2} \sum_{n\in\mathbb{Z}} e^{2\pi in\tau/\beta} \sum_k \lam_{n,k} f_k(x) \; . \ee
Note the multi-valued part $\frac{2\pi m\tau}{q\beta}$, and that $\lam_{0,0} \sim \lam_{0,0} + \frac{2\pi}{q}\sqrt{\beta V_\Sigma}$ with $V_\Sigma$ the volume of $\Sigma$ while the other $\lam_{n,k}$ are single-valued. Finally we also expand
\be A_\tau(\tau,x) = \beta^{-1/2} \sum_{n\in\mathbb{Z}} e^{2\pi in\tau/\beta} \sum_k A_{\tau,n,k} f_k(x) \; . \ee
Note that with $\dim M = D$ we have dimensions $[\phi] = [\lam] = \frac{D-4}{2}$ and $[A_\tau] = \frac{D-2}{2}$, and therefore $[\phi_{n,k}] = [\lam_{n,k}] = -2$ and $[A_{\tau,n,k}] = -1$. Their measures are
\be \mc{D}\phi = \prod_{n\in\mathbb{Z},k\ne 0} \frac{\mu^2}{\sqrt{2\pi}} d\phi_{n,k} \; , \qquad \mc{D}\lam = \prod_{n\in\mathbb{Z},k} \frac{\mu^2}{\sqrt{2\pi}} d\lam_{n,k} \; , \qquad \mc{D}A_\tau = \prod_{n\in\mathbb{Z},k} \frac{\mu}{\sqrt{2\pi}} dA_{\tau,n,k} \ee
where $\mu$ is a factor with dimensions of mass, needed to keep the measure dimensionless. The Jacobian for our Coulomb decomposition is
\be \ba \mc{D} A & = \mc{D} A_\tau \mc{D} A_i \\
& = \mc{D} A_\tau \mc{D} \tilde A_i \mc{D}(\hat\nabla_i \phi) \\
& = \mc{D} A_\tau \mc{D} \tilde A_i \mc{D}\phi \, {\rm det}'(\mu^{-2}\Delta_\Sigma)^{\infty/2} \; . \ea \ee
We find a factor of the spatial Laplacian determinant, with zero mode omitted, for each Fourier mode $n$ on $S^1$. See appendix \ref{appendix:nophase} for other details about the measure. The full measure is
\be \frac{\mc{D} A}{\mc{G}_M} = \frac{1}{\mc{G}_M^0} \frac{\mc{D} A_\tau}{\mc{G}_M^{\rm multi}} \mc{D} \tilde A_i \, {\rm det}'(\mu^{-2}\Delta_\Sigma)^{\infty/2} \ee
where $\mc{G}_M^0$ captures the spatially constant and single-valued, i.e. $k=m=0$, part of $\lam$ and $\mc{G}_M^{\rm multi}$ captures the multi-valued, i.e. $\frac{2\pi m\tau}{q\beta}$, part of $\lam$. One can show that $A_\tau$ and $\tilde A_i$ decouple in the action. See appendix \ref{appendix:nophase} for details. The $A_\tau$ term's kinetic operator is the spatial scalar Laplacian. Integrating out the spatially non-constant, i.e. $k\ne 0$, part of $A_\tau$ then gives a determinant for each Fourier mode $n$, cancelling the Jacobian determinant already present. The remaining spatially constant part of $A_\tau$ drops out of the action and is integrated freely. We parametrize the $n\ne 0$ part of the $k=0$ part of $A_\tau$ as the time derivative of the auxiliary field $\sigma = \beta^{-1/2} \sum_{n\ne 0} e^{2\pi in\tau/\beta} \sig_n$. The associated Jacobian is ${\rm det}'(\mu^{-1} \p_\tau)$, whose zeta regularization is simply $\mu\beta$, and we can cancel all of $\sig$ against the $n\ne 0$ part of the $k=0$ part of $\lambda$. This leaves
\be Z(\beta) = \left(\frac{\mu\beta \int_{\mathbb R} \frac{\mu \, dA_{\tau,0,0}}{\mc{G}_M^{\rm multi}}}{\int_{U(1)} \mu^2 d\lam_{0,0}}\right) \int \mc{D}\tilde A_i \, e^{-S[\tilde A_i]} \; . \ee
The action of $\mc{G}_M^{\rm multi}$ quotients the integral of $dA_{\tau,0,0}$ over $\mathbb{R}$ to one over $U(1)$, represented as $0 \le A_{\tau,0,0} < \frac{2\pi}{q\beta} \sqrt{\beta V_\Sigma}$. Recalling also from above that $\lam_{0,0} \sim \lam_{0,0} + \frac{2\pi}{q}\sqrt{\beta V_\Sigma}$, we see that the factor in parentheses cancels out, leaving
\be Z(\beta) = \int \mc{D}\tilde A_i \, e^{-S[\tilde A_i]} \ee
with no prefactor. This is precisely the expression one would get from the covariant phase space formalism. Since it is manifestly equal to the configuration space path integral formulation of the partition function, the covariant phase space formalism apparently ``knows" about the constant gauge transformation.

\section{Phase space path integral derivation of $Z_{\rm edge}(\beta)$}
\label{app:PIZedge}

In this appendix we compute $Z_{\rm edge}(\beta)$ for Maxwell on a static manifold with the DEM boundary condition using the phase space path integral, recovering the answer obtained in the body of the paper using a canonical trace. Note that in Euclidean signature we have $\p M = S^1 \times \p\Sigma$. Previously we mostly focused on a single Cauchy slice, but in this context the fields $\alpha, E_\perp$ are functions on $\p M$ satisfying $\int_{\p\Sigma} \alpha = \int_{\p\Sigma} E_\perp = 0$ at every fixed time. The Euclidean action in Hamiltonian form is
\be S_{\rm edge} = \int d\tau \int_{\p\Sigma} \left( -i E_\perp \dot\alpha + \half E_\perp \frac{1}{K} E_\perp \right) \ee
and the path integral measure is read off from the symplectic form. The partition function is then
\be Z_{\rm edge}(\beta) = \int \mc{D} \alpha(\tau,x) \, \mc{D} E_\perp(\tau,x) \, \exp[-\int d\tau \int_{\p\Sigma} \left( -i E_\perp \dot\alpha + \half E_\perp \frac{1}{K} E_\perp \right)] \, . \ee
We have explicitly included the arguments of $\alpha, E_\perp$ in the measure as a reminder that they are functions on $\p M = S^1 \times \p\Sigma$. After integrating $\p_\tau$ by parts to replace $E_\perp \dot\alpha \to -\dot E_\perp\alpha$ we see that the temporally non-constant part of $\alpha$ acts as a Lagrange multiplier setting $\dot E_\perp=0$, and the temporal zero mode of $\alpha$ is integrated over freely, giving a factor of $|\mc{G}'_{\p\Sigma}|$. This leaves
\be Z_{\rm edge}(\beta) = |\mc{G}'_{\p\Sigma}| \int \mc{D} E_\perp(x) \, \exp[-\half\beta \int_{\p\Sigma} E_\perp \frac{1}{K} E_\perp ] \ee
where now $E_\perp$ only depends on $x$. This is now recognizable as $\Tr e^{-\beta H_{\rm edge}}$. Compare with formulas in section \ref{sec:EdgeP}.

\section{Scalar $Z(S^2)$ without zero mode}
\label{appendix:zmode}

In this appendix we study the 2D minimally coupled scalar and explicitly demonstrate the effect on $Z(S^2)$ of removing the zero mode from the theory. We take the sphere to have radius $R$, and the scalar to have periodicity $\phi \sim \phi + \frac{2\pi}{q}$. The action is
\be S[\phi] = -\half\int_{S^2} \phi\nabla^2\phi \; . \ee
To compute the partition function 
\be \ba Z(S^2) & = \int \mc{D}\phi \, e^{-S[\phi]} \; , \label{appeq:compactPI}\ea \ee
we orthonormally expand our field
\be \phi = \sum_{l=0}^\infty \sum_{m=-l}^l\phi_{lm} Y_{lm} \label{appeq:Ylmexpand}\ee
in terms of the usual spherical harmonics 
\be -\nabla^2 Y_{lm} = \lam_l Y_{lm}\; , \quad  \lambda_l=l(l+1)\; , \qquad \int_{S^2} Y_{lm}Y_{l'm'} = \delta_{ll'} \delta_{mm'} \; . \ee
We have picked a basis where $Y_{lm}$ are real. The normalized zero mode is
\be Y_{00} = \frac{1}{\sqrt{4\pi R^2}} \, . \ee
The mode coefficients $\phi_{lm}$ in \eqref{appeq:Ylmexpand} are real-valued except for $\phi_{00}$, which is periodic:
\be \phi_{00} \sim \phi_{00} + \frac{2\pi}{q} \sqrt{4\pi R^2} \; . \label{appeq:period} \ee
The action in terms of these modes is
\be S = \half\sum_{l=0}^\infty \sum_{m=-l}^l \lam_l \phi_{lm}^2 \; . \label{appeq:actionmode}\ee
We use the path integral measure
\be \mc{D}\phi \equiv \prod_{l=0}^\infty \prod_{m=-l}^l\frac{\mu}{\sqrt{2\pi}} d\phi_{lm} \label{appeq:compactmeasure}\ee
where $\mu$ is a mass scale introduced to keep the measure dimensionless. Using \eqref{appeq:period}, \eqref{appeq:actionmode} and \eqref{appeq:compactmeasure}, we compute the partition function \eqref{appeq:compactPI} to obtain
\be \ba Z(S^2) = \left( \mu \int_0^{\frac{2\pi}{q}\sqrt{4\pi R^2}} \frac{d\phi_{00}}{\sqrt{2\pi}} \right) \prod_{l\ne 0} \sqrt{\frac{\mu^2}{\lam_l^{2l+1}}}  = \frac{\mu}{q} \sqrt{8\pi^2 R^2} \, {\rm det}' \left( \frac{-\nabla^2}{\mu^2} \right)^{-1/2} \ea \ee
where the prime in $\det'$ indicates omission of the zero mode from the determinant. In this form one sees clearly that the zero mode contributes $\frac{\mu}{q}\sqrt{8\pi^2 R^2}$ to the partition function. The functional determinant can be evaluated in various ways. Zeta regularization gives \cite{Weisberger:1987kh}
\be -\half \log {\rm det}' \left( \frac{-\nabla^2}{\mu^2} \right) = -\frac{2}{3} \log \mu R + \text{non-universal} \; . \ee
This is the log term for the theory with the zero mode omitted. Combining with the zero mode contribution gives the full partition function,
\be \ba \log Z(S^2) & = \log \frac{\mu R}{q} - \frac{2}{3} \log \mu R + \text{non-universal} \\
& = \frac{1}{3} \log \mu R + \log \frac{1}{q} + \text{non-universal} \; . \ea \label{appeq:compactlog}\ee
We see that the overall coefficient of $\log\mu R$ is $\frac{1}{3}$, which is consistent with the general rule for 2D CFT that $\log Z(S^2) \sim \frac{c}{3} \log \mu R$ where $c$ is the central charge. 


\section{$dS_D$ static patch and $S^{D}$ partition functions with no brick wall}\label{app:sphere_PI}

In this appendix, we extract the relevant calculations in \cite{Anninos:2020hfj} and adapt them to our notations. For notational simplicity, we will drop the superscripts ``ADLS" that were present in section \ref{sec:MaxwellSD}.

\subsection{(Quasi)canonical bulk partition function in $dS_D$ static patch}\label{app:quasicanonical}

An object that plays a prominent role in \cite{Anninos:2020hfj} is the Harish-Chandra character $\chi(g)$, defined as a trace of a group element $g$ over the representation space associated with unitary irreducible representations (UIRs) of the isometry group $SO(1,D)$ of $dS_{D}$. The important point is that the character associated with the $dS$ boost, $g=e^{-i \hat H t}$, admits an expansion in terms of quasinormal modes (QNMs) in the static patch \cite{Sun:2020sgn}
\begin{align}\label{eq:character}
    \chi(t) = \sum_z D_z \, e^{-iz|t|} \; .
\end{align}
Here $z$ labels QNM frequencies and $D_z$ their degeneracies. Here are some examples that are relevant to our discussions in the main text:
\begin{itemize}
    \item Scalar with mass $m^2 R^2 = \Delta \bar \Delta = \Delta \left( D-1-\Delta\right) >0$ 
    \begin{align}\label{appeq:massivescalarchar}
    \chi(t) =  \frac{e^{-\Delta \frac{t}{R}}+e^{-\bar \Delta \frac{t}{R}}}{\left|1-e^{-\frac{t}{R}}\right|^{D-1}}
    \end{align}

    \item Massless scalar 
    \begin{align}\label{appeq:masslesschar}
    \chi(t) =  \frac{1+e^{-(D-1) \frac{t}{R}}}{\left|1-e^{-\frac{t}{R}}\right|^{D-1}} -1 
    \end{align}

    \item Massless spin-1 field
    \begin{align}\label{appeq:maxchar}
    \chi(t)= (D-1) \frac{e^{-\frac{t}{R}}+e^{-(D-2)\frac{t}{R}}}{\left|1-e^{-\frac{t}{R}}\right|^{D-1}}-\frac{1+e^{-(D-1) \frac{t}{R}}}{\left|1-e^{-\frac{t}{R}}\right|^{D-1}}+1
    \end{align}

    \item Spin-1 field with mass $m^2 R^2 = (\Delta -1)(\bar \Delta -1)= (\Delta -1)(D-2- \Delta)$
    \begin{align}\label{appeq:procachar}
    \chi(t) = (D-1) \frac{e^{-\Delta \frac{t}{R}}+e^{-\bar \Delta \frac{t}{R}}}{\left|1-e^{-\frac{t}{R}}\right|^{D-1}}
    \end{align}
\end{itemize}
For the scalar case, the characters such as \eqref{appeq:massivescalarchar} or \eqref{appeq:masslesschar} have recently been understood in terms of real-time correlators in the static patch \cite{Grewal:2024emf}. 

A key observation in \cite{Anninos:2020hfj} is that the Fourier transform 
\begin{align}\label{introeq:doschar}
    \tilde\rho^\text{dS}(\omega)\equiv \int_{-\infty}^\infty \frac{dt}{2\pi}e^{i\omega t} \chi(t) 
\end{align}
can be interpreted as a spectral density for the single-particle static patch Hamiltonian, leading to a definition of the thermal canonical partition function
\begin{align}\label{appeq:idealgas}
    \log Z_{\rm bulk} \equiv \log \widetilde \Tr \, e^{-2\pi R \hat H} \equiv -\int_0^\infty d\omega \, \tilde\rho^\text{dS}(\omega) \log \left(e^{\pi R \omega }-e^{-\pi R \omega } \right) 
\end{align}
for free bosonic fields in a static patch. Substituting \eqref{introeq:doschar} into \eqref{appeq:idealgas} gives the formula
\begin{align}\label{appeq:Zbulk}
    \log Z_\text{bulk} = \int_0^\infty \frac{dt}{2t}\frac{1+e^{-\frac{t}{R}}}{1-e^{-\frac{t}{R}}} \chi(t) \; . 
\end{align}
This integral is UV-divergent in the region $t\to 0$, which can be regulated in various ways. As explained in the appendix C in \cite{Anninos:2020hfj}, the scheme-independent part of \eqref{appeq:Zbulk} can be evaluated as\footnote{For massless fields, there could be additional IR divergences in the region $t\to \infty$, which we will postpone discussing until appendix \ref{app:compactSD}.}
\begin{align}\label{appeq:ZbulkIR}
    \log Z_\text{bulk} = \zeta'(0)+\alpha^{\rm bulk}_{D} \log \mu R \; , \qquad \zeta(z)\equiv \frac{1}{\Gamma(z)} \int_0^\infty \frac{dt}{2t}t^z \frac{1+e^{-\frac{t}{R}}}{1-e^{-\frac{t}{R}}} \chi(t) \; . 
\end{align}
In \eqref{appeq:ZbulkIR}, $\alpha_{D}$ is read off as the coefficient of $\frac{1}{t}$ in the small $t$ expansion of the integrand \eqref{appeq:Zbulk}:
\begin{align}\label{appeq:readoff}
    \log Z_\text{bulk} = \int_0^\infty dt \left( \cdots + \frac{\alpha^{\rm bulk}_{D}}{t}+\cdots \right) \, . 
\end{align}
In even $D$, the first term $\zeta'(0)$ in \eqref{appeq:ZbulkIR} can be absorbed into the second term by a redefinition of $\mu$, making the former ambiguous, so that the coefficient $\alpha^{\rm bulk}_{D}$ is the scheme-independent quantity. In odd $D$, we have $\alpha^{\rm bulk}_{D}=0$, so that the first term is unambiguously defined. 

For the examples \eqref{appeq:massivescalarchar}-\eqref{appeq:procachar}, one can easily check using Mathematica that $\alpha^{\rm bulk}_{D}=0$ in odd $D$, while for a compact scalar or Maxwell on $S^D$, one obtains in even $D$
\begin{center}
	\begin{tabular}{ |c|c|c|c|c|c| } 
		\hline
		$D$ & 2 & 4 & 6 & 8 &10  \\
		\hline
		$\alpha_D^{\rm compact, bulk}$ & $\frac13$ & $\frac{29}{90}$ & $\frac{1139}{3780}$ & $\frac{32377}{113400}$& $\frac{2046263}{7484400}$\\ 
        \hline
        $\alpha_D^{\rm Max, bulk}$ & &$-\frac{16}{45}$ & $-\frac{331}{945}$ & $-\frac{1592}{4725}$ & $-\frac{303601}{935550}$\\ 
		\hline
	\end{tabular}
\end{center}
Note that in $D=2$, $\alpha_2^{\rm compact, bulk}$ coincides with the negative of the coefficient \eqref{eq:Sedgerequired} or \eqref{appeq:compactlog}, while in $D=4$, $\alpha_4^{\rm Max, bulk}$ coincides with the coefficient \eqref{eq:Sbulk}. We did not include $\alpha_2^{\rm Max, bulk}$ because the $D=2$ case exhibits special features and therefore we treat it separately in section \ref{sec:2DMaxwell}.

\subsection{Compact scalar on $S^{D-2}$ ($D\geq 4$)}\label{app:compactSD}

In this appendix, we consider a free compact scalar with periodicity
\begin{align}
    \phi \sim \phi +\frac{2\pi }{q} \; .
\end{align}
The $S^{D-2}$ partition function for such a scalar is \cite{Law:2020cpj}
\begin{align}\label{appeq:compactDPI}
    Z_{\rm PI}^{\rm compact} \left( S^{D-2}\right)= {\rm det}' \left( - \frac{\nabla_{(0)}^2}{\mu^2} \right)^{-1/2}  {\rm Vol}(G)_{\rm PI}\; , \qquad {\rm Vol}(G)_{\rm PI} = \frac{\sqrt{2\pi \mu^2 R^{D-2} {\rm Vol}\left(S^{D-2}\right)} }{q} \; . 
\end{align}
Here $R$ is the radius of the round $S^{D-2}$ and $\text{Vol}(S^{D-2}) = \frac{2\pi^{\frac{D-1}{2}}}{\Gamma\left( \frac{D-1}{2}\right)}$ is the volume for a unit round $S^{D-2}$. The Laplacian $-\nabla_{(0)}^2$ on $S^{D-2}$ has eigenvalues and degeneracy 
\begin{align}\label{appeq:scalarlapeigen}
    \lambda_l = \frac{l(l+D-3)}{R^2} \; , \qquad D^{SO(D-1)}_l = \frac{D+2l-3}{D-3} \binom{D+l-4}{D-4} \; , \qquad l\geq 0 \; . 
\end{align}
The notation $D^{SO(D-1)}_l$ means that this is the dimension for a $SO(D-1)$ UIR corresponding to Young diagrams with a single row of $l$ boxes. Prime means that we exclude the zero mode $l=0$ from the determinant. For the purpose of comparing with \cite{Anninos:2020hfj}, we put \eqref{appeq:compactDPI} into the heat kernel regularized form \cite{Vassilevich:2003xt}
\begin{align}\label{eq:ZPIcompactheatkernel}
	\log  Z_{\rm PI}^{\rm compact} \left( S^{D-2}\right) = \log {\rm Vol} (G)_{\rm PI} + \int_0^\infty  \frac{d\tau}{2\tau} e^{-\epsilon^2/4\tau}\Tr' \, e^{-\left( -\nabla_{(0)}^2\right) \tau  } \; . 
\end{align}
Focusing on quantities that are independent of the regularization scheme, we can simply identify the parameter $\mu$ in \eqref{appeq:compactDPI} with the heat kernel regulator through $\mu = \frac{2e^{-\gamma}}{\epsilon}$. Following the analogous manipulations in appendices G.2 and G.3 in \cite{Anninos:2020hfj}, we can recast \eqref{eq:ZPIcompactheatkernel} into the form
\begin{align}
     Z_{\rm PI}^{\rm compact} \left( S^{D-2}\right) = Z_{\rm G} Z_{\rm char}
\end{align}
where (formally taking $\epsilon=0$)
\begin{align}
    \log Z_{\rm G} =& -\log \frac{ q}{\sqrt{2\pi \mu^2 R^{D-2} {\rm Vol}\left(S^{D-2}\right)} }-\int_0^\infty \frac{dt}{2t} \left[ 1+e^{- (D-3)\frac{t}{R}}\right] \nn \, , \\
    \log Z_{\rm char} =& \int_0^\infty\frac{dt}{2t}\frac{1+e^{-\frac{t}{R}}}{1-e^{-\frac{t}{R}}} \frac{1+e^{-(D-3) \frac{t}{R}}}{\left( 1-e^{-\frac{t}{R}}\right)^{D-3}} \; .
\end{align}
Rearranging, we have
\begin{align}\label{appeq:compactPIresult}
    \log  Z_{\rm PI}^{\rm compact} \left( S^{D-2}\right) =&\int_0^\infty\frac{dt}{2t}\frac{1+e^{-\frac{t}{R}}}{1-e^{-\frac{t}{R}}}\left[  \frac{1+e^{-(D-3) \frac{t}{R}}}{\left( 1-e^{-\frac{t}{R}}\right)^{D-3}} -1\right] \nn\\
    &-\log \frac{ q}{\sqrt{2\pi \mu^2 R^{D-2} {\rm Vol}\left(S^{D-2}\right)} } + \int_0^\infty\frac{dt}{2t}\left[\frac{1+e^{-\frac{t}{R}}}{1-e^{-\frac{t}{R}}} -1 -e^{-(D-3)\frac{t}{R}}\right] \; . 
\end{align}
The quantity in brackets on the first line is the character \eqref{appeq:masslesschar} for a massless scalar in $dS_{D-2}$. Although the compact scalar has no IR divergences, the individual integrals in \eqref{appeq:compactPIresult} do. We can temporarily regulate them by inserting a factor of $e^{-\delta t}$ into their integrands, and letting $\delta \to 0$. Since the total integral is free of such divergences, the $\delta$-dependence must drop out at the end. For the UV-divergences, one can check that upon evaluating the integrals using the prescriptions \eqref{appeq:ZbulkIR}, the $\mu^2 R^2$ factors on the second line of \eqref{appeq:compactPIresult} from both terms cancel.

\subsection{Maxwell path integral on $S^{D}$ ($D\geq 3$)}\label{app:ADLSmax}

The path integral for Maxwell on $S^D$ in general $D\geq 3$ is \cite{Giombi:2015haa}
\begin{align}\label{appeq:MaxPID}
	Z_\text{PI}=\frac{1}{{\rm Vol}(G)_{\rm PI}} \frac{ \det'\left(\frac{-\nabla_{(0)}^2}{\mu^2}\right)^{1/2}}{\det(\frac{-\nabla_{(1)}^2 + D-1}{\mu^2})^{1/2}} \;, \qquad  \frac{1}{{\rm Vol}(G)_{\rm PI}}= \frac{q}{ \sqrt{2\pi \mu^4 R^D\text{Vol}(S^D)}} \; .
\end{align}
Here $R$ is the radius of the round sphere, $\text{Vol}(S^D) = \frac{2\pi^{\frac{D+1}{2}}}{\Gamma\left( \frac{D+1}{2}\right)}$ is the volume for a unit round $S^D$, $q$ the fundamental charge, $\mu$ is a parameter with dimension of mass. $\nabla_{(0)}^2$ and $\nabla_{(1)}^2$ are the scalar and transverse vector Laplacians on the unit round $S^D$. The eigenvalues and degeneracies of the former are given by shifting $D\to D+2$ in \eqref{appeq:scalarlapeigen}, while those for the latter are \cite{Rubin:1983be}
\begin{align}\label{appeq:veceigen}
    \lambda_{l,1} = \frac{l(l+D-1)-1}{R^2} \; , \quad D^{SO(D+1)}_{l,1} = \frac{l (D+l-1) (D+2 l-1) (D+l-3)!}{(D-2)! (l+1)!}\; , \quad l\geq 1 \; . 
\end{align}
These notations originate from the fact that the eigenfunctions of the transverse vector Laplacian on $S^D$ furnish $SO(D+1)$ UIRs corresponding to two-row Young diagrams with $l$ boxes in the first row and one box in the second row. 

Following appendices G.2 and G.3 in \cite{Anninos:2020hfj}, the heat kernel regularized form for \eqref{appeq:MaxPID} 
\begin{align}\label{eq:ZPIheatkernel}
	\log Z_\text{PI} = \log \frac{1}{{\rm Vol} (G)_{\rm PI}} + \int_0^\infty  \frac{d\tau}{2\tau} e^{-\epsilon^2/4\tau}\left[ \Tr\, e^{-\left( -\nabla_{(1)}^2 + D-1\right) \frac{\tau}{R^2} } - \Tr' \, e^{-\left( -\nabla_{(0)}^2\right) \frac{\tau}{R^2}  }\right]  
\end{align}
(again identifying $\mu = \frac{2e^{-\gamma}}{\epsilon}$), can be cast into the form
\begin{align}\label{appeq:MaxPIseparata}
     Z_\text{PI}=&  Z_\text{char}Z_\text{G} = Z_\text{bulk}   Z_\text{edge}Z_\text{G} \; . 
\end{align}
In this expression,
\begin{align}\label{appeq:Zbulkorigin}
    \log Z_\text{bulk} = \int_0^\infty \frac{dt}{2t}\frac{1+e^{-\frac{t}{R}}}{1-e^{-\frac{t}{R}}} \left[ (D-1) \frac{e^{-\frac{t}{R}}+e^{-(D-2)\frac{t}{R}}}{\left( 1-e^{-\frac{t}{R}}\right)^{D-1}}-\frac{1+e^{-(D-1) \frac{t}{R}}}{\left( 1-e^{-\frac{t}{R}}\right)^{D-1}}+1\right]
\end{align}
is the quasicanonical bulk partition function \eqref{appeq:idealgas} with the massless spin-1 character \eqref{appeq:maxchar}. 

The factor $Z_\text{edge}$ in \eqref{appeq:MaxPIseparata} is defined as the reciprocal of that in \cite{Anninos:2020hfj}, i.e. 
\begin{align}\label{appeq:Zedgeorigin}
    \log Z_\text{edge} = -\int_0^\infty\frac{dt}{2t}\frac{1+e^{-\frac{t}{R}}}{1-e^{-\frac{t}{R}}} \left[ \frac{1+e^{-(D-3) \frac{t}{R}}}{\left( 1-e^{-\frac{t}{R}}\right)^{D-3}}-1\right] \; .
\end{align}
Note that there is an overall minus sign. The factor
\begin{align}\label{appeq:Gfactor}
    \log Z_\text{G} =& \log \frac{1}{{\rm Vol}(G)_{\rm PI}} + \int_0^\infty\frac{dt}{2t} \left[ -2\cdot \frac{1+e^{-\frac{t}{R}}}{1-e^{-\frac{t}{R}}}+2+e^{- (D-1)\frac{t}{R}}+e^{- (D-3)\frac{t}{R}} \right] 
\end{align}
is associated with the constant gauge transformation. 

To proceed, we partially compute the integrals using the prescriptions \eqref{appeq:ZbulkIR} to rewrite 
\begin{align}\label{appeq:Gfactorscompute}
    \log Z_\text{G} =& \log \frac{q}{ \sqrt{ 2\pi \mu^2 R^{D-2}\text{Vol}(S^{D-2})}}- \int_0^\infty\frac{dt}{2t} \left[\frac{1+e^{-\frac{t}{R}}}{1-e^{-\frac{t}{R}}}- 1-e^{- (D-3)\frac{t}{R}}\right]
\end{align}
where we have used $\text{Vol}(S^{D})=\frac{2\pi}{D-1}\text{Vol}(S^{D-2})$. Here these equalities are up to scheme-dependent terms. If we combine \eqref{appeq:Zedgeorigin} and \eqref{appeq:Gfactorscompute} so that
\begin{align}\label{appeq:Zedgeredefine}
    \tilde Z_\text{edge} = Z_\text{edge} Z_\text{G} \; ,
\end{align}
we would have
\begin{align}
    \log \tilde Z_\text{edge} =&-\int_0^\infty\frac{dt}{2t}\frac{1+e^{-\frac{t}{R}}}{1-e^{-\frac{t}{R}}}\left[  \frac{1+e^{-(D-3) \frac{t}{R}}}{\left( 1-e^{-\frac{t}{R}}\right)^{D-3}} -1\right] \nn\\
    &+\log \frac{ q}{\sqrt{2\pi \mu^2 R^{D-2} {\rm Vol}\left(S^{D-2}\right)} } - \int_0^\infty\frac{dt}{2t}\left[\frac{1+e^{-\frac{t}{R}}}{1-e^{-\frac{t}{R}}} -1 -e^{-(D-3)\frac{t}{R}}\right] \; ,
\end{align}
which is the negative of \eqref{appeq:compactPIresult}. In other words, the redefined edge partition function \eqref{appeq:Zedgeredefine} is exactly the same as the $S^{D-2}$ partition function for a free compact scalar with period $\frac{2\pi}{q}$, i.e. 
\begin{align}
    \tilde Z_\text{edge} = \frac{1}{Z_{\rm PI}^{\rm compact} \left( S^{D-2}\right)} \;. 
\end{align}

\subsection{Proca path integral on $S^{D}$ ($D\geq 3$)}\label{appsec:proca}

The path integral for Proca with mass $m^2R^2=(\Delta-1)(\bar\Delta-1)= (\Delta -1)(D-2- \Delta )$ on $S^D$ is \cite{Law:2020cpj}
\begin{align}\label{appeq:ProcaPI}
 	Z_\text{PI}=\frac{m}{\mu} \det\left(\frac{-\nabla_{(1)}^2 +m^2 + \frac{D-1}{R^2}}{\mu^2}\right)^{-1/2} \; .
\end{align}
Here $\nabla_{(1)}^2$ is the transverse vector Laplacians on the round $S^D$ with radius $R$, whose eigenvalues and degeneracies are given in \eqref{appeq:veceigen}. Notice that there is an overall factor of $m$.

Similar to the case of a compact scalar or Maxwell, starting with a heat-kernel regularized form, we can massage \eqref{appeq:ProcaPI} (following appendix F in \cite{Anninos:2020hfj}) into the form
\begin{align}
    Z_\text{PI}= Z_{\rm bulk}Z_{\rm edge}\;.
\end{align}
Here $Z_{\rm bulk}$ is the quasicanonical partition function \eqref{appeq:idealgas} with the massive spin-1 character \eqref{appeq:procachar}; $Z_{\rm edge}$ takes the form of a path integral for a ghost scalar with mass $m^2$ on $S^{D-2}$:
\begin{gather}
     \log  Z_{\rm edge} = -\int_0^\infty \frac{dt}{2t}\frac{1+e^{-\frac{t}{R}}}{1-e^{-\frac{t}{R}}}\frac{e^{-(\Delta-1) \frac{t}{R}}+e^{-(\bar \Delta-1) \frac{t}{R}}}{\left(1-e^{-\frac{t}{R}}\right)^{D-3}} \; .
 \end{gather}


\bibliographystyle{utphys}
\bibliography{ref}

\end{document}